\documentclass[11pt,a4paper]{article}
\pdfoutput=1
\usepackage{jheppub}
\usepackage{amsthm,amsbsy,amsfonts,mathrsfs,enumerate,float,wrapfig,amsmath}
\usepackage[utf8]{inputenc}
\usepackage{subfigure}

\newcommand{\N}{\mathcal{N}}

\newcommand{\nn}{\nonumber}
\newcommand{\be}{\begin{equation}}
\newcommand{\ee}{\end{equation}}
\newcommand{\bea}{\begin{eqnarray}}
\newcommand{\eea}{\end{eqnarray}}
\newcommand{\ba}{\begin{align}}
\newcommand{\ea}{\end{align}}
\newcommand{\F}{\mathcal{F}}

\newcommand{\relu}[1]{\big\lceil\!\big|\, {#1} \,\big|\!\big\rceil} 
\newcommand{\ia}{\tilde{\mathfrak{a}}}

\usepackage{tikz}
\usetikzlibrary{positioning}
\usetikzlibrary{arrows}
\title{Complete prepotential for 5d $\mathcal{N}=1$ superconformal field theories}
\author[a]{Hirotaka Hayashi,}
\author[b]{Sung-Soo Kim,}
\author[c]{Kimyeong Lee,}
\author[d]{and Futoshi Yagi}
\affiliation[a]{Department of Physics, School of Science, Tokai University,\\ 4-1-1 Kitakaname, Hiratsuka-shi, Kanagawa 259-1292, Japan}
\affiliation[b]{School of Physics, University of Electronic Science and Technology of China, \\
No.4, Section 2, North Jianshe Road, Chengdu, Sichuan 610054, China}
\affiliation[c]{School of Physics, Korea Institute for Advanced Study, \\
85 Hoegi-ro Dongdaemun-gu, Seoul 02455, Korea}
\affiliation[d]{School of Mathematics, Southwest Jiaotong University,\\ 
West zone, High-tech district, Chengdu, Sichuan 611756, China}
\emailAdd{h.hayashi@tokai.ac.jp}
\emailAdd{sungsoo.kim@uestc.edu.cn}
\emailAdd{klee@kias.re.kr}
\emailAdd{futoshi\_yagi@swjtu.edu.cn}
\abstract{
For any 5d ${\cal N}=1$  superconformal field theory, we propose a ``complete'' prepotential which reduces to the perturbative prepotential for any of its  possible gauge theory realizations, manifests its global symmetry when written in terms of  the invariant Coulomb branch parameters, and is valid for the whole parameter region.  
 As concrete examples, we consider $SU(2)$ gauge theories with up to 7 flavors, $Sp(2)$ gauge theories with up to 9 flavors, and $Sp(2)$ gauge theories with 1 antisymmetric tensor and up to 7 flavors,  as well as their dual gauge theories. 
}
\begin{document}
\preprint{KIAS-1962}
\maketitle
\section{Introduction}\label{sec:intro}
A large class of five-dimensional (5d) $\mathcal{N}=1$ gauge theories have non-trivial fixed points. Novel phenomena such as dualities and global symmetry enhancements emerge at such conformal fixed point~\cite{Seiberg:1996bd, Morrison:1996xf,Douglas:1996xp, Intriligator:1997pq}. Two or more theories of different gauge theory descriptions can flow to the same fixed point and thus they are UV-dual in the sense that they flow to the same superconformal field theory (SCFT) at UV. For instance, 5d $Sp(N)$ gauge theories with $N_f$ hypermultiplets in the fundamental representations (flavors) and 5d $SU(N+1)_\kappa$ gauge theories of Chern-Simons (CS) level $\kappa=N+3-N_f/2$ with $N_f$ flavors are a typical example of such UV-duality~\cite{Gaiotto:2015una}. In particular, 5d $\mathcal{N}=1$ gauge theories of rank-2 gauge groups are completely classified with their dual partners and field contents~\cite{Jefferson:2018irk,Jefferson:2017ahm,Hayashi:2018lyv,Apruzzi:2019opn, Apruzzi:2019enx,Bhardwaj:2019jtr}. Such dual theories enjoy intriguing enhanced global symmetry, whose symmetry structure arises through non-trivial interplays between instanton particles and hypermultiplets, and is often checked from the index functions like superconformal index or partition function with shifted Coulomb branch parameters. 

Prepotential captures low energy effective descriptions of these SCFTs in Coulomb branch where gauge group is completely broken to $U(1)^r$ where $r$ is the rank of gauge group. Intriligator, Morrison, and Seiberg (IMS) proposed the explicit form of the prepotential, which is one-loop exact and at most cubic~\cite{Intriligator:1997pq}. We refer to this perturbative prepotential as the IMS prepotential. The IMS prepotential is readily determined from the gauge groups, the CS levels (if exit), and hypermultiplet contents. Thus it respects perturbative global symmetry from the hypermultiplets. The first derivative of the prepotential with respect to Coulomb branch moduli yields monopole string tension, and also the second derivative describes the effective coupling, which plays a role of the Coulomb branch metric. As there are no instanton contributions, the IMS prepotential is insensitive to the global symmetry enhancements.

As many of such 5d $\mathcal{N}=1$ theories can be engineered via Type IIB 5-brane webs \cite{Aharony:1997ju, Aharony:1997bh} or M-theory on Calabi-Yau (CY) threefold \cite{Katz:1996fh, Katz:1997eq}, brane configurations also provide a direct description of the prepotential. For instance, one can study CY geometry of 5d gauge theories to obtain their triple intersections which yield the prepotential of the 5d theories or one can also scan possible gauge theory descriptions from the geometry which lead to a classification of the UV-dual theories~\cite{Jefferson:2018irk, Bhardwaj:2018yhy, Bhardwaj:2018vuu, Apruzzi:2018nre, Bhardwaj:2019jtr, Apruzzi:2019vpe, Apruzzi:2019opn}. Though Type IIB 5-branes can be understood as dual description of the geometry, not all CY geometries can be realized as a 5-brane web. For those theories whose 5-brane configurations exist, UV-dual structure is more intuitive as they can be realized as an S-duality or resolution of orientifold planes, followed by Hanany-Witten transitions~\cite{Hanany:1996ie}. Enhanced global symmetry can be also read off from the 7-brane analysis~\cite{Gaberdiel:1997ud,Gaberdiel:1998mv, DeWolfe:1999hj}. Areas of the compact faces of a given 5-brane web correspond to monopole string tensions, from which one can readily reconstruct the prepotential. Equivalence of the areas of the compact faces from 5-brane webs and the prepotential for the corresponding theories is a necessary condition to find a new 5-brane diagram. See recent proposals of 5-brane webs for $SO(N)$ (7 $\le N \le$ 12) gauge theories with spinor matter~\cite{Zafrir:2015ftn}, for $G_2$ gauge theories with flavors~\cite{Hayashi:2018bkd,Hayashi:2018lyv}, for $SU(6)$ gauge theories with hypermultiplets in the rank-3 antisymmetric representation \cite{Hayashi:2019yxj} and also for 6d D-Type conformal matter on a circle~\cite{Hayashi:2015zka, Hayashi:2015vhy, Kim:2019dqn}.

A 5-brane web contains much more than just the prepotential as one can compute the partition functions \cite{Aganagic:2002qg, Iqbal:2007ii} (also leading to Gopakumar-Vafa (GV) invariants~\cite{Gopakumar:1998jq, Huang:2013yta}) and Seiberg-Witten curves based on it. 5-brane webs can be understood as a dual CY toric(-like \cite{Kim:2014nqa, Benini:2009gi}) diagram \cite{Leung:1997tw} and hence inherits M-theory configurations. One can deform 5-brane webs using the Hanany-Witten transitions and the flop transitions, and also apply $SL(2, \mathbb{Z})$ transformations. Through such transitions, one can reach all the parameter regions of the theories. 5-brane webs hence naturally capture both perturbative and non-perturbative aspects of the theory. On the other hand, since the IMS prepotential is valid in the perturbative regime where the gauge coupling is small, the IMS prepotential is incomplete in describing the theories as a whole including other phases of parameter regions.

In this paper, we attempt to extend and generalize the IMS prepotential to include non-perturbative regime and also to capture other allowed parameter regions where dual gauge theories are naturally realized. In other words, we construct ``{\it complete}'' prepotential over the extended K\"ahler cone\footnote{Here, the extended K\"ahler cone \cite{Witten:1996qb} refers to the enlarged K\"ahler moduli space \cite{Aspinwall:1993nu}.}, such that (i) it reduces to the perturbative prepotential of its gauge theory description if exists, (ii) it manifests its global symmetry, and (iii) it is valid for the whole parameter region.  

To this end, we use 5-brane webs as they inherit all the parameter regions. In particular, we introduce invariant Coulomb branch parameters that are invariant under the enhanced global symmetry \cite{Mitev:2014jza}. The complete prepotential expressed in terms of the invariant Coulomb branch parameters is manifestly invariant under the enhanced global symmetry. The mass parameters together with the instanton mass form invariant polynomials of the representation of the enhanced global symmetry. Along the way, we introduce a new notation in terms of the step function which is useful to keep track of flop transitions on a 5-brane web, motivated by the expression in \cite{Closset:2018bjz}. By considering all possible flops,\footnote{Flop invariant property is explored in a geometric setup based on the combined fiber diagram  \cite{Apruzzi:2019kgb}.} one can cover all the parameter regimes, which leads to the complete prepotential. In the week coupling limit, of course, this complete prepotential naturally reduces to the IMS prepotential. To exhaust the form of the complete prepotentials, we discuss different approaches and test against various consistency checks to support our complete prepotentials.

The organization of the paper is as follows. In section \ref{sec:rank1}, we demonstrate how to construct the complete prepotential for the rank-1 theory, the $SU(2)$ gauge theory with $N_f\le 7$ flavors, which is expressed as the representation of $E_{N_f+1}$ symmetry. In subsequent sections, we apply our method to rank-2 theories. As representative examples, we consider the $Sp(2)$ gauge theory with $N_f\le 9$ flavors in section \ref{sec:rank2} and the $Sp(2)$ gauge theory with one antisymmetric and $N_f\le 7$ flavors in section \ref{sec:rank2AS}. In these sections, we discuss other ways of obtaining the complete prepotential from geometry and from the GV invariants to support the form of the complete prepotentials. We also consider various consistency check like duality and RG flows. We conclude with possible applications, generalizations and restrictions. In Appendices we list the explicit form of the complete prepotentials that are mentioned in the main text. 
 
\bigskip
\section{Prepotential for Rank-1 theories}\label{sec:rank1}

In this section we first start from determining complete prepotentials for simple examples, namely rank-1 theories. 

\subsection{Complete prepotential}
\label{sec:rk1F}
A 5d $\mathcal{N}=1$ supersymmetric gauge theory with a gauge group $G$ has a Coulomb branch which is parametrized by the real scalar field $\phi$ in the vector multiplet. On the Coulomb branch, the gauge group is broken to $U(1)^{r_G}$ where $r_G$ is the rank of $G$. The prepotential governing the low-energy abelian theory is given by \cite{Seiberg:1996bd, Morrison:1996xf, Intriligator:1997pq}
\begin{align}
\mathcal{F}(\phi) = \frac{1}{2}m_0h_{ij}\phi_i\phi_j + \frac{\kappa}{6}d_{ijk}\phi_i\phi_j\phi_k + \frac{1}{12}\bigg(\sum_{r\in\text{roots}}\left|r\cdot \phi\right|^3 - \sum_f\sum_{w \in R_f}\left|w\cdot \phi +m_f\right|^3\bigg), \label{eq:prepotential}
\end{align}
where  $m_0$ is the inverse of the gauge coupling squared, $\kappa$ is the classical Chern-Simons level and $m_f$ is a mass parameter for the matter $f$. $r$ is a root of the Lie algebra $\mathfrak{g}$ associated to $G$ and $w$ is a weight of the representation $R_f$ of the Lie algebra $\mathfrak{g}$. Here, $h_{ij} = \text{Tr}(T_iT_j),\,  d_{ijk} = \frac{1}{2}\text{Tr}\left(T_i\{T_j, T_k\}\right)$ where $T_i$ are the Cartan generators of $\mathfrak{g}$. The terms with the absolute values in the prepotential are the one-loop exact quantum contributions. 
The first derivative of the prepotential $\frac{\partial \mathcal{F}}{\partial \phi_i}$ gives the monopole string tensions $T_i$, and a second derivative  $\frac{\partial^2 \mathcal{F}}{\partial \phi_i\partial \phi_j}$ yields the effective coupling $\tau_{\rm eff}$ which is the metric on the Coulomb branch. We call this prepotential the Intriligator-Morrison-Seiberg (IMS) prepotential for later convenience. 

As the IMS prepotential is a perturbative quantity, it is insensitive to non-perturbative phenomena such as global symmetry enhancements and UV dualities.  The discrete theta angle of 5d $Sp(N)$ theory is also not captured in the IMS prepotential. As a concrete example, consider the 5d $\mathcal{N}=1$ $SU(2)$ gauge theory with $N_f\le 7$ hypermultiplets in the fundamental representation (flavors). It has a perturbative global symmetry of $SO(2N_f)\times U(1)_I$, where $SO(2N_f)$ comes from $N_f$ flavor symmetry and $U(1)_I$ corresponds to the conserved symmetry of the instanton particle. In the infinite coupling limit, the theory becomes a superconformal field theory and enjoys the enhanced global symmetry $E_{N_f+1}\supset SO(2N_f)\times U(1)_I$ \cite{Seiberg:1996bd}.  For pure $SU(2)=Sp(1)$ gauge theory, manifest global symmetry is $U(1)_I$, and there are two distinct theories which differ by the theta angle $\theta =0, \pi$, called $SU(2)_0$ and $SU(2)_\pi$ gauge theories, respectively. Though they do not have any hypermultiplet, these two theories have different global symmetries at the conformal fixed point: the global symmetry for the $SU(2)_0$ gauge theory is enhanced to $SU(2) \supset U(1)_I$,  while that for the $SU(2)_\pi$ gauge theory remains as $U(1)_I$. They are also often referred to as the $E_1$ and $\widetilde{E}_1$ theories, respectively \cite{Morrison:1996xf}. The IMS prepotentials for the $E_1$ and $\widetilde{E}_1$ theories are same and given by 
\begin{align}\label{eq:pureSU2prep}
	\mathcal{F}_{SU(2)}=\frac12 m_0 \,a^2+
\frac43 a^3,
\end{align} 
where $m_0\ge 0$ is the instanton mass and 
the Coulomb parameter $a=\phi$ lies in the Weyl chamber $a\ge 0$.\footnote{Throughout this paper, we choose a naturally ordered Coulomb branch parameters with a judicious choice of Weyl chamber.}
As non abelian global symmetry 
is realized as the Weyl reflection on the mass parameters, 
one can see that the IMS prepotential \eqref{eq:pureSU2prep} does not show the $SU(2)$ symmetry manifestly.\footnote{In other words, the IMS prepotential \eqref{eq:pureSU2prep} is not manifestly invariant under $m_0\leftrightarrow - m_0$.} Rather it is of a $U(1)_I$ symmetry associated with $m_0$. The IMS prepotential \eqref{eq:pureSU2prep} hence does not capture the symmetry enhancement as it is. 
The corresponding monopole string tension $T$ is given by 
\begin{align}\label{eq:pureSU2T}
	T= \, m_0\, a + 4 \,a^2,
\end{align}
and the effective coupling $\tau_{\rm eff}$ is 
\begin{align}\label{eq:su2EffCoupling}
	\tau_{\rm eff}=\, m_0 + 8\, a\  .
\end{align} 

\begin{figure}[t]
\centering
\includegraphics[width=12cm]{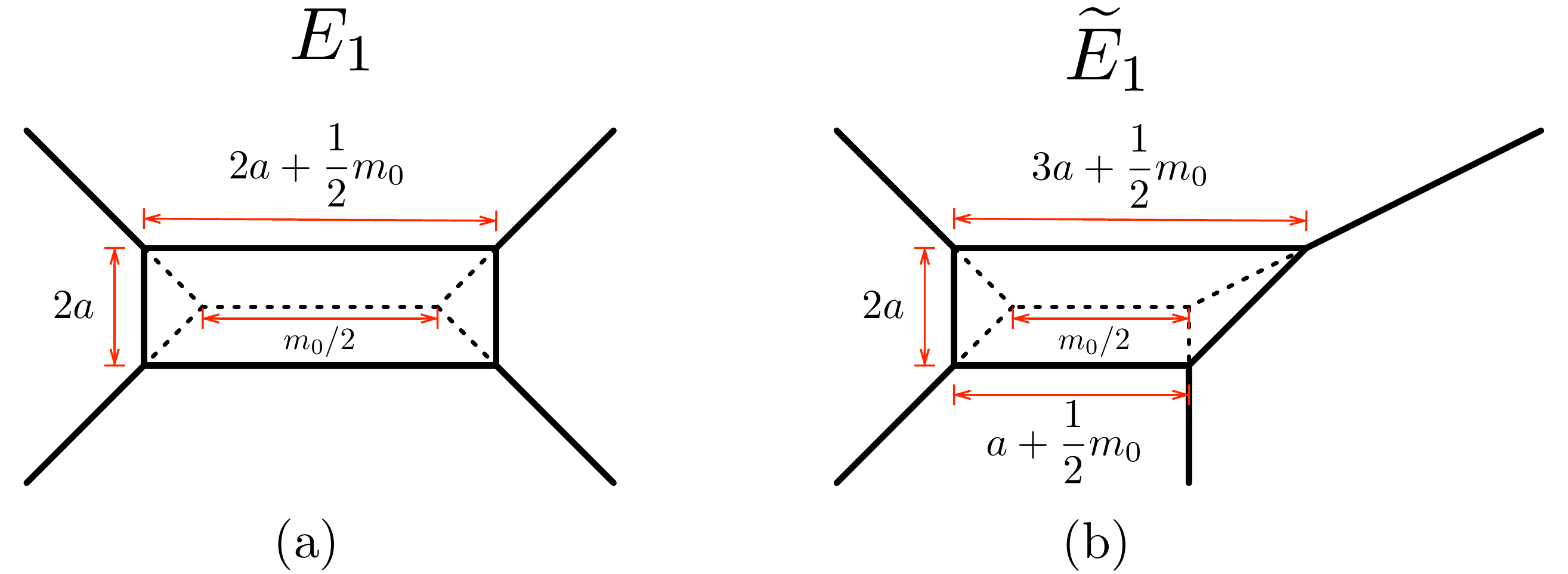}
\caption{(a) A 5-brane web for pure $SU(2)_0$ gauge theory (or $E_1$ theory). (b) A 5-brane web for pure $SU(2)_\pi$ gauge theory (or $\widetilde{E}_1$ theory).}
\label{fig:E1E1tilde}
\end{figure}
From the perspective of 5-brane webs in Type IIB string theory, it is also straightforward to obtain the prepotential. 
As a BPS configuration, 5-brane web is made of 5-branes of $(p,q)$ charges forming edges and faces as shown in Figure~\ref{fig:E1E1tilde}. Given a 5-brane web, one can associate the parameters of the 5-brane web diagram with the gauge theory parameters, $m_0, m_f,$ and $\phi_i$ (or equivalently $a_i$). Areas of the compact faces of a 5-brane correspond to the monopole string tension for the theory. As one can express the areas of the compact faces in terms of the gauge theory parameters, one can readily obtain the prepotential from a given 5-brane web. For instance, the 5-brane webs in Figure~\ref{fig:E1E1tilde} are pure $SU(2)$ theories with different discrete theta angles. It is easy to check that the area of the compact faces of these 5-branes in Figure ~\ref{fig:E1E1tilde} is the same as \eqref{eq:pureSU2T} and therefore yields the same prepotential as \eqref{eq:pureSU2prep}. 

5-brane configurations, of course, convey more information than just a prepotential. Though the IMS prepotential for the $E_1$ theory and that for the $\widetilde E_1$ theory are the same, their 5-brane webs are different in the sense that they are not related by any continuous deformations. In \cite{Mitev:2014jza, Aharony:1997bh}, it was discussed that the enhanced global symmetry can be captured in 5-brane webs as the fiber-base duality, with the introduction of the \emph{invariant} Coulomb branch parameter, which is the shifted Coulomb branch parameter such that it is invariant under the exchange of fiber and base K\"ahler parameters. For instance, one can see that the $SU(2)_0$ 5-brane web given in Figure \ref{fig:E1E1tilde}(a) can be symmetric under the reflection with respect to the $(1,1)$ 5-brane when the fiber K\"ahler parameter $Q_{\rm F}$ and the base K\"ahler parameter $Q_{\rm B}$ are suitably chosen such that $Q_{\rm F} \leftrightarrow Q_{\rm B}$. This fiber-base duality hence leads to the global symmetry enhancement to $E_1=SU(2)$. 

For the pure $SU(2)_0$ gauge theory, the invariant Coulomb branch parameter $\ia$ is given by \cite{Mitev:2014jza}
\begin{align}
\ia = a  + \frac{1}{8}\,m_0\ , 
\end{align}
with which the effective coupling \eqref{eq:su2EffCoupling} becomes independent of $m_0$. 
The prepotential for the $E_1$ theory is then expressed as
\begin{align}\label{eq:InvPrep4E1}
	\mathcal{F}_{E_1} = -\frac{1}{16}m_0^2 \,\ia + \frac43\, \ia^3 \ ,
\end{align}
where irrelevant constants are/will be neglected. 
It is clear that this new form of the prepotential \eqref{eq:InvPrep4E1} is invariant under $SU(2)$ Weyl reflection $m_0 \leftrightarrow -\, m_0$. The new prepotential, expressed in terms of the invariant Coulomb branch parameter, hence makes $SU(2)$ enhanced global symmetry manifest. The prepotential \eqref{eq:InvPrep4E1} is also valid for all possible phases of $m_0$. 
\begin{figure}[t]
\centering
\includegraphics[width=11cm]{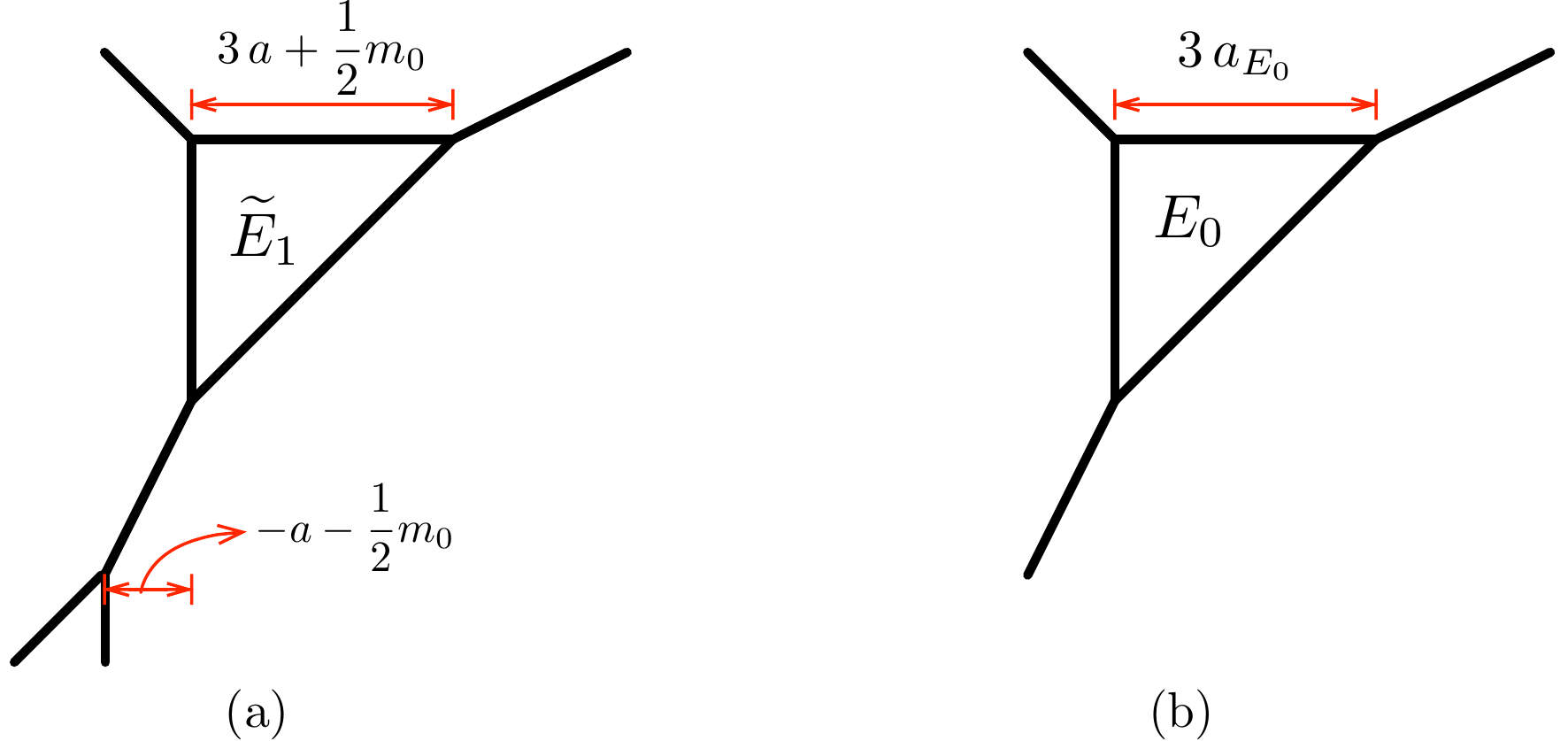}
\caption{(a) $\widetilde{E}_1$ theory with $a+\frac12m_0<0$. (b) $E_0$ theory}
\label{fig:E1tildeE0}
\end{figure}

For the pure $SU(2)_\pi$ gauge theory, as there is no enhanced global symmetry, no invariant Coulomb branch parameter is needed, and hence, for $m_0>0$, the prepotential \eqref{eq:pureSU2prep} remains unaltered.  For $m_0<0$, on the other hand, the theory can become non-Lagrangian as depicted in Figure \ref{fig:E1tildeE0}(a). In this phase, the IMS description is not valid. The prepotential can be nevertheless read off from the 5-brane web. 
The area of the compact face, corresponding to the monopole string tension, is 
\begin{align}
	T = \frac92 \big( \,a + \frac16 m_0 \big)^2 \ .
\end{align}
We note here that the Coulomb branch parameter $a$ is further restricted since the monopole tension $T$ should be non-negative. $\it i.e.$, $a> -\frac16 m_0$ for $m_0<0$. (See also \cite{Hayashi:2017btw}.) We call these allowed parameter regions the physical Coulomb branch \cite{Jefferson:2017ahm}.
The prepotential in the parameter region is then given by 
\begin{align}\label{eq:Prep4E1t}
	\mathcal{F}= \frac{3}{2} a^3 + \frac34 m_0\, a^2 + \frac18 m_0^2 \,a \ .
\end{align}
One can see that it is not invariant under $m_0\leftrightarrow-m_0$ and thus the corresponding global symmetry is still $U(1)_I$, as expected. 

As the $\widetilde{E}_1$ theory allows the decoupling (RG flows) to the $E_0$ theory, we can further check whether the prepotential \eqref{eq:Prep4E1t} reproduces that for the $E_0$ theory. By taking $m_0\to -\infty$, keeping $3a+\frac12 m_0$ finite, in Figure \ref{fig:E1tildeE0}(a), we find that the prepotential for the $E_0$ theory, obtained from the resulting web diagram  Figure \ref{fig:E1tildeE0}(b) is given by
\begin{align}\label{eq:InvPrep4E0}
	\mathcal{F}_{E_0} = \frac32\, (a_{E_0})^3 \ ,
\end{align}
where irrelevant constant is dropped and 
\begin{align}
	a^{E_0} = a_{\widetilde{E}_1} + \frac16 m_0{}_{\widetilde{E}_1}\ .
\end{align}

In order to incorporate all the regimes of the parameters $a, m_0$, we found that it is convenient to introduce the following new symbol, motivated by \cite{Closset:2018bjz},
\begin{align}\label{eq:doubleAbs}
\relu{x} 
	\equiv \theta (-x) \cdot x = \left\{ \begin{array}{cc}
		0 &\quad x >0\,; \\
		x &\quad x < 0\, , 
	\end{array}\right. 
\end{align}
with the Heaviside step function $\theta(x)$, which is related to the absolute value as
\begin{align}
\big|\,x\,\big| =x - 2\relu{x}\  \quad {\rm or}\quad  \big|\,x\,\big| =-x - 2\relu{-x}\ .
\end{align}
It is then straightforward to see that prepotential that includes all ranges of $m_0$ for the $\widetilde{E}_1$ theory takes the form
\begin{align}\label{eq:InvPrep4E1tilde}
	\mathcal{F}_{\widetilde{E}_1} = \frac{1}{2}m_0 \,a^2 +\frac43\, a^3+ \frac16 \relu{ a+\frac12 m_0 }^3  .
\end{align}
 The prepotential \eqref{eq:InvPrep4E1tilde} then becomes the perturbative prepotential for pure $SU(2)_\pi$ theory \eqref{eq:pureSU2prep} when $a\ge0$ and $a + \frac12 m_0 \ge 0$. \eqref{eq:InvPrep4E1tilde} gives rise to that of non-Lagrangian theory \eqref{eq:Prep4E1t} when $a + \frac12 m_0 \le 0$ and $a+\frac16 m_0\ge0$. These ranges come from the condition that each length of the edges in Figure \ref{fig:E1E1tilde}(b) and Figure \ref{fig:E1tildeE0}(a), or equivalently from each monopole string tension being positive. 


The prepotentials for the pure $SU(2)$ gauge theories with $\theta=0,\pi$, \eqref{eq:InvPrep4E1} and  \eqref{eq:InvPrep4E1tilde}, therefore, cover not only the usual perturbative (IMS prepotential) regime but also include non-perturbative regime. They are expressed in terms of the invariant Coulomb moduli and thus manifest in enhanced global symmetry. We call such new prepotentials {\it complete} prepotentials. Notice here that the first two terms in the complete prepotential for the $\widetilde{E}_1$ theory \eqref{eq:InvPrep4E1tilde} are just the IMS prepotential \eqref{eq:pureSU2prep}, and the last term is given as the 5-brane length cubed associated with the flop transitions in Figure \ref{fig:E1tildeE0}(a). 
The last term is interpreted as the contribution from the light instanton particle.
It is in fact true that the complete prepotential can be read off by carefully tracing out all possible flops in a given 5-brane web. 
\begin{figure}[t]
\centering
\includegraphics[width=12cm]{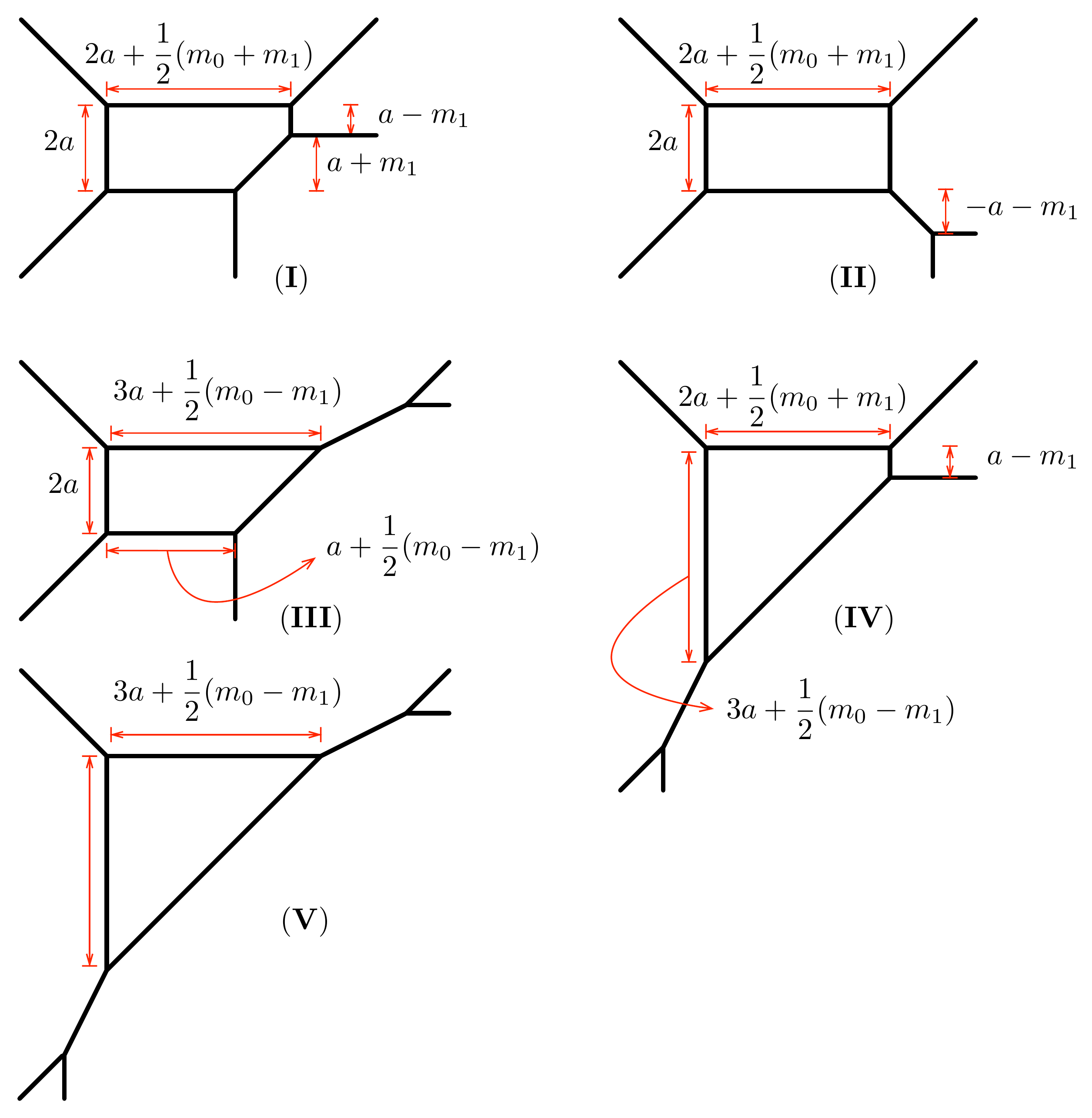}
\caption{All possible distinct phases of $E_2$ web diagram.}
\label{fig:E2phases}
\end{figure}
Let us now consider cases with flavors. For $N_f=1$ flavor, the IMS prepotential is given as
\begin{align}
	\mathcal{F}_{\rm IMS} &= \frac{1}{2}m_0 \,a^2 +\frac43\, a^3- \frac1{12} |a \pm m_1|^3 \crcr
	&= \frac{1}{2}m_0 \,a^2 +\frac76\, a^3  - \frac12 m_1^2\, a+ \frac16 \relu{a \pm m_1}^3,
\label{eq:IMSE2}
\end{align}
where we introduced the shorthand notation $\pm$ for a double sum, e.g., $|x \pm y|^3 \equiv |x + y|^3 + |x - y|^3$.
Possible phases of the 5-brane webs for the theory with $N_f=1$ flavor are depicted in Figure \ref{fig:E2phases} \cite{Morrison:1996xf, Aharony:1997ju, Hayashi:2017btw}.
All these phases in the parameter space of the $E_2$ theory,
which we identify with the K\"ahler cone of the corresponding geometry,
is depicted in Figure \ref{fig:E2Kahler}, 
which adds more quantitative information to the figure that appears in \cite{Morrison:1996xf}.
\begin{figure}[t]
\centering
\includegraphics[width=14cm]{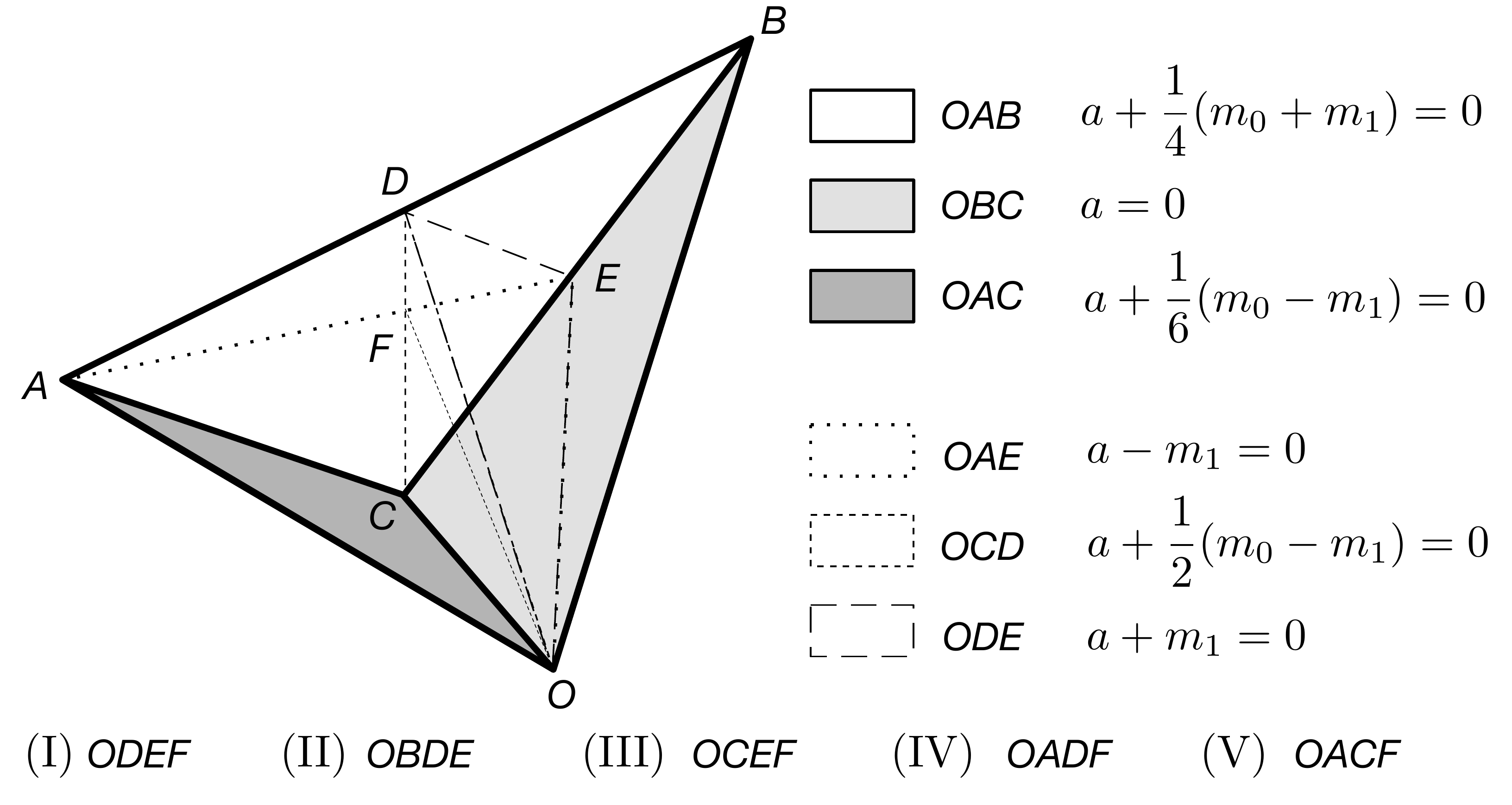}
\caption{The parameter space of the $E_2$ theory, which is identified as the K\"ahler cone of the corresponding geometry. The allowed parameter region is the space which is surrounded by the three planes represented by $OAB$, $OBC$, and $OCD$, respectively. If we fix the mass parameters $m_0$ and $m_1$, this allowed region is identified as the physical Coulomb moduli.
This allowed parameter region is divided into five phases due to the three internal ``walls'', which are represented by $OAE$, $OCD$, and $ODE$, respectively. These walls are the place where the flop transition occurs. Especially, Phase ${\rm (I)}$ is identified as the region surrounded by these three internal walls, which we represent as $ODEF$. }
\label{fig:E2Kahler}
\end{figure}

From these phases given in Figure \ref{fig:E2phases}, one can compute the area $T_i$ of each phase and then obtain the corresponding prepotential $\F_i$:\\ 
Phase (I): $\{ a+m_1>0~\&~ a-m_1>0 ~\&~a+ \frac12 (m_0-m_1)>0 \}$
\begin{align}\label{eq:FE2-1}
	\F_{\rm (I)} &= \frac76 a^3 + \frac12 m_0\, a^2 -\frac{1}{2}m_1^2 \,a\ . 
\end{align}
Phase (II): $\{ a>0~\&~ a+ \frac14 (m_0+m_1)>0~\&~a+m_1<0 \}$
\begin{align}\label{eq:FE2-2}
	\F_{\rm (II)} &= \frac43 a^3 + \frac12 (m_0 + m_1)\,a^2 \ . 
\end{align}
Phase (III):$\{ a>0~\&~a-m_1<0~\&~a+ \frac12 (m_0-m_1)>0 \}$
\begin{align}\label{eq:FE2-3}
	\F_{\rm (III)} &= \frac43 a^3 + \frac12 (m_0 - m_1)\,a^2 \ . 
\end{align}
Phase (IV): $\{ a+\frac14 (m_0+m_1)>0~\&~ a-m_1>0 ~\&~a+ \frac12 (m_0-m_1)<0 \}$
\begin{align}\label{eq:FE2-4}
	\F_{\rm (IV)} &= \frac43 a^3 + \frac14 (3m_0 - m_1)\,a^2 +\frac18 (m_0+m_1) (m_0-3m_1)\, a\ . 
\end{align}
Phase (V): $\{ a+\frac16 (m_0-m_1)>0~\&~ a-m_1<0 ~\&~a+ \frac12 (m_0-m_1)<0 \}$
\begin{align}\label{eq:FE2-5}
	\F_{\rm (V)} &= \frac32 a^3 + \frac34 (m_0 - m_1)\,a^2 +\frac18 (m_0 - m_1)^2\, a\ . 
\end{align}
These prepotentials for all phases can be put in a single short expression, 
\begin{align}\label{eq:F4E2}
	\mathcal{F}_{E_2} = &\,\frac76\, a^3 +\frac1{2} m_0 \,a^2-\frac12m_1^2\,a 
	+\frac16 \relu{a \pm m_1}^3 
	+\frac16 \relu{a+\frac12 (m_0-m_1)}^3\ ,
\end{align}
which is valid in all region and so it is the complete prepotential of the $E_2$ theory.

Among five different phases in Figure \ref{fig:E2phases}, (I), (II), and (III) are the phases that the IMS prepotential for the $SU(2)$ gauge theory with one flavor has, and 
thus the prepotential covering all the allowed phases is expressed as  
\begin{align}
	\mathcal{F}_{E_2}=& ~\mathcal{F}_{\rm IMS}  +\frac16 \relu{a+\frac12 (m_0-m_1)}^3\ ,
\end{align}
where $\mathcal{F}_{\rm IMS}$ is given in \eqref{eq:IMSE2} and the last term is responsible for the phases where $m_0<-2a +m_1$. Some of such phases can on the other hand related to the IMS phases by S-duality. For instance, the phase (IV) is S-dual of the phase (III).

The complete prepotential \eqref{eq:F4E2} incorporating all the parameter regions is then clearly expressed  
 as the IMS prepotential plus the S-dual pairs or the contributions of all possible flops,
\begin{align}
	\mathcal{F}_{E_2} = &\,\F_{\rm CFT}+\frac16 \sum_{\rm flops} \relu{\rm flops}^3 \ , 
\end{align}
where $\F_{\rm CFT}$ is $\F_{(\rm I)}$ in \eqref{eq:FE2-1}, 
and there are three different types of flops, as depicted in  \ref{fig:E2phases}, which accounts for the last three terms in \eqref{eq:F4E2}. Note that $\F_{\rm CFT}$ is the prepotential on a phase of the parameter region where the Coulomb branch modulus $a$ can be much larger than the other mass parameters including the inverse coupling constant squared $m_0$. In this paper, we call this phase 
``the CFT phase,"
which is compatible with the parameter region where all the mass parameters are turned off. 

One can see that the RG flows to one less flavor cases naturally connect to the complete prepotentials for $E_1$ and $\widetilde{E}_1$ theories.
 Namely, by taking $m_1\to -\infty$ and $m_0\to +\infty$ while $m^{E_1}_0=m_0+m_1$ is kept fixed, one readily finds that \eqref{eq:F4E2} becomes \eqref{eq:pureSU2prep}, {\it i.e.,}
$\F_{E_2}\to \F_{E_1}$. Likewise, 
by taking $m_1\to +\infty~\&~ m_0\to \infty$ while $m^{\widetilde{E}_1}_0=m_0-m_1$ is kept fixed \cite{Hayashi:2017btw}, one readily finds that \eqref{eq:F4E2} again becomes \eqref{eq:Prep4E1t
}, {\it i.e.,}
$\F_{E_2}\to \F_{\widetilde{E}_1}$.  


The perturbative global symmetry for the $SU(2)$ theory with $N_f=1$ flavor is $SO(2)\times U(1)_I$. On the other hand, one can read off the enhanced $E_2=SU(2)\times U(1)$ symmetry from  \eqref{eq:F4E2} by introducing the invariant Coulomb branch parameter $\ia$. 
For the $SU(2)$ gauge theory with $N_f=1$ flavor, the invariant Coulomb branch parameter is given by \cite{Mitev:2014jza},
\begin{align}
\ia = a  + \frac{1}{7}\,m_0\ . 
\end{align}
The complete prepotential \eqref{eq:F4E2} is then expressed as
\begin{align}\label{eq:invF4E2}
	\mathcal{F}_{E_2} = &\,\frac76\, \ia^3 - \big(x^2+\frac17\, y^2\big) \ia 
	+\frac16 \relu{\ia+ \frac47 y}^3+\frac16 \relu{\ia \pm x -\frac37 y}^3 \ ,
\end{align}
where we have introduced $E_2=SU(2)\times U(1)$ symmetry parameters, $x,y$ \cite{Kim:2012gu} 
\begin{align}\label{eq:E2para}
	x= \frac14 m_0 + \frac14 m_1\ ,\qquad 	
	y=- \frac14 m_0 + \frac74 m_1\ ,
\end{align}
so that the last two terms in \eqref{eq:invF4E2} become an $SU(2)$ doublet, invariant under $x\leftrightarrow -x$, and the $U(1)$ parameter is denoted by $y$. 
As a consequence, a particular set of flop transitions is closely related to the Weyl reflections 
of enhanced global symmetry. 
It means that one can obtain the complete prepotential by taking into account all possible flop transitions, which enables one to span  all the parameter regions, or equivalently by considering  the Weyl reflections 
for the corresponding representation of enhanced global symmetry. The latter would be more systematic when the rank of gauge group or the number of hypermultiplet is large.

The same complete prepotential \eqref{eq:invF4E2} can be obtained by the following systematic procedure: (i) Start with the IMS prepotential for a given theory in a Weyl chamber with certain phases for masses. (ii) Rewrite the IMS prepotential in terms of the invariant Coulomb branch parameter $\ia$. (iii) Then apply the Weyl reflections 
for enhanced global symmetry. 
For instance, from (i) and (ii), we get the prepotential for $SU(2)$ gauge theory with $N_f=1$ flavor is given as
\begin{align}
	\mathcal{F} &= \frac76\, \ia^3  - \Big(\frac12 m_1^2+ \frac{1}{14} m_0^2 \Big)\, \ia+ \frac16\relu{\ia-\frac17 m_0 \pm m_1}^3.
\end{align}
Using \eqref{eq:E2para} and also applying the Weyl reflection 
of $SU(2)$, one reproduces \eqref{eq:invF4E2}. 

One can easily repeat the procedure and get the complete prepotential for $SU(2)$ gauge theory with $N_f\le 7$ flavors. 
Here, the invariant Coulomb branch parameter $\ia$ 
for the $SU(2)$ theory with $N_f$ flavors is then given by\footnote{
Given the IMS prepotential, one can take a limit, called CFT phase, where the Coulomb branch parameters $a_i$ are much larger than mass parameters $m_j$ including the instanton mass $m_0$, subject to $a_i\gg m_0 \gg m_j$. In this phase, we see that the effective coupling $\tau_{ij}=\frac{\partial^2\F}{\partial \phi_i\partial \phi_j}$ to be only a function of $a$ and $m_0$.  For instance, the prepotential for 5d $SU(2)$ gauge theory with $N_f$ flavors in the CFT phase is given by
\begin{align*}
	\mathcal{F}^{SU(2)+N_f{\bf F}}_{\rm CFT}=\frac12 m_0 \,a^2+ \frac16(8-N_f) \, a^3 .
\end{align*}
Form this, one can consider the second derivatives of $\mathcal{F}_{\rm CFT}$, which gives the effective coupling in the CFT phase,
$\tau_{\rm eff}=m_0 + (8-N_f) a. $ 
 In the CFT phase, we require that no mass parameter appears so as not to be transformed under the Weyl reflection of the enhanced global symmetry. In other words, by shifting the Coulomb branch parameter, we introduce the invariant Coulomb branch parameter $\ia$ \eqref{eq:invCBpara} so that $\mathcal{F}_{\rm CFT}$ respects the enhanced global symmetry. 
}
\begin{align}\label{eq:invCBpara}
\ia = a  + \frac{1}{8-N_f}\,m_0\ ,
\end{align}
which is equivalent to \cite{Mitev:2014jza}. 
As an example, we write the prepotential the $SU(2)$ theory with $N_f=7$ flavors. First, the IMS prepotential for the $SU(2)$ theory with $N_f=7$ flavors 
\begin{align}
		6\,\mathcal{F}_{\rm IMS} = &\ a^3+3 m_0 a^2- 3\sum_{k=0}^{7} m_k^2  a 
	+\sum_{k=1}^7 \relu{a \pm m_k}^3 \ .
\end{align}
With the invariant Coulomb branch parameter $\ia=a+m_0$, one applies the Weyl reflection 
of $E_8$ given in Appendix \ref{sec:conventionE8} to get the complete prepotential 
\begin{align}\label{eq:invFforE8-a}
	6\,\mathcal{F}^{SU(2)+7{\bf F}} &= \  - 3\sum_{k=0}^{7} m_k^2  \,\ia +\ia^3
	+\sum_{k=1}^7\relu{\ia -m_0 \pm m_k}^3 
	\\
	&
+\sum_{k=1}^7 \relu{\ia +  m_0 \pm m_k}^3
	+\sum_{1 \le i < j \le 7}\relu{\ia \pm m_i \pm m_j}^3 
	\crcr
	&
	+\sum_{\substack{\{s_i=\pm 1\}\\\text{even} +}}\relu{\ia - \frac12 m_0+\frac12\sum_{k=1}^7 s_k m_k}^3
		+\sum_{\substack{\{s_i=\pm 1\}\\\text{odd} +}}\relu{\ia + \frac12 m_0+\frac12\sum_{k=1}^7 s_k m_k}^3.
		\nonumber
\end{align}
Here, the summation symbol in the term corresponding to the (conjugate) spinor representation 
\begin{align}
\sum_{\substack{\{s_i=\pm 1\}\\\text{even} +}}
\quad \text{and} \quad
\sum_{\substack{\{s_i=\pm 1\}\\\text{odd} +}}
\end{align}
denotes that we sum over all possible combination of $s_i$ such that even (odd) number of $s_i$ take the value $s_i=+1$, while the remaining take the value $s_i=-1$. 
It is straightforward to check that this complete prepotential goes back to the IMS prepotential in the weak coupling region $m_0 \gg |m_f|, |a|$.
We note that linear combinations of $m_i$ in $\relu{~}^3$ form $SO(14)$ representation and the coefficient of $m_0$ is the $U(1)_I$ charge. 
We can put the terms in \eqref{eq:invFforE8-a} to re-express the prepotential as an $SO(16)$ manifest form
$\F = \F_{\rm CFT}+ \F^{SO(16)}_{\bf 120} +\F^{SO(16)}_{\bf 128},$
and they can be further reorganized as the adjoint representation of $E_8$
\begin{align}
	E_8 &~\supset~ SO(16)~ \supset~ SO(14)\times U(1)_I\crcr
	{\bf 248} &={\bf 120}+ {\bf 128} ={\bf 1}_0+{\bf 14}_{-2}+ {\bf 14}_2 +{\bf 91}_0 +{\bf 64}_{-1}+\overline{\bf 64}_{1} .
\end{align}
Therefore we get a complete prepotential which is $E_8$ manifest, $\mathcal{F}_{E_8} = \F_{\rm CFT}+ \F^{E_8}_{\bf 248}$, 
\begin{align}\label{eq:invFforE8}
	6\,\mathcal{F}_{E_8} = - 3\sum_{k=0}^{7} m_k^2  \,\ia +\ia^3+\sum_{w_{E_8}\in \bf{248}} \relu{\ia + \vec{w}_{E_8}\cdot \vec{m}}^3 \ .
\end{align}
We note that in counting the dimension of representations, we added $8\relu{\ia}^3$ by hand to account for the 8 Cartans, which did not necessarily appear in \eqref{eq:invFforE8-a} since $\relu{\ia}^3=0$ for $\ia>0$.

For less flavors, one can decouple flavors one by one from the $\F_{E_8}$ to obtain the complete prepotentials for the $SU(2)$ gauge theory with $N_f\le 7$ flavors. They show the $E_{N_f+1}$ enhanced global symmetry, and the explicit expressions are listed in Appendix \ref{sec:app-rk1Prep}. 
\subsection{Prepotential from partition function}
\label{sec:Rank1GV-prep}

Suppose that we compactify a 
5d $\mathcal{N}=1$ gauge theory on $S^1$ with radius $R$.
Or equivalently, we consider a 
4d $\mathcal{N}=2$ gauge theory with Kaluza-Klein modes.
It is well known that the prepotential $\F_{\mathbb{R}^{3,1} \times S^1}$ of this theory \cite{Nekrasov:1996cz, Lawrence:1997jr} is obtained from the 5d Nekrasov partition function $Z_{\mathbb{R}^{3,1} \times S^1}$ by taking the limit \cite{Nekrasov:2002qd, Losev:2003py, Nekrasov:2003rj} 
\begin{align}\label{eq:S1prep}
\F_{\mathbb{R}^{3,1} \times S^1} 
=  \lim_{\epsilon_1, \epsilon_2 \to 0} \epsilon_1 \epsilon_2 \log Z_{\mathbb{R}^{3,1} \times S^1}.
\end{align}
This prepotential, of course, includes instanton contribution as well as 1-loop perturbative part.
It is known that IMS prepotential can be reproduced from the decompactification limit of this prepotential 
\begin{align}\label{eq:IMS_lim}
\lim_{R \to \infty} \frac{1}{R} \F_{\mathbb{R}^{3,1} \times S^1}  \Rightarrow \F_{\text{IMS}},
\end{align}
It has been discussed that the instanton contribution vanishes in this limit because the instanton factor $e^{-R m_0}$ is exponentially suppressed. This claim is correct as long as we consider the weak coupling region $m_0 \gg |m_f|, |a|$. 

However, in this paper, we are considering all the possible parameter region, even including the region $m_0<0$,
where the original gauge theory description is not valid any more.
In this case, it is not correct to claim that the instanton contribution still vanishes in the limit.
We claim that the same limit indeed reproduces our complete prepotential if we admit all the parameter region:
\begin{align}\label{eq:decomp-limit}
\F_{\text{Complete}} = 
\lim_{R \to \infty} \frac{1}{R} \F_{\mathbb{R}^{3,1} \times S^1}.
\end{align}
That is, the instanton contribution does remain in the decompactification limit in a way similar to perturbative part.
This claim, of course, goes back to the traditional claim \eqref{eq:IMS_lim}, when we consider the weak coupling region, 
where our complete prepotential reproduces the IMS prepotential.

In order to see this, let us first start from Nekrasov partition function.
Nekrasov partition function is identified as the refined topological string partition function on the corresponding Calabi-Yau 3-fold, which is expressed in the following form by using the Gopakumar-Vafa invariants $N_C^{(j_L, j_R)}$ \cite{Gopakumar:1998ii, Iqbal:2007ii, Gopakumar:1998jq}:
\begin{align}\label{eq:GV-general}
Z_{\mathbb{R}^{3,1} \times S^1} 
&= Z_0 \exp \left( \sum_{C \in H_2(X,\mathbb{Z})} \sum_{j_L,j_R} \sum_{n=1}^{\infty} \frac{N_C^{(j_L, j_R)}  [j_L, j_R]_{t^n, q^n}}{n \left( t^{\frac{n}{2}} - t^{-\frac{n}{2}} \right) \left( q^{\frac{n}{2}} - q^{-\frac{n}{2}} \right) }    e^{- n  R  \, T_C} \right)
\cr
&= Z_0 \text{PE} \left( \sum_{C \in H_2(X,\mathbb{Z})} \sum_{j_L,j_R} 
\frac{N_C^{(j_L, j_R)}  [j_L, j_R]_{t, q}}{\left( t^{\frac{1}{2}} - t^{-\frac{1}{2}} \right) \left( q^{\frac{1}{2}} - q^{-\frac{1}{2}} \right) } e^{-  R  \, T_C} \right)
\end{align}
where PE stands for the plethystic exponential and 
\begin{align}
[j_L, j_R]_{t,q} := (-1)^{j_L+j_R+1}
 \left( (t q)^{-j_L} + \cdots (t q)^{j_L} \right)
 \left( (t q^{-1})^{-j_R} + \cdots (t q^{-1})^{j_R} \right),
\end{align}
with $t=e^{+R \, \epsilon_1}$, $q=e^{-R \, \epsilon_2}$. 
The variable $R \, T_C$ is the K\"ahler parameters corresponding to the the 2-cycle $C$ and can be interpreted as the
masses of M2-branes wrapped on $C$, which is rescaled by the radius $R$ for later convenience.
The prefactor $Z_0$ is the part which does not vanish in the limit $R  \, T_C \to \infty$.

When we take the limit as in \eqref{eq:S1prep}, we obtain
\begin{align}
\F_{\mathbb{R}^{3,1} \times S^1} 
&= R \F_0 - \frac{1}{R^2} \sum_{C \in H_2(X,\mathbb{Z})} \sum_{j_L,j_R} (-1)^{j_L+j_R+1} (2j_L+1) (2j_R+1) N_C^{(j_L, j_R)} 
\mathrm{Li}_3 \left( e^{- R  \, T_C} \right)
\end{align}
where $R \F_0 = \lim_{\epsilon_1, \epsilon_2 \to 0} \epsilon_1 \epsilon_2 \log Z_{0}$
corresponds to the triple-intersection term in the CFT phase.
$\mathrm{Li}_3$ is the polylogarithm function defined as $\mathrm{Li}_s (z) = \sum_{k=1}^{\infty} \frac{z^k}{k^s}$.
When we finally take the decompactification limit, we obtain 
\begin{align}\label{eq:infinite-prep}
\F_{\text{Complete}} 
= \F_0 - \frac{1}{6} \sum_{C \in H_2(X,\mathbb{Z})} \sum_{j_L,j_R}  (-1)^{j_L+j_R+1} (2j_L+1) (2j_R+1) N_C^{(j_L, j_R)} \relu{ T_C } ^3,
\end{align}
where we used
\begin{align}
\lim_{R \to + \infty} 
\frac{1}{R^3}
\mathrm{Li}_3 \left( e^{- Rx} \right) = 
\frac{1}{6} \relu{x}^3.
\end{align}
At this stage, it looks as if our complete prepotential includes infinitely many terms
corresponding to all the 2-cycles $C \in H_2(X,\mathbb{Z})$. 
Note, however, that 
the Coulomb branch parameter $a$ must be inside the physical Coulomb moduli.
Therefore, if $T_C>0$ is always satisfied for certain $C$ in the physical Coulomb moduli, 
it indicates that the corresponding term vanishes by definition \eqref{eq:doubleAbs}: $\relu{ T_C } = 0$. 
That is how we obtain the finitely many terms.

In the following, we see this phenomena in the specific example: $Sp(1)$ gauge theory with $N_f=7$ flavors.
This theory enjoys $E_8$ global symmetry at the UV fixed point.
We follow the convention for the representations of $E_8$ summarized in Appendix \ref{sec:conventionE8}.
The Calabi-Yau geometry corresponding to the 
$SU(2)$ gauge theory with $N_f=7$ flavors
is the 8th local del Pezzo surface
and its Gopakumar-Vafa invariants are computed in \cite{Huang:2013yta} explicitly.
Their results indicates that the corresponding 5d Nekrasov partition function can be written in the following form:
\begin{align}\label{eq:FE8}
 \frac{Z_{\mathbb{R}^{3,1} \times S^1} }{Z_0}  = 
\text{PE} & \Bigg[
\frac{1}{ ( t^{\frac{1}{2}} - t^{-\frac{1}{2}} ) ( q^{\frac{1}{2}} - q^{-\frac{1}{2}} ) }
\cr
& \times \Bigg\{ 
	~  \bigg[ [0,0] { \chi_{\mu_8}}+ [\frac12,\frac12]  \bigg]  \tilde A
	+ \bigg[[0,\frac12] ({ \chi_{\mu_1}}+1)+  [\frac12,1]{ \chi_{\mu_8}}+ [1,\frac32] \bigg] \tilde{A}^2 \cr
	& \quad +\bigg[[0,0]\big(\chi_{\mu_8} + { \chi_{\mu_7}}\big) + [0,1]\big(\chi_{\mu_8} + { \chi_{\mu_2}}+\chi_{\mu_1}+1\big) + [0,2] \chi_{\mu_8} \cr
	& \quad  \quad +[\frac12,1] \big(\chi_{\mu_8}+{ \chi_{\mu_1}}+1\big)+ [\frac12, \frac32] 
	\big(\chi_{\mu_8}+{ \chi_{\mu_7}}+\chi_{\mu_1}+1\big) 
	+ [\frac12,\frac52] + [1,0] 
	\cr
	&\quad  \quad + [1,1] \chi_{\mu_8} + [1,2]\big(\chi_{\mu_8}+{ \chi_{\mu_1}}+1\big) 
+[\frac32, \frac32] + [\frac32, \frac52] \chi_{\mu_8} +[2,3]\bigg]\tilde{A}^3
	\cr
& \quad + \mathcal{O}(\tilde{A}^4) 	\Biggr\} \Bigg].  
\end{align}
where we denote $\tilde{A} = e^{- R \ia}$. 
and we abbreviated the label $t,q$ in $[j_L, j_R]_{t,q}$ for simplicity.

In Appendix \ref{sec:PhysCB}, we have derived that the physical Coulomb moduli is given as
\begin{align}\label{eq:physCB-E8}
&2 \ia - w \cdot \vec{m} \ge0 , \quad \forall w \in \mu_1,
\cr
&3 \ia - w \cdot \vec{m} \ge0 , \quad \forall w \in \mu_2.
\end{align}
Also, we discussed that the following inequalities are satisfied from these conditions: 
\begin{align}\label{eq:ineq-E8}
&n_0 \ia \ge 0, \qquad \forall n_0 \ge 1
\cr
&n_1 \ia - w \cdot \vec{m} \ge0 , \quad \forall w \in \mu_1, \quad \forall n_1 \ge 2,
\cr
&n_2 \ia - w \cdot \vec{m} \ge0 , \quad \forall w \in \mu_2, \quad \forall n_2 \ge 3,
\cr
&n_7 \ia - w \cdot \vec{m} \ge0 , \quad \forall w \in \mu_7, \quad \forall n_7 \ge 3,
\cr
&n_8 \ia - w \cdot \vec{m} \ge0 , \quad \forall w \in \mu_8, \quad \forall n_8 \ge 2.
\end{align}
These inequalities indicate that all the terms in \eqref{eq:FE8} of the form
\begin{align}
& 1 \cdot \tilde{A}^{n_0} \quad (n_0 \ge 1), \qquad
{ \chi_{\mu_1}} \tilde{A}^{n_1} \quad (n_1 \ge 2), \qquad
{ \chi_{\mu_2}} \tilde{A}^{n_2} \quad (n_2 \ge 3), \qquad
\cr
& { \chi_{\mu_7}} \tilde{A}^{n_7} \quad (n_7 \ge 3), \qquad
{ \chi_{\mu_8}} \tilde{A}^{n_8} \quad (n_8 \ge 2)
\end{align}
vanish after the decompactification limit \eqref{eq:decomp-limit}
because they produce the terms proportional to $\relu{ \text{(LHS of \eqref{eq:ineq-E8})} }^3 = 0$
in the complete prepotential \eqref{eq:infinite-prep}.
Only the exceptional term 
in \eqref{eq:FE8} is 
\begin{align}\label{eq:BPSmu8}
[0,0]{ \chi_{\mu_8}}  \tilde{A},
\end{align}
which contributes as 
\begin{align}
\frac{1}{6} \sum_{w \in \bf{\mu_8}} \relu{ \ia + w \cdot \vec{m} }^3.
\end{align}
If we identify the triple-intersection term as%
\footnote{
This part cannot be read off from the Gopakumar-Vafa invariants and should be computed separately. 
Here, we simply imported the results from \eqref{eq:invFforE8}. 
}
\begin{align}
\F_0 = - \frac12 \sum_{k=0}^{7} m_k^2  \,\ia + \frac16 \ia^3,
\end{align}
we reproduce our complete prepotential \eqref{eq:invFforE8} from the partition function \eqref{eq:FE8}.

Through the computation above, we have an interesting observation. 
The physical Coulomb moduli \eqref{eq:physCB-E8} can be regarded as the condition that the K\"ahler parameters $T_C$ corresponding to the terms in
\begin{align}\label{eq:mu12}
[0,\frac12] { \chi_{\mu_1}} \tilde{A}^2,
\qquad
[0,1]  { \chi_{\mu_2}} \tilde{A}^3,
\end{align}
which appear in \eqref{eq:FE8}, 
should be all positive.
%
\begin{figure}
\centering
\begin{minipage}[b]{0.45\hsize}
\centering
\includegraphics[width=
0.25 \hsize]{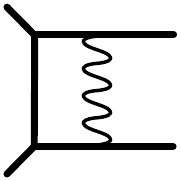}
\caption{String generating BPS particle with spin $(j_L,j_R) = (0,\frac12)$}
\label{fig:BPS-0-half}
\end{minipage}
\hspace{1mm}
\begin{minipage}[b]{0.45\hsize}
\centering
\includegraphics[width=
0.4\hsize]{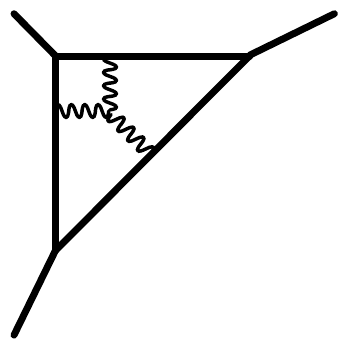}
\caption{String junction generating BPS particle with spin $(j_L,j_R) = (0,1)$}
\label{fig:BPS-0-1}
\end{minipage}
\end{figure}
%
The first one in \eqref{eq:mu12} corresponds to the W-boson and their partners related by the Weyl reflections. 
The example of the string generating such particles is depicted in Figure \ref{fig:BPS-0-half}.
The second one in \eqref{eq:mu12} includes the term which remains after the flavor decoupling down to the $E_0$ theory. 
It implies that such term corresponds to the string junction which appears in the web diagram as depicted in Figure \ref{fig:BPS-0-1}.
At the boundary of the physical Coulomb moduli, where the magnetic monopole tension vanishes, either of these two types of BPS particle becomes massless.%
\footnote{These two cases are studied in \cite{Witten:1996qb} as examples of ``a divisor collapsing to a curve'' and ``a divisor collapsing to a point'', respectively.}
%
Here, we summarize our observation:
\begin{flushleft}
{ \bf Observation 1}: 
Whenever the monopole tension goes to zero,
at least one BPS particle either with spin $(j_L,j_R)=(0,\frac{1}{2})$ or with spin $(j_L,j_R)=(0,1)$ also becomes massless
and vice versa.
\end{flushleft}
We expect that it is true for more generic examples.
If this is true, we can reproduce the physical Coulomb moduli from the Gopakumar-Vafa invariants by imposing 
$T_C \ge 0$ for all the 2-cycle $C$ satisfying $N^{(j_L,j_R)}_C > 0$ with $(j_L,j_R)=(0,\frac{1}{2})$ or $(j_L,j_R)=(0,1)$.

\begin{figure}
\centering
\includegraphics[width=
0.5\hsize]{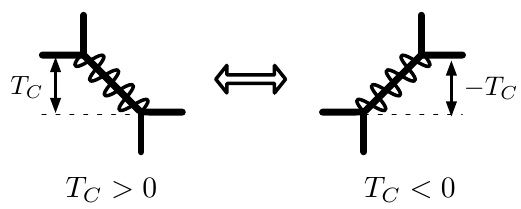}
\caption{String generating BPS particle with spin $(j_L,j_R) = (0,0)$, which can realize $T_C<0$.}
\label{fig:BPS0-0}
\end{figure}

Moreover, we have further observation from \eqref{eq:BPSmu8}
that the BPS particle which contributed to the complete prepotential 
is the contribution from the hypermultiplets and their partners.
Such BPS particle can realize negative K\"ahler parameter $T_C < 0$ by the transition as depicted in Figure \ref{fig:BPS0-0}.
The brane configuration in Figure \ref{fig:BPS0-0} can be interpreted as the flop transition of the rational curves with the normal bundle $\mathcal{O}(-1) \oplus \mathcal{O}(-1)$. 
We summarize our observation as follows:
\begin{flushleft}
{ \bf Observation 2}:
The negative K\"ahler parameters $T_C<0$ which contribute to the complete prepotential can be realized only from the rational curves with the normal bundle $\mathcal{O}(-1) \oplus \mathcal{O}(-1)$.
\footnote{
For example, the cycle with the K\"ahler parameter $2a$, which is the rational curve with the normal bundle $\mathcal{O}(-2) \oplus \mathcal{O}(0)$, does not contribute to the complete prepotential in our convention because we chose the Weyl chamber $a>0$. The contribution from the cycle with the K\"ahler parameter of the form $m_i - m_j$ also vanishes because we remove such contribution from the the partition function as the``extra factor'' \cite{Konishi:2006ya, Bergman:2013ala,  Bao:2013pwa, Hayashi:2013qwa, Bergman:2013aca, Hwang:2014uwa}.}
\end{flushleft}
Again, we expect that it is true for other examples.
Note that M2-brane wrapping such a curve corresponds to the BPS particle with spin $(j_L,j_R)=(0,0)$.
Therefore, this 
 implies that the BPS particle with $(j_L,j_R) \neq (0,0)$ cannot contribute 
to the complete prepotential and thus, \eqref{eq:infinite-prep} simplifies as
\begin{align}
\F_{\text{Complete}} 
= \F_0 + \frac{1}{6} \sum_{C \in H_2(X,\mathbb{Z})} N_C^{(0,0)} \relu{ T_C } ^3.
\end{align}
Only some of the BPS particles with spin $(j_L,j_R) = (0,0)$ can contribute to the complete prepotential.

If these two observations would be applicable for generic gauge theories of any rank,%
\footnote{It would be interesting to find more precise mathematical expressions for these observations,
which  might have appeared in the math literature.}  
they would give a systematic method to obtain the complete prepotential 
as well as the physical Coulomb moduli from the list of Gopakumar-Vafa invariants.
We will use them later to study $Sp(2)$ gauge theory.
 
%
%
%
%
%
%
%
%

\subsection{Prepotential from Mori cone generators 
}\label{sec:rk1geometry}

We have seen in section \ref{sec:rk1F} that the effective prepotentials for the $SU(2)$ gauge theories receive novel contributions from light instanton particles. 
In section \ref{sec:Rank1GV-prep} the contributions were interpreted in terms of BPS particles wrapped on holomorphic curves in the local Calabi-Yau threefold dual to the 5-brane web for the $SU(2)$ gauge theory with $7$ flavors. 
In this subsection, we will identify the holomorphic curves more explicitly in the dual geometry.


We focus on the 5d $SU(2)$ gauge theory with $7$ flavors and hence on the 8th del Pezzo surface $dP_8$ which is the base manifold of the dual local Calabi-Yau threefold. We first determine relations between the volume of curves in $dP_8$ and the parameters of the $SU(2)$ gauge theory with $7$ flavors. Any element in the second homology of $dP_8$ is generated by the hyperplane class $L$ and the exceptional curve classes $X_i \; (i=0, \cdots, 7)$. The intersection numbers are given by
\begin{align}
L\cdot L = 1, \qquad X_i \cdot X_j = - \delta_{ij}, \qquad L\cdot X_i = 0,\label{dPintersection}
\end{align}
for all $i, j = 0, \cdots, 7$ and the canonical divisor class is 
\bea
K_{dP_8} = -3L + \sum_{i=0}^7X_i.
\eea
In order to see the structure of the $SU(2)$ gauge theory with $7$ flavors, it is useful to regard $dP_8$ as a blow up of $dP_1$ at $7$ points. The $dP_1$ part gives the $SU(2)$ gauge theory and the $7$ points blow ups introduce $7$ flavors. Let the $7$ exceptional curves associated to the $7$ blow ups be $X_i \; (i=1, \cdots, 7)$. The fiber class $F$ of $dP_1$ is then given by $F = L - X_0$. The volume of the fiber class is related to the Coulomb branch modulus $a$ of the $SU(2)$ gauge theory and we have
\be
\text{vol}(F) = 2a. \label{LX0}
\ee

\begin{figure}[t]
\centering
\includegraphics[width=10cm]{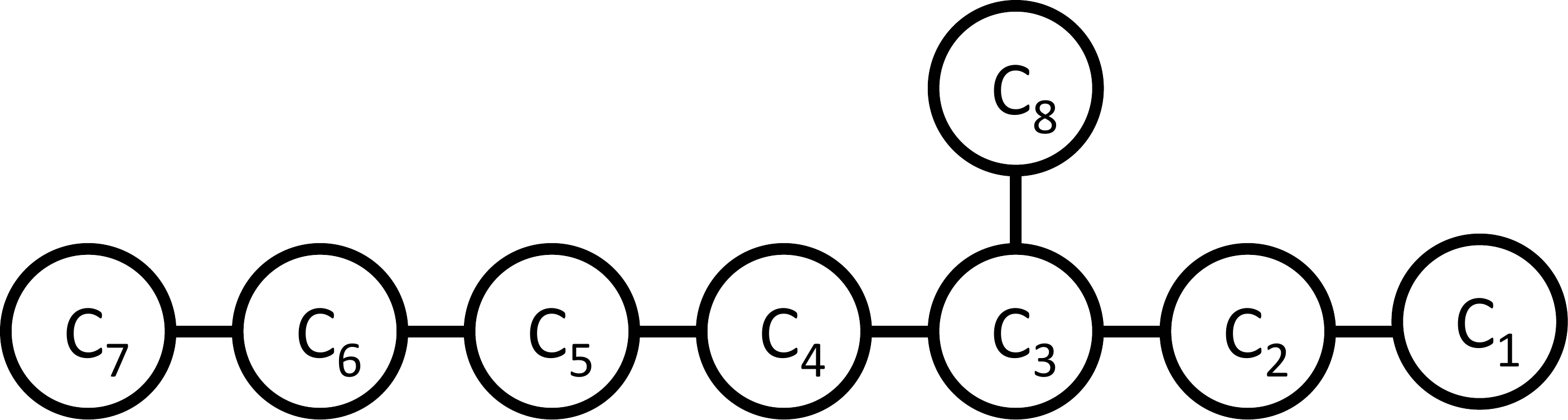}
\caption{The intersection structure of the curves in \eqref{E8root}.}
\label{fig:E8dynkin}
\end{figure}

In order to see the other relations, we can make use of the fact that the $E_8$ root lattice is a subset in $H_2(dP_8, \mathbb{Z})$. The $E_8$ root lattice in $H_2(dP_8, \mathbb{Z})$ is given by
\begin{align}
R_{dP_8} = \left\{C \in H_2(dP_8, \mathbb{Z}) \; \middle| \; C\cdot C = -2, C\cdot K_{dP_8} = 0 \right\}
\end{align}
Then $R_{dP_8}$ is generated by 
\begin{align}
C_{i} = X_{i-1} - X_{i}\; (i=1, \cdots, 7), \quad C_8 = L - (X_1 + X_2 + X_3). \label{E8root}
\end{align}
The matrix $C_i \cdot C_j$ yields the negative of the Cartan matrix of the $E_8$ Lie algebra and the intersection structure of $C_i\; (i=1, \cdots 8)$ forms the Dynkin diagram of $E_8$ as in Figure \ref{fig:E8dynkin}. Namely the curves $C_1, \cdots, C_8$ correspond to the simple roots of $E_8$. 

The appearance of the $E_8$ root system is related to the fact that the $SU(2)$ gauge theory with $7$ flavors shows an $E_8$ global symmetry at the UV fixed point. The mass parameters $m_i,\; (i=0, \cdots, 7)$ are associated with the Cartan part of the $E_8$ global symmetry. Here $m_0$ is the inverse of the gauge coupling squared. The other $m_i$'s are the mass of the $7$ flavors and they are associated with the Cartan part of the perturbative $SO(14)$ flavor symmetry. The curves $C_2, \cdots, C_8$ form the $D_7$ Dynkin diagram and hence they are related to the simple roots of $SO(14)$. Therefore the volume of the curves can be parameterized by
\begin{align}
\text{vol}(C_i) = m_i - m_{i-1} \; (i=2, \cdots, 7), \qquad \text{vol}(C_8) = m_2 + m_1.\label{C2to8}
\end{align}
On the other hand, the $E_8$ root corresponding to $C_1$ is related to a spinor weight of $SO(14)$ and its volume is given by
\begin{align}
\text{vol}(C_1) = \frac{1}{2}(m_0 + m_1 - m_2 - m_3 - m_4 - m_5 - m_6 - m_7). \label{C1}
\end{align}
Then solving \eqref{LX0}, \eqref{C2to8}, \eqref{C1} in terms of $L, X_i$ for $i = 0, \cdots, 7$ yields
\begin{equation}\label{dPvolume}
\begin{split}
\text{vol}(L )&= 3a + \frac{1}{2}(m_0 - m_1 - m_2 - m_3 - m_4 - m_5 - m_6 - m_7),\\
\text{vol}(X_0) &= a + \frac{1}{2}(m_0 - m_1 - m_2 - m_3 - m_4 - m_5 - m_6 - m_7),\\
\text{vol}(X_i) &=a - m_i \qquad (i=1, \cdots, 7).
\end{split}
\end{equation}

We would like to identify the holomorphic curves explicitly in the $dP_8$ which yield the contributions given by $\relu{}$ in the complete prepotential of \eqref{eq:invFforE8}. When one of $\relu{}$'s vanishes, the corresponding BPS particle becomes massless. This occurs on a codimension-$1$ boundary of the K\"ahler cone of the geometry. In other words, the BPS particle arises from an M2-brane wrapped on a curve which corresponds to a generator of the Mori cone. Therefore we will concentrate on the generators of the Mori cone of the geometry.
Furthermore, as mentioned in Observation 2 in the previous subsection, we expect 
that the BPS particles which contribute to the effective prepotential come from M2-branes wrapped on rational curves with the normal bundle $\mathcal{O}(-1) \oplus \mathcal{O}(-1)$.
Hence we consider rational curves with the self-intersection number $-1$ in the generators of the Mori cone of $dP_8$. In general we need to see 
all such curves of all the local Calabi-Yau threefolds that are related by flop transitions. This is because the complete prepotential is valid in the whole parameter region which corresponds to the extended K\"ahler cone. 
However for the rank-1 case, flopping such a curve inside $dP_8$ corresponds to a blow down of $dP_8$ and introduces a flopped curve outside the compact surface. The blow down of $dP_8$ is $dP_n$ with $0 \leq n < 8$ or $\mathbb{F}_0$. The rational curves with the self-intersection number $-1$ in the generators of their Mori cones are a subset of those of $dP_8$. This can be expected since the blow down geometry is related to the $SU(2)$ gauge theory with less flavors.
Therefore it is enough to look at 
the generators of the Mori cone of $dP_8$.\footnote{
The situation is different for higher rank cases. For a higher rank case, the geometry may be given by gluing compact complex surfaces. Then flopping a curve in one surface may introduce a new rational curve with the self-intersection number -1 as a generator of the Mori cone of a different surface, which could yield a BPS particle in a different representation when an M2-brane is wrapped on the curve. Or sometimes a blow down geometry may have a new type of curves in the generators of the Mori cone. We will encounter those cases in section \ref{sec:rk2geometry}.}

The 
rational curves with the self-intersection number $-1$ in the generators of the Mori cone of $dP_8$ are given by 
\begin{equation}\label{dP8Mori}
\begin{split}
&X_i, \quad L - X_i - X_j, \quad 2L - \sum_{j=1}^5X_{i_j}, \quad 3L - 2X_{i_1} - \sum_{j=1}^6X_{i_{j+1}}, \\
&4L - 2\sum_{j=1}^3X_{i_j} - \sum_{j=1}^5X_{i_{j+3}}, \quad 5L - 2\sum_{j=1}^6X_{i_j} - \sum_{j=1}^2X_{i_{j+6}},\quad 6L - 3X_{i_1} - \sum_{j=1}^72X_{i_{j+1}}.
\end{split}
\end{equation}
It is also possible to compute the volume of 
the curves in \eqref{dP8Mori} using the parameterization \eqref{dPvolume}, Then the volume of 
the curves is
\begin{equation}
\begin{split}
V_1 = \ia \pm m_i \pm m_j, \quad V_2 = \ia + \frac{1}{2}(\pm m_0 \pm m_1  \pm m_2  \pm m_3 \pm m_4 \pm m_5 \pm m_6 \pm m_7),
\end{split}
\end{equation}
where $\ia = a + m_0$ and the number of plus signs in $V_2$ is even. These are nothing but the mass of the particles which contribute to the complete prepotential \eqref{eq:invFforE8} for the $SU(2)$ gauge theory with $7$ flavors. Hence 
we are able to identify the holomorphic curves in $dP_8$ which yield the contributions given by $\relu{}$ of the complete prepotential \eqref{eq:invFforE8}.

\section{Prepotential for Rank-2 theories: $Sp(2)$ gauge theory with $9$ flavors}\label{sec:rank2}

In section \ref{sec:rank1} we determined the complete prepotentials including the effects of instanton particles for the $Sp(1)$ gauge theory with $7$ flavors. In this section we generalize the analysis to the $Sp(2)$ gauge theory with $9$ flavors. We will see that the enhanced flavor symmetry is not enough to determine the prepotential completely and there are more terms which are not connected to contributions by perturbative particles by the Weyl reflections of the enhanced global symmetry group. We will identify the new contributions from the corresponding brane web and perform various consistency check with dualities and geometries.

\subsection{Complete prepotential}
\label{sec:rank2prep}
We start from the effective prepotential for the $Sp(2)$ gauge theory with $9$ flavors on a gauge theory phase. 
On this phase, the effective prepotential in \cite{Intriligator:1997pq} is valid and it is given by
\begin{equation}\label{rk2IMS}
\begin{split}
\mathcal{F}_{\text{IMS}} =& \frac{1}{2}m_0(a_1^2 + a_2^2) + \frac{1}{6}\left[(a_1 - a_2)^3 + (a_1 + a_2)^3 + 8(a_1^3 + a_2^3)\right]
- \frac{1}{12} \sum_{i=1}^2\sum_{j=1}^9 |a_i \pm m_j|^3 \\
=& \frac{1}{2}m_0(a_1^2 + a_2^2) + \frac{1}{6}(a_1^3-a_2^3) + a_1a_2^2 
- \frac{1}{2}  \sum_{I=1}^2 \sum_{i=1}^9m_i^2a_I
+ \frac{1}{6} \sum_{I=1}^2 \sum_{i=1}^9 \relu{a_I \pm m_i}^3,
\end{split}
\end{equation}
where we chose the Weyl chamber $a_1 \ge a_2 \ge 0$.

The $Sp(2)$ gauge theory with $9$ flavors exhibits an $SO(20)$ global symmetry at the UV fixed point. Therefore we first extend the prepotential \eqref{rk2IMS} by making it invariant under the Weyl group symmetry of $SO(20)$. One subtlety is that the Coulomb branch moduli $a_1$ and $a_2$ also transform under the Weyl reflections. 
Hence we first need to determine the Coulomb branch moduli that are invariant under 
the enhanced global symmetry.  
It is then useful to look at the effective coupling on the CFT phase. On the CFT phase all the mass parameters including $m_0$ will be treated equally and the effective coupling is also invariant under the Weyl reflections. 
 Let us first temporarily assume that \eqref{rk2IMS} is valid on the CFT phase and consider a region where the Coulomb branch moduli are larger than the mass parameters, namely $a_1, a_2 > |m_i|$ for $i = 1, \cdots, 9$. The prepotential \eqref{rk2IMS} reduces to 
\begin{equation}\label{rk2IMS1}
\begin{split}
\mathcal{F}_{\text{IMS}} = \frac{1}{6}(a_1^3 - a_2^3) + a_1a_2^2 + \frac{1}{2}m_0(a_1^2 + a_2^2) - \frac{1}{2}\sum_{i=1}^9m_i^2(a_1 + a_2).
\end{split}
\end{equation}
Then taking a derivative of the prepotential \eqref{rk2IMS1} with respect to Coulomb branch moduli twice yields 
\begin{align}\label{rk2eff1}
\frac{\partial \mathcal{F}_{\text{IMS}} }{\partial a_1^2} = a_1 + m_0, \quad \frac{\partial \mathcal{F}_{\text{IMS}} }{\partial a_2^2} = 2a_1 - a_2 + m_0, \quad \frac{\partial \mathcal{F}_{\text{IMS}} }{\partial a_1a_2} = 2a_2.
\end{align}
Then it is impossible to redefine $a_1$ and $a_2$ to make three combinations in \eqref{rk2eff1} invariant under the Weyl reflections of the enhanced global symmetry,  
as $m_0$ is not invariant. 

\begin{figure}[t]
\centering
\subfigure[]{\label{fig:Sp2v1}
\includegraphics[width=7cm]{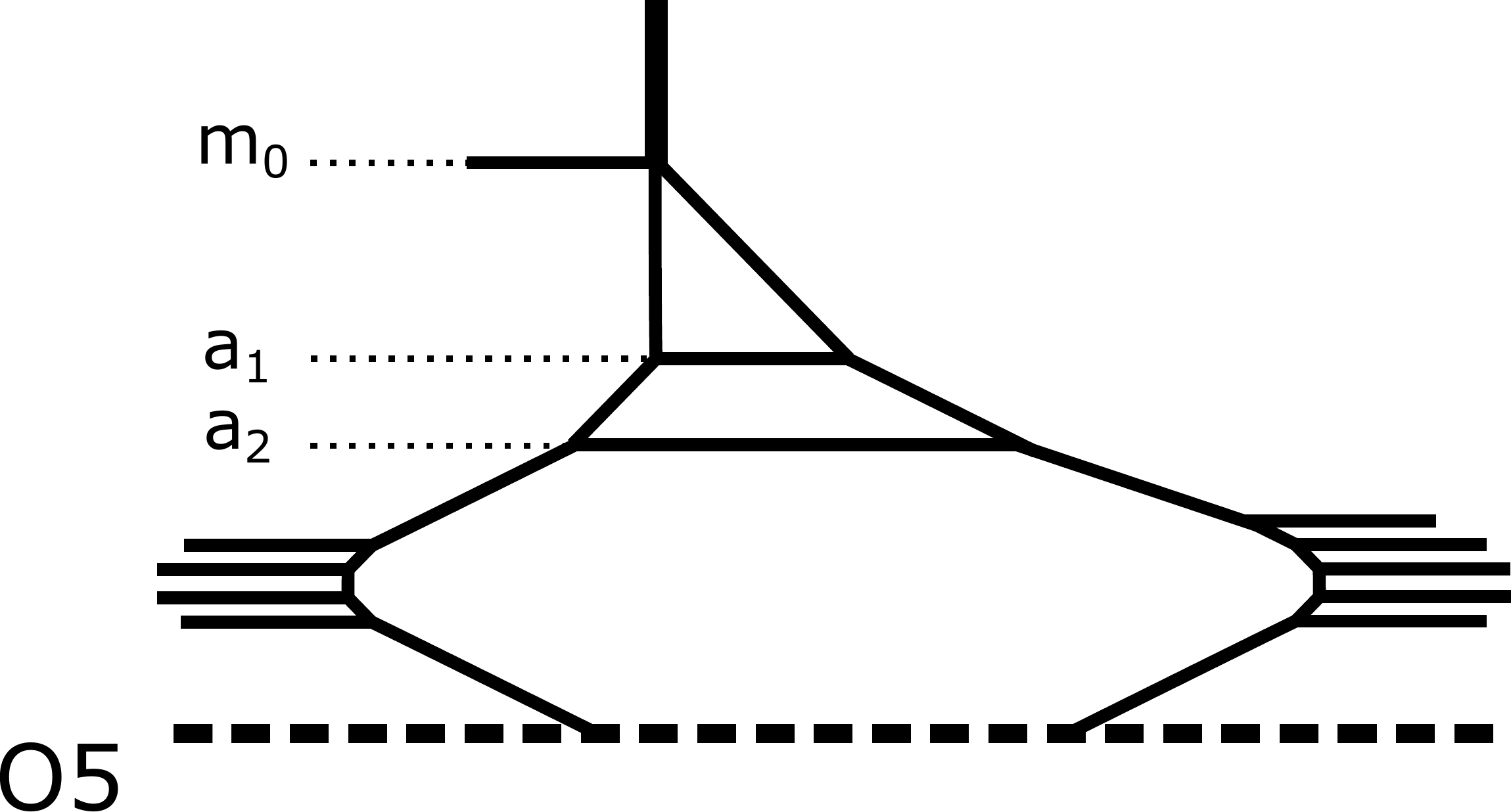}}
\subfigure[]{\label{fig:Sp2v2}
\includegraphics[width=7cm]{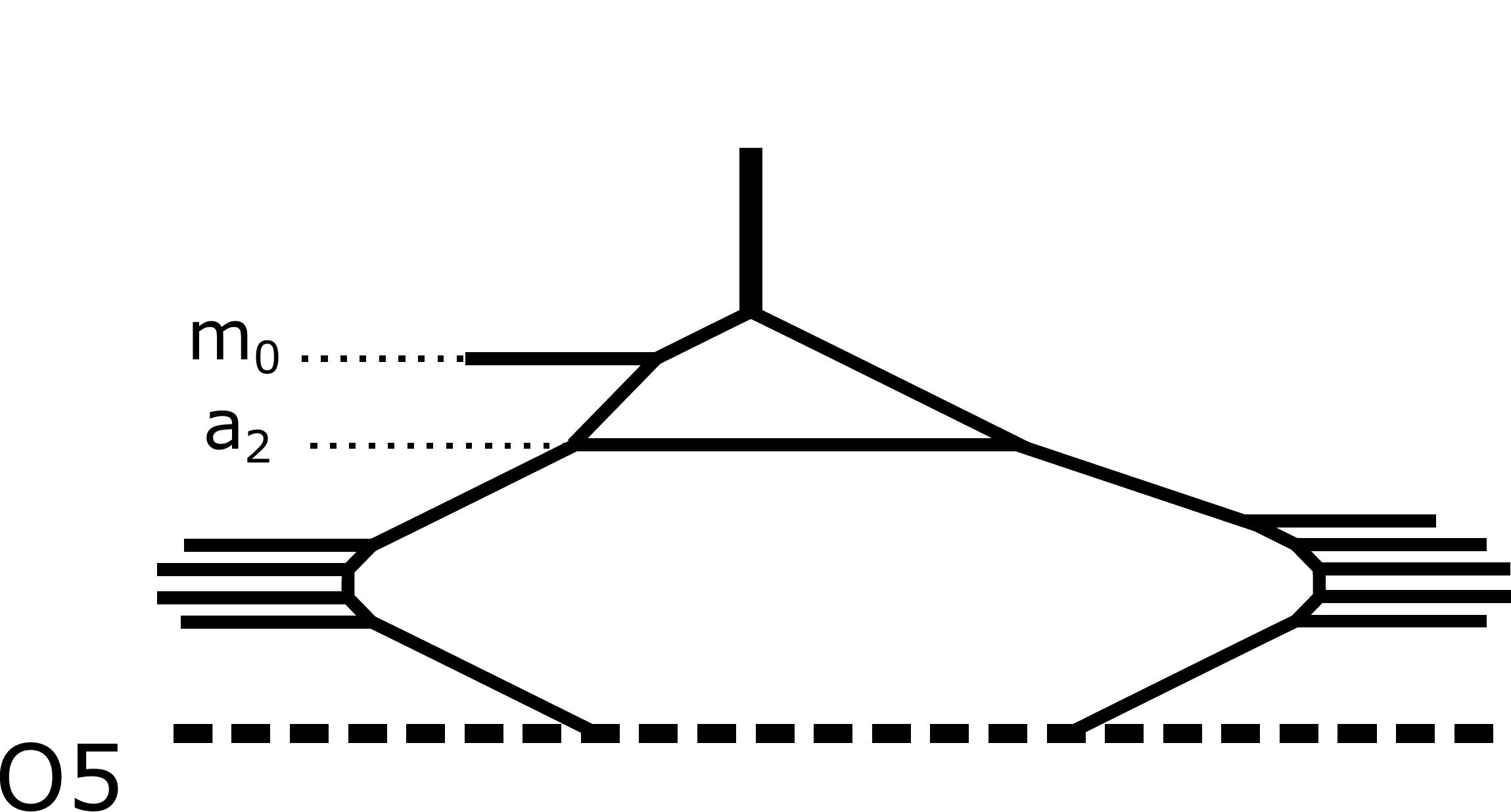}}
\subfigure[]{\label{fig:Sp2v3}
\includegraphics[width=7cm]{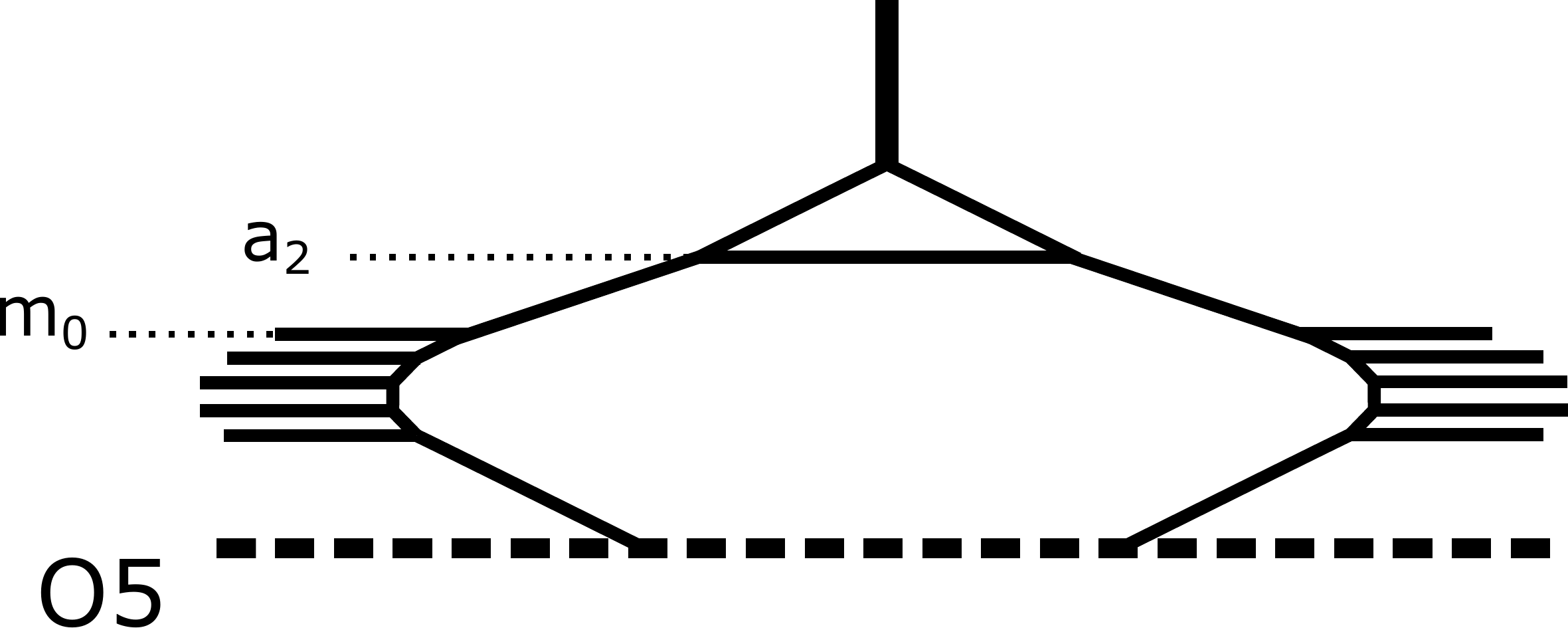}}
\caption{(a). The 5-brane web diagram for the $Sp(2)$ gauge theory with nine flavors. (b). Lowering the left top D5-brane in Figure \ref{fig:Sp2v1} until the its height is lower than the that of the top color brane. (c). The diagram after performing a flop transition with respect to a line of the length $m_0 - a_2$ in Figure \ref{fig:Sp2v2}.}
\label{fig:Sp2web}
\end{figure}
Since there is no shift of $a_1$ and $a_2$ which makes \eqref{rk2eff1} invariant under the Weyl reflection,  
we claim that \eqref{rk2IMS} is actually not valid on the CFT phase.
In fact it is natural that the prepotential \eqref{rk2IMS} is not valid on the CFT phase since \eqref{rk2IMS} assumes $m_0$ is larger than the other parameters. There can be further phase transition when $m_0$ is smaller than the Coulomb branch moduli. In order to see it we make use of a 5-brane web diagram for the $Sp(2)$ gauge theory with $9$ flavors in Figure \ref{fig:Sp2v1}. The diagram describes the theory in the parameter region $m_0 > a_1 > a_2 > |m_f| > 0$.  Then it is possible to change $m_0$ into a smaller value and the diagram in Figure \ref{fig:Sp2v1} becomes the one in Figure \ref{fig:Sp2v2} when $a_1 > m_0 > a_2 > |m_f| > 0$. We can further move to the parameter region $a_1 > a_2 > m_0 > |m_f| > 0$ by performing a flop transition with respect to a line of the length $m_0 - a_2$, which yields the diagram in Figure \ref{fig:Sp2v3}. Therefore the prepotential on the CFT phase should be given by
\begin{equation}\label{rk2CFT}
\begin{split}
\mathcal{F}_{\text{CFT}} =& \mathcal{F}_{\text{IMS}} - \frac{1}{6}(a_2 - m_0)^3\\
=& \frac{1}{6}(a_1^3 - 2a_2^3) + a_1a_2^2 + \frac{1}{2}m_0(a_1^2 + 2a_2^2) - \frac{1}{2}\sum_{i=1}^9m_i^2a_1 - \frac{1}{2}\sum_{i=0}^9m_i^2a_2 + \frac{1}{6}m_0^3. 
\end{split}
\end{equation}
Then taking a derivative of the prepotential \eqref{rk2IMS1} with respect to Coulomb branch moduli twice yields 
\begin{align}\label{rk2eff1-a}
\frac{\partial \mathcal{F}_{\text{CFT}} }{\partial a_1^2} = a_1 + m_0, \quad \frac{\partial \mathcal{F}_{\text{CFT}} }{\partial a_2^2} = 2a_1 - 2a_2 + 2m_0, \quad \frac{\partial \mathcal{F}_{\text{CFT}} }{\partial a_1a_2} = 2a_2,
\end{align}
and we can define invariant Coulomb branch moduli $\ia_1, \ia_2$ as
\begin{equation}\label{rk2invCB}
\ia_1 = a_1 + m_0, \qquad  \ia_2 = a_2.
\end{equation}
Indeed when we rewrite the prepotential \eqref{rk2CFT} in terms of the invariant Coulomb branch moduli, the prepotential becomes 
\begin{equation}
\mathcal{F}_{\text{CFT}} =  \frac{1}{6}(\ia_1^3 - 2\ia_2^3) + \ia_1\ia_2^2 + \frac{1}{2}\sum_{i=0}^9m_i^2(\ia_1 - \ia_2)+ \frac{1}{2}m_0\sum_{i=0}^9m_i^2,
\end{equation}
which is invariant under the Weyl group symmetry of $SO(20)$ up to the constant term which only depends on the mass parameters. Since the constant term does not affect any physical quantity  we will use the following effective prepotential on the CFT phase,
\begin{equation}\label{rk2CFT1}
\mathcal{F}_{\text{CFT}} =  \frac{1}{6}(\ia_1^3 - 2\ia_2^3) + \ia_1\ia_2^2 - \frac{1}{2}\sum_{i=0}^9m_i^2(\ia_1 + \ia_2).
\end{equation}

Using the effective prepotential on the CFT phase the effective prepotential we have now is given by 
\begin{equation}\label{rk2FSpv1}
\begin{split}
\mathcal{F} = 
& \mathcal{F}_{\text{CFT}}
+ \frac{1}{6}\sum_{i=1}^9\relu{\ia_1 - m_0 \pm m_i}^3 + \frac{1}{6}\sum_{i=1}^9\relu{\ia_2 \pm m_i}^3 + \frac{1}{6}\relu{\ia_2 - m_0}^3.
\end{split}
\end{equation}
It is straightforward to make the prepotential \eqref{rk2FSpv1} invariant under the Weyl group  symmetry of $SO(20)$. The Weyl reflection invariant prepotential is then 
\begin{equation}\label{rk2FSpv2}
\begin{split}
\mathcal{F} = & \mathcal{F}_{\text{CFT}}
+ \frac{1}{6}\sum_{0 \leq i < j \leq 9}\relu{\ia_1 \pm m_i \pm m_j}^3 + \frac{1}{6}\sum_{i=0}^9\relu{\ia_2 \pm m_i}^3 .
\end{split}
\end{equation}

\begin{figure}[t]
\centering
\subfigure[]{\label{fig:Sp2v4}
\includegraphics[width=7cm]{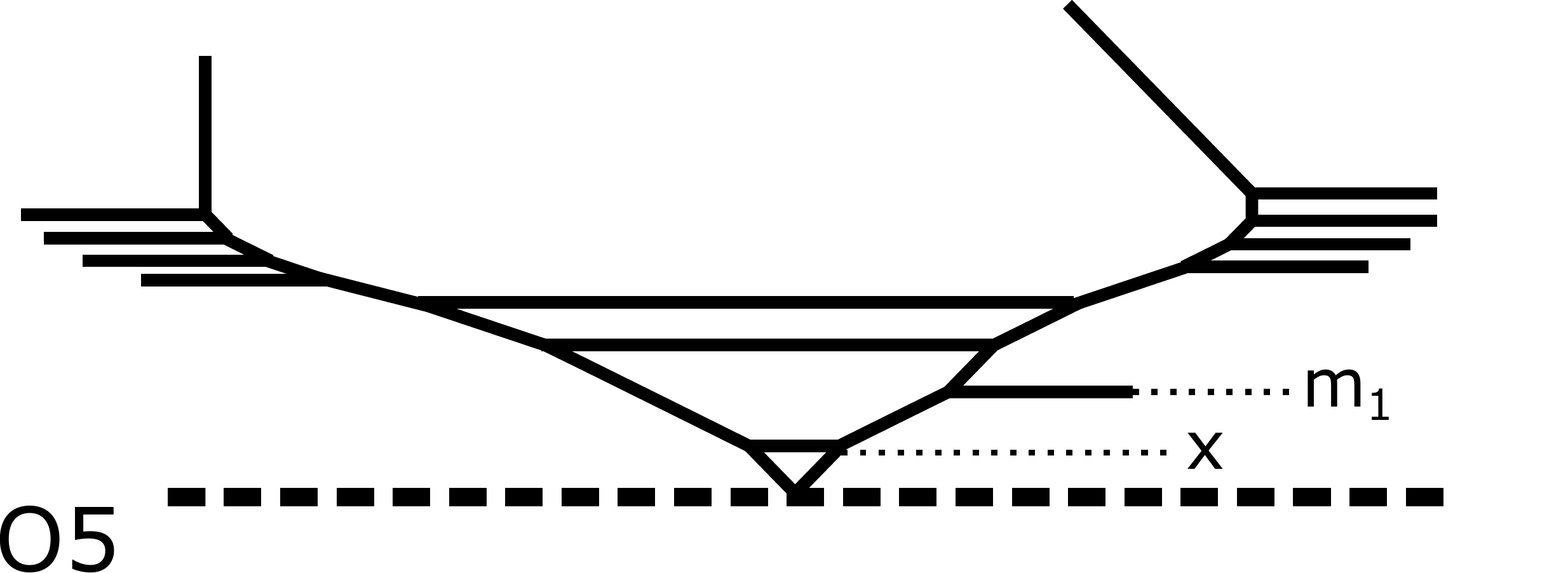}}
\subfigure[]{\label{fig:Sp2v5}
\includegraphics[width=7cm]{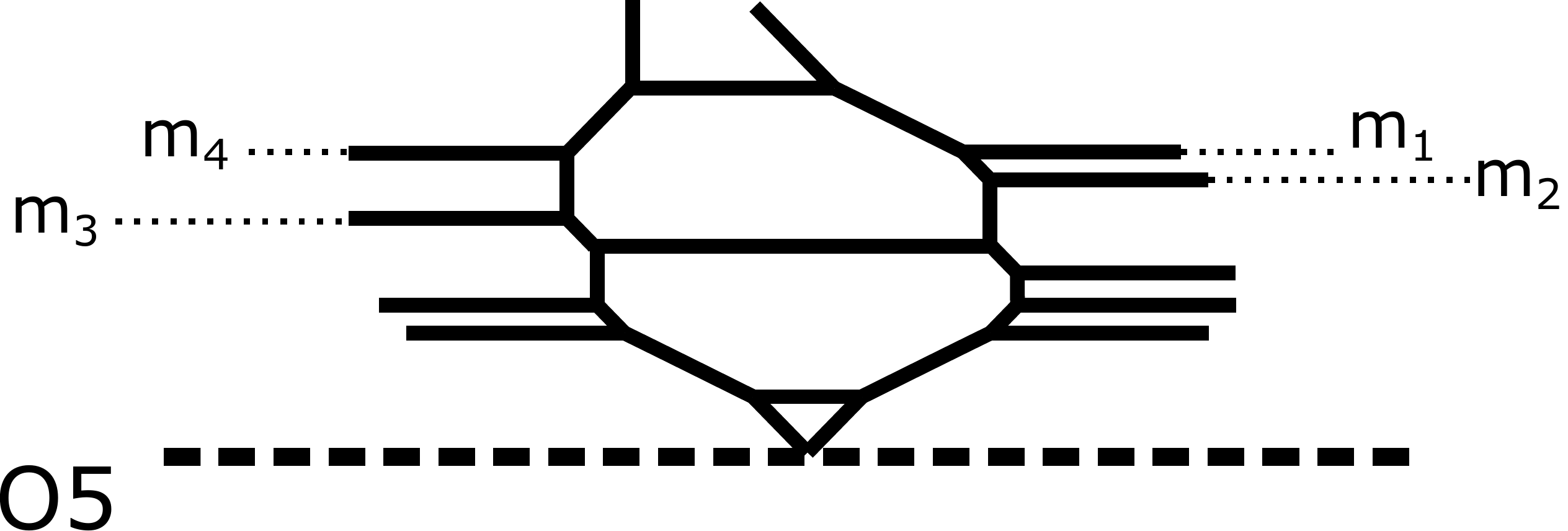}}
\caption{(a). The 5-brane web diagram for the $Sp(2)$ gauge theory with nine flavors in a phase where the line of the length \eqref{rk2matter1} can be seen. (b). The 5-brane web diagram for the $Sp(2)$ gauge theory with nine flavors in a phase where the line of the length \eqref{rk2matter2} can be seen.}
\label{fig:Sp2flop}
\end{figure}
In fact it turns out that the prepotential \eqref{rk2FSpv2} is not complete. The 5-brane web diagrams in certain phases imply that there are lines which can be flopped with length which does not appear in \eqref{rk2FSpv2}. The web diagram in such a phase is depicted in Figure \ref{fig:Sp2v4}. Then there is a floppable line with the length 
\begin{equation}\label{rk2matter1}
m_1 - x = \ia_1 + \ia_2 -\frac{1}{2}m_0 + \frac{1}{2}m_1  - \frac{1}{2}\sum_{i=2}^9m_i.
\end{equation}
We can also consider the 5-brane web in Figure \ref{fig:Sp2v5}. Then the length of the middle internal D5-brane is 
\begin{equation}\label{rk2matter2}
2a_1 + a_2 + m_0 - \sum_{i=1}^4m_i = 2\ia_1 + \ia_2 - \sum_{i=0}^4m_i,
\end{equation}
which can be also flopped. Then including the contributions \eqref{rk2matter1} and \eqref{rk2matter2} and making the prepotential invariant under the Weyl group symmetry of $SO(20)$ is 
\begin{equation}\label{rk2FSp}
\begin{split}
\mathcal{F} = & \frac{1}{6}(\ia_1^3 - 2\ia_2^3) + \ia_1\ia_2^2 - \frac{1}{2}\sum_{f=0}^9m_f^2(\ia_1 + \ia_2)\\
&+ \frac{1}{6}\sum_{0 \leq i < j \leq 9}\relu{\ia_1 \pm m_i \pm m_j}^3 + \frac{1}{6}\sum_{i=0}^9\relu{\ia_2 \pm m_i}^3\\
&+\frac{1}{6}\sum_{\substack{\{ s_i = \pm 1\} \\\text{odd $+$}}} 
\relu{\ia_1 + \ia_2 + \frac{1}{2}\sum_{i=0}^9s_im_i}^3\\
& + \frac{1}{6}\sum_{0\leq i_1 < i_2 < i_3 < i_4 < i_5 \leq 9}\relu{2\ia_1 + \ia_2 \pm m_{i_1} \pm  m_{i_2} \pm m_{i_3} \pm m_{i_4} \pm m_{i_5}}^3.
\end{split}
\end{equation}
We argue that the prepotential \eqref{rk2FSp} is the complete prepotential of the $Sp(2)$ gauge theory with $9$ flavors. 
In other words, the prepotential can be used in any region in the physical Coulomb branch moduli space of the theory. 
It is also straightforward to check that it reduces to the IMS prepotential \eqref{rk2IMS} in the weak coupling region
$m_0 \gg |m_f|, |a_I|$. 
In the rest of this section we will provide more support for \eqref{rk2FSp} and give a physical explanation of the necessity of the new terms. 

\subsection{Consistency with dualities}
\label{sec:rk2dualities}

We have obtained the complete prepotential of the $Sp(2)$ gauge theory with $9$ flavors \eqref{rk2FSp} by including the region where instanton particles become massless. The $Sp(2)$ gauge theory with $9$ flavors is dual to the $SU(3)$ gauge theory with $9$ flavors and the CS level $\kappa = \pm\frac{1}{2}$ and also to the $[4]- SU(2) - SU(2) - [3]$ quiver theory. Therefore the complete prepotential \eqref{rk2FSp} should reproduce the IMS prepotentials of the $SU(3)$ gauge theory and the $SU(2) \times SU(2)$ gauge theory in the weak coupling limit. We will see that the IMS prepotentials of the dual theories are indeed reproduced from the complete prepotential of the $Sp(2)$ gauge theory. In fact, the additional terms which we added by hand from the 5-brane webs are necessary to realize the IMS prepotentials of the dual theories. 

Let us first consider the dual $SU(3)$ gauge theory with $9$ flavors and the CS level $\kappa = \pm \frac{1}{2}$. The duality map in the case of $10$ flavors has been obtained in \cite{Hayashi:2016abm, Yun:2016yzw}. It is straightforward to obtain the duality map after decoupling one flavor and it is given by 
\begin{equation}\label{SpSUmap}
\begin{split}
m_0^{Sp} &=\frac{3}{4} m_0^{SU} + \frac{1}{4}\sum_{i=1}^9m_i^{SU},\\
m_i^{Sp} &= m_i^{SU} + \frac{1}{4}(m_0^{SU} - \sum_{i=1}^9m_i^{SU}), \quad (i=1, \cdots, 9)\\
a_i^{Sp} &= a_i^{SU} + \frac{1}{4}(m_0^{SU} - \sum_{i=1}^9m_i^{SU}), \quad (i=1, 2).
\end{split}
\end{equation}
By substituting the duality map \eqref{SpSUmap} into complete prepotential \eqref{rk2FSp}, the complete prepotential is now written in terms of the parameters of the $SU(3)$ gauge theory. For obtaining the IMS prepotential we need to consider 
the weak coupling region $m_0 \gg |m_f|, |a_I|$. 
Then most of the terms disappear and the remaining terms are 
\begin{align}
\sum_{0 \leq i < j \leq 9}&\relu{\ia_1^{Sp} \pm m_i^{Sp} \pm m_j^{Sp}}^3\to \sum_{i=1}^9\relu{a_1^{SU} - m_i^{SU}}^3,\label{SpSUmap1}\\
\sum_{i=0}^9&\relu{\ia_2^{Sp} \pm m_i^{Sp}}^3\to \left(a_2^{SU} - \frac{1}{2}\sum_{f=0}^9m_f^{SU}\right)^3 + \sum_{i=1}^9\relu{a_2^{SU} - m_i^{SU}}^3,\label{SpSUmap2}\\
\sum_{\substack{ \{ s_i = \pm 1 \} \\\text{odd $+$}}}&\relu{\ia_1^{Sp} + \ia_2^{Sp} + \frac{1}{2}\sum_{i=0}^9s_im_i^{Sp}}^3\to\sum_{i=1}^9\relu{a_1^{SU} + a_2^{SU} + m_i^{SU}}^3\nn\\
&\hspace{4.8cm}=\sum_{i=1}^9\relu{-a_3^{SU} + m_i^{SU}}^3,\label{SpSUmap3}\\
\sum_{0\leq i_1 < i_2 < i_3 < i_4 < i_5 \leq 9}&\relu{2\ia_1^{Sp} + \ia_2^{Sp} \pm m_{i_1}^{Sp} \pm  m_{i_2}^{Sp} \pm m_{i_3}^{Sp} \pm m_{i_4}^{Sp} \pm m_{i_5}^{Sp}}^3\to 0. \label{SpSUmap4}
\end{align}
Summing all the terms \eqref{SpSUmap1}-\eqref{SpSUmap4} with the prepotential on the CFT phase \eqref{rk2CFT1} yields 
\begin{equation}\label{FSptoSU}
\begin{split}
\mathcal{F}_{\text{weak}}^{SU(3)} 
&=\frac{1}{6}\left((a_1^{SU})^3-(a_2^{SU})^3\right) + a_1^{SU}\left(a_2^{SU}\right)^2 + \frac{1}{2}m_0^{SU}\left((a_1^{SU})^2 + a_1^{SU}a_2^{SU} +(a_2^{SU})^2\right)\\
&\quad-\frac{1}{2}\sum_{i=1}^9m_i^{SU}a_1^{SU}a_2^{SU} -\frac{1}{2}\sum_{i=1}^9(a_1^{SU} + a_2^{SU})\left(m_i^{SU}
 \right)^2\\
&\quad+\frac{1}{6}\sum_{i=1}^9\left(\relu{a_1^{SU} - m_i^{SU}}^3 +\relu{a_2^{SU} - m_i^{SU}}^3 + \relu{-a_3^{SU} + m_i^{SU}}^3\right),
\end{split}
\end{equation}
where we omitted constant terms which do not depend on the Coulomb branch moduli of the $SU(3)$ gauge theories. 
Using the identity \eqref{eq:doubleAbs},
we find that this agree with the IMS prepotential
\begin{equation}\label{FSU}
\begin{split}
\mathcal{F}_{\text{IMS}}^{SU(3)} &= \frac{1}{4}m_0^{SU}\left((a_1^{SU})^2 + (a_2^{SU})^2 + (a_3^{SU})^2\right) + \frac{1}{12}\left((a_1^{SU})^3 + (a_2^{SU})^3 + (a_3^{SU})^3\right) \\
&\quad+\frac{1}{6}\left((a_1^{SU} - a_2^{SU})^3 + (a_1^{SU} - a_3^{SU})^3 + (a_2^{SU} - a_3^{SU})^3\right) \\
&\quad-\frac{1}{12}\sum_{i=1}^3\sum_{j=1}^9\left| a_i^{SU} - m_j^{SU} \right|^3
\end{split}
\end{equation}
for the $SU(3)$ gauge theory with $9$ flavors and the CS level $\kappa = \frac{1}{2}$ up to the constant terms.

We can also see from \eqref{SpSUmap3} that a part of the terms associated with the conjugate spinor representation gives perturbative contributions in the $SU(3)$ gauge theory. 
Therefore, the contribution in the conjugate spinor representation, which was added by hand from the discussion with the 5-brane web diagram in section \ref{sec:rank2prep}, turns out to be necessary to make the duality hold.

\begin{figure}[t]
\centering
\subfigure[]{\label{fig:Sp1Sp1v1}
\includegraphics[width=7cm]{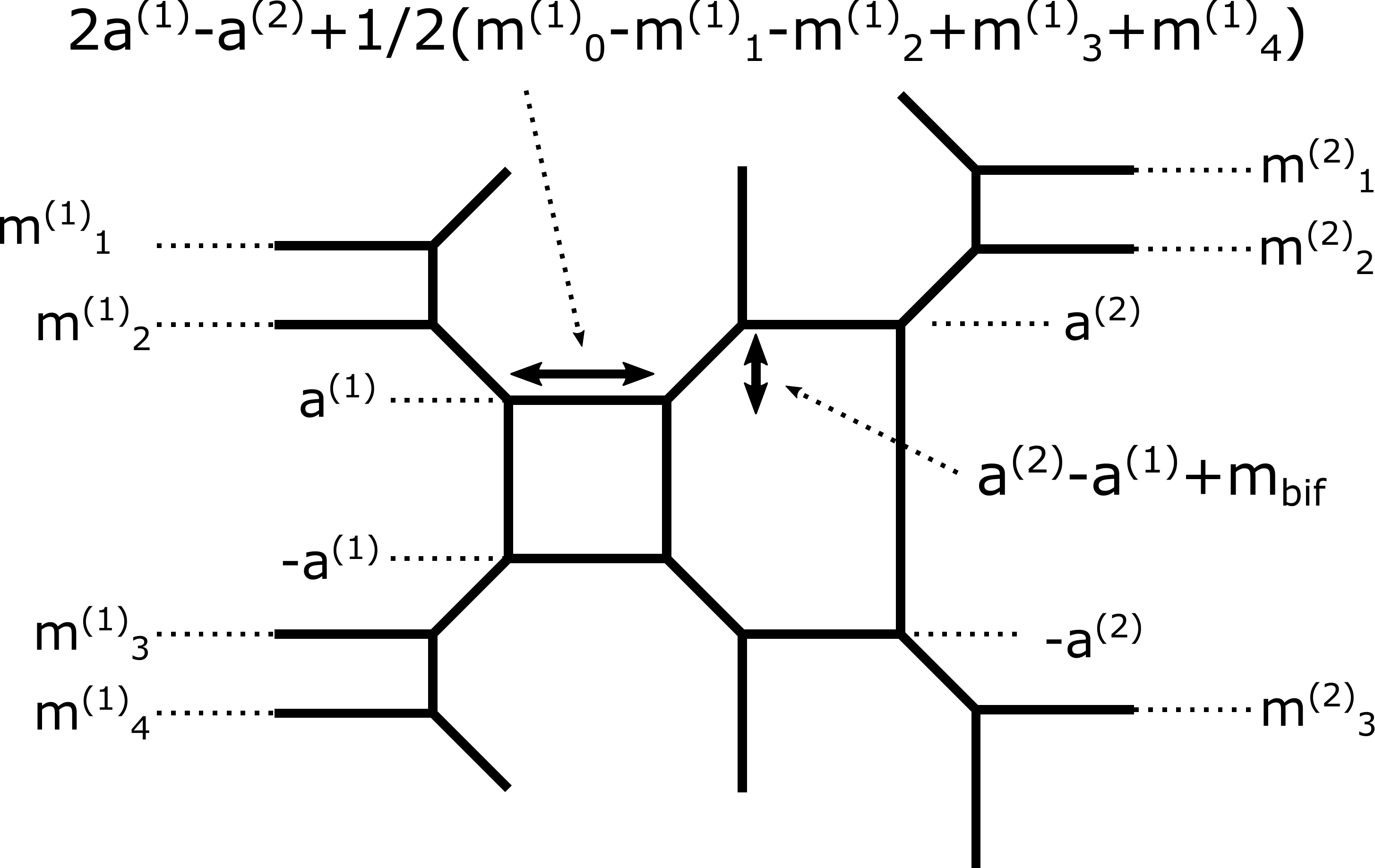}}
\subfigure[]{\label{fig:Sp1Sp1v2}
\includegraphics[width=5cm]{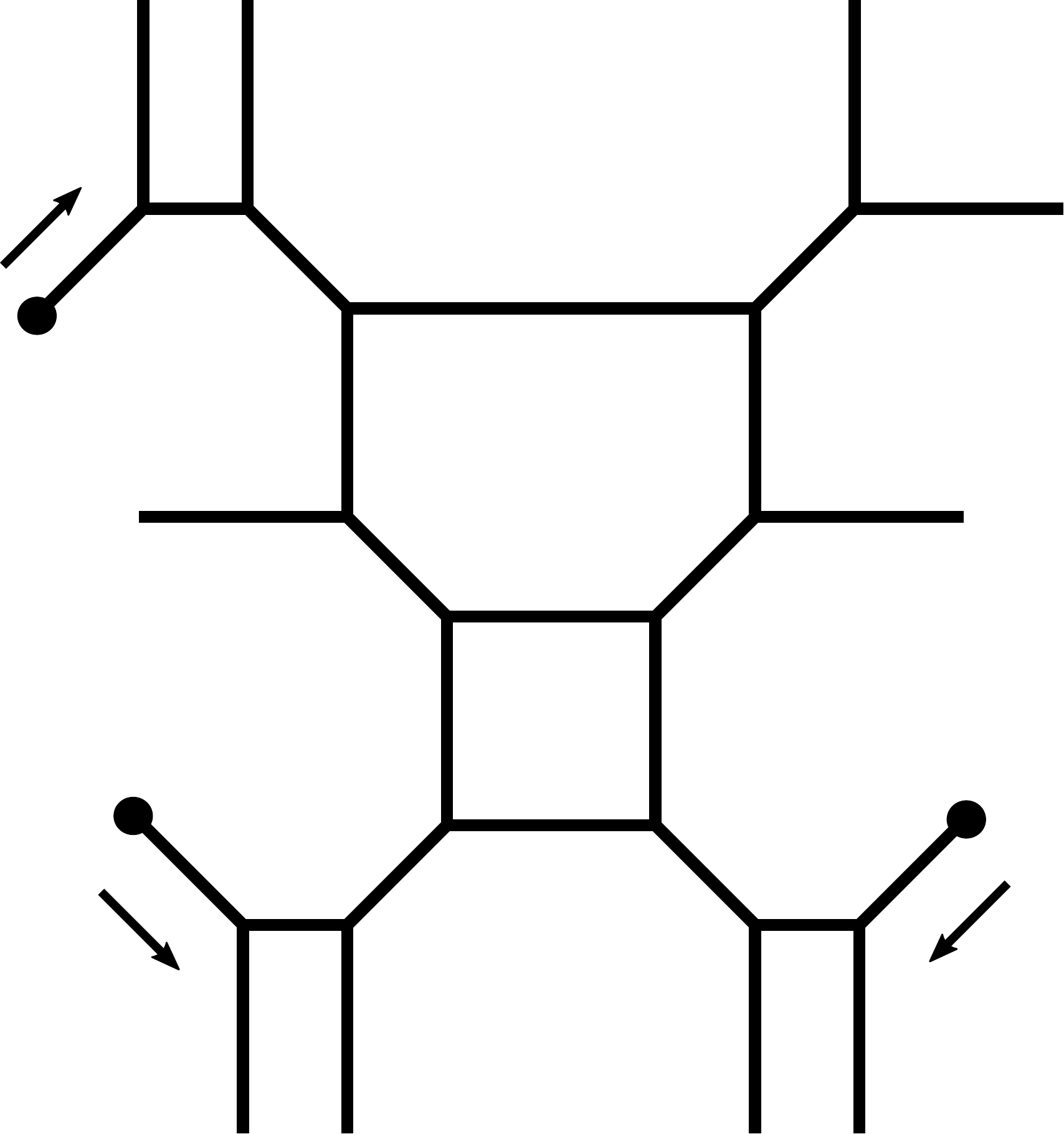}}
\subfigure[]{\label{fig:Sp1Sp1v3}
\includegraphics[width=5cm]{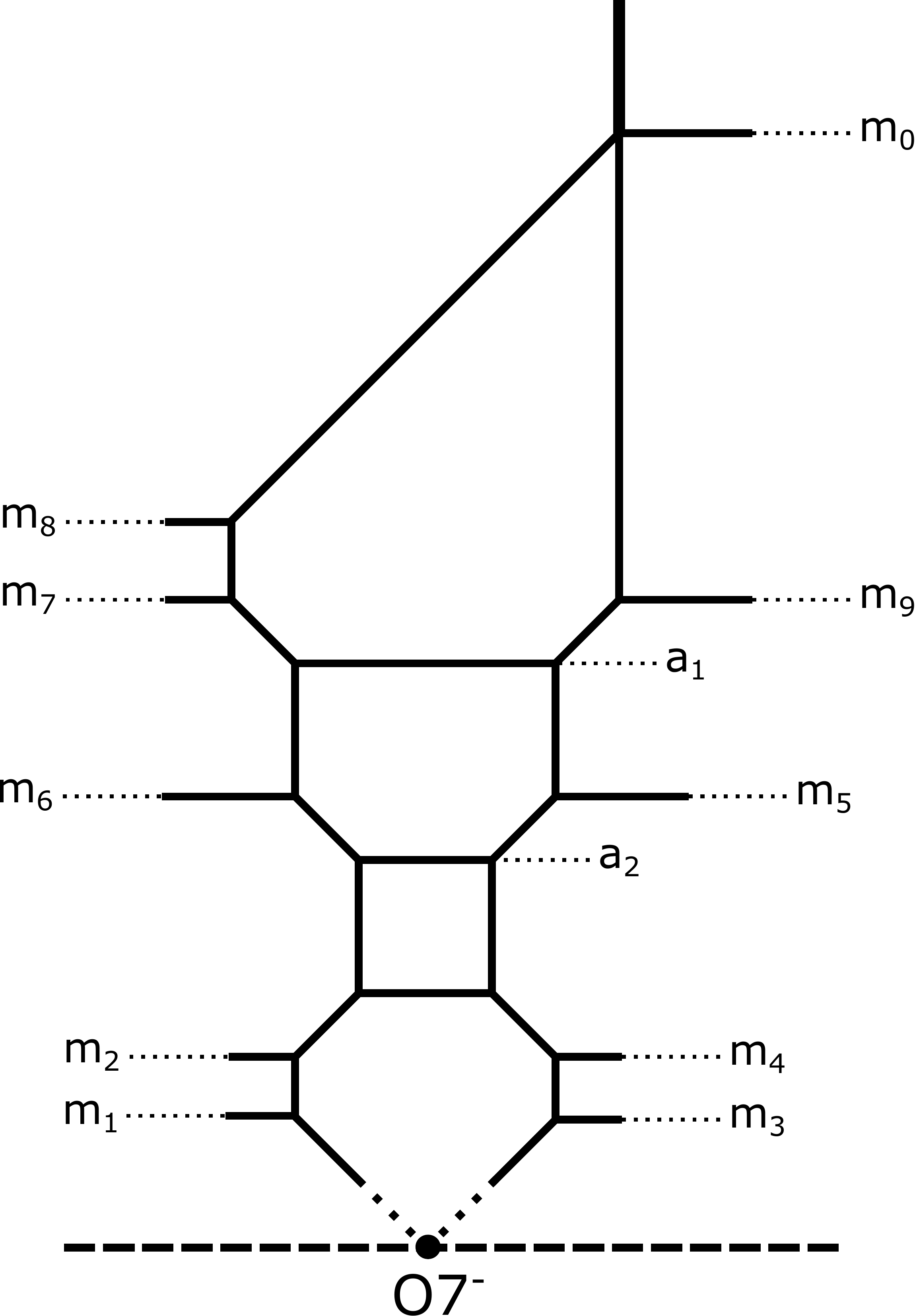}}
\caption{(a). The 5-brane web diagram for the quiver theory $[4] - SU(2) - SU(2) - [3]$ and the gauge theory parameterization. (b). S-dual to the diagram in Figure \ref{fig:Sp1Sp1v1}. The black dots represent 7-branes. (c). The 5-brane web diagram for the $Sp(2)$ gauge theory with $9$ flavors and the gauge theory parameterization. The black dot in this figure represents an O7$^-$-plane.}
\label{fig:Sp1Sp1web}
\end{figure}
The $Sp(2)$ gauge theory with $9$ flavors has another dual frame given by the quiver gauge theory $[4] - SU(2) - SU(2) - [3]$. It is also possible to obtain the duality map using 5-brane webs in Figure \ref{fig:Sp1Sp1web}. The diagram in Figure \ref{fig:Sp1Sp1v1} shows the parameterization for the $SU(2) \times SU(2)$ quiver theory. The S-dual of the diagram is simply given by rotating the diagram by $\pi$ as in Figure \ref{fig:Sp1Sp1v2}. We further move 7-branes in Figure \ref{fig:Sp1Sp1v2} in the directions indicated by the arrows in the figure so that we can read off the parameterization of the $Sp(2)$ gauge theory. The resulting diagram after moving the 7-branes is given in Figure \ref{fig:Sp1Sp1v3}. In order to see the parameterization in terms of the $Sp(2)$ gauge theory, the flavor mass parameters and the Coulomb branch moduli should be measured from the location of an O7$^-$-plane. The parameterization of the $Sp(2)$ gauge theory can be read off as in Figure \ref{fig:Sp1Sp1v3}. Comparing the two parameterizations, the explicit map between mass parameters is given by
\begin{equation}\label{Spquivermap1}
\begin{split}
m_1^{Sp} =& \frac{1}{2}(-m_1^{(1)} + m_2^{(1)} - m_3^{(1)} - m_4^{(1)}), \quad m_2^{Sp} = \frac{1}{2}(m_1^{(1)} - m_2^{(1)} - m_3^{(1)} - m_4^{(1)}),\\
m_3^{Sp} =& \frac{1}{2}(m_1^{(1)} + m_2^{(1)} - m_3^{(1)} + m_4^{(1)}), \quad m_4^{Sp} = \frac{1}{2}(m_1^{(1)} + m_2^{(1)} + m_3^{(1)} - m_4^{(1)}),\\
m_5^{Sp} =& \frac{1}{2}m_0^{(1)} - m_{\text{bif}}, \quad m_6^{Sp} = \frac{1}{2}m_0^{(1)} + m_{\text{bif}},\\
m_7^{Sp} =& \frac{1}{2}(m_0^{(1)} + m_0^{(2)} - m_1^{(2)} + m_2^{(2)} + m_3^{(2)}), \quad m_8^{Sp} = \frac{1}{2}(m_0^{(1)} + m_0^{(2)} + m_1^{(2)} - m_2^{(2)} + m_3^{(2)}),\\
m_9^{Sp} =& \frac{1}{2}(m_0^{(1)} + m_0^{(2)} - m_1^{(2)} - m_2^{(2)} - m_3^{(2)}), \quad m_0^{Sp} = \frac{1}{2}(m_0^{(1)} + m_0^{(1)} + m_1^{(2)} + m_2^{(2)} - m_3^{(2)}).
\end{split}
\end{equation}
Also the map for the Coulomb branch moduli is given by
\begin{equation}\label{Spquivermap2}
\begin{split}
a_1^{Sp} =& a^{(2)} + \frac{1}{2}(m_0^{(1)} + m_0^{(2)} - m_1^{(2)} - m_2^{(2)} + m_3^{(2)}),\\
a_2^{Sp} =& a^{(1)} - a^{(2)} + \frac{1}{2}m_0^{(1)}.
\end{split}
\end{equation}

We can then substitute the duality map \eqref{Spquivermap1} and \eqref{Spquivermap2} into complete prepotential \eqref{rk2FSp}. For obtaining the IMS prepotential we need to consider 
the weak coupling region $m_0^{(1)}, m_0^{(2)} \gg |m_f^{(1)}|, |m_f^{(2)}|, |a^{(1)}|, |a^{(2)}|$.
Then most of the terms again disappear and the remaining terms are 
\begin{align}
\sum_{0 \leq i < j \leq 9}&\relu{\ia_1^{Sp} \pm m_i^{Sp} \pm m_j^{Sp}}^3 \to \sum_{i=1}^3 \relu{a_2^{(2)} \pm m_i^{(2)}}^3 ,\label{Spx3map1}\\
\sum_{i=0}^9&\relu{\ia_2^{Sp} \pm m_i^{Sp}}^3\to \left((a^{(1)} + a^{(2)}) + \frac{1}{2}\left(-m_0^{(2)} + m_1^{(2)} + m_2^{(2)} + m_3^{(2)}\right)\right)^3\nn \\
&\hspace{3cm}+\left((a^{(1)} + a^{(2)}) + \frac{1}{2}\left(-m_0^{(2)} + m_1^{(2)} - m_2^{(2)} - m_3^{(2)}\right)\right)^3\nn\\
&\hspace{3cm}+\left((a^{(1)} + a^{(2)}) + \frac{1}{2}\left(-m_0^{(2)} - m_1^{(2)} + m_2^{(2)} - m_3^{(2)}\right)\right)^3\nn\\
&\hspace{3cm}+\left((a^{(1)} + a^{(2)}) + \frac{1}{2}\left(-m_0^{(2)} - m_1^{(2)} - m_2^{(2)} + m_3^{(2)}\right)\right)^3\nn\\
&\hspace{3cm} + \relu{a^{(1)}-a^{(2)} \pm m_{\text{bif}}}^3,\label{Spx3map2}\\
\sum_{\substack{\{ s_i = \pm 1 \}\\\text{odd $+$}}}&\relu{\ia_1^{Sp} + \ia_2^{Sp} + \frac{1}{2}\sum_{i=0}^9s_im_i^{Sp}}^3\to\sum_{i=1}^4 \relu{a^{(1)} \pm m_i^{(1)}}^3 \label{Spx3map3}\\
\sum_{0\leq i_1 < i_2 < i_3 < i_4 < i_5 \leq 9}&\relu{2\ia_1^{Sp} + \ia_2^{Sp} \pm m_{i_1}^{Sp} \pm  m_{i_2}^{Sp} \pm m_{i_3}^{Sp} \pm m_{i_4}^{Sp} \pm m_{i_5}^{Sp}}^3\to \relu{a^{(1)}+a^{(2)} \pm m_{\text{bif}}}^3. \label{Spx3map4}
\end{align}

Summing all the terms \eqref{Spx3map1}-\eqref{Spx3map4} with the prepotential on the CFT phase \eqref{rk2CFT1} yields 
\begin{equation}\label{FSptoSpx2}
\begin{split}
\mathcal{F}_{\text{weak}}^{SU(2)\times SU(2)} 
=&\frac{1}{6}\left(2(a^{(1)})^3 + 5(a^{(2)})^3 -6a^{(1)}(a^{(2)})^2\right) + \frac{1}{2}m_0^{(1)}(a^{(1)})^2 + \frac{1}{2}m_0^{(2)}(a^{(2)})^2\\
&-\frac{1}{2}a^{(1)}\sum_{i=1}^4(m_i^{(1)})^2-\frac{1}{2}a^{(2)}\sum_{i=1}^3(m_i^{(2)})^2 -a^{(1)}(m_{\text{bif}})^2 \\
&+\frac{1}{6}\sum_{i=1}^4 \relu{a^{(1)} \pm m_i^{(1)}}^3
+\frac{1}{6}\sum_{i=1}^3 \relu{a^{(2)} \pm m_i^{(2)}}^3 \\
&+\frac{1}{6} \relu{a^{(1)} \pm a^{(2)} \pm  m_{\text{bif}}}^3,
\end{split}
\end{equation}
where we again ignore terms which do not depend on the Coulomb branch moduli of the $SU(2) \times SU(2)$ gauge theory. 
We find that this agrees with the IMS prepotential 
\begin{equation}\label{FSpx2IMS}
\begin{split}
\mathcal{F}_{\text{IMS}}^{SU(2)\times SU(2)} 
=& \frac{1}{2}\left(m_0^{(1)}(a^{(1)})^2 + m_0^{(2)}(a^{(2)})^2\right) + \frac{4}{3}\left((a^{(1)})^3 + (a^{(2)})^3\right) \\
&-\frac{1}{12}\sum_{j=1}^4 \left|a^{(1)} \pm m_j^{(1)}\right|^3
-\frac{1}{12}\sum_{j=1}^3 \left|a^{(2)} \pm m_j^{(2)}\right|^3\\
&-\frac{1}{12} \left|a^{(1)} \pm a^{(2)} \pm m_{\text{bif}}\right|^3.
\end{split}
\end{equation}
for the $[4] - SU(2) - SU(2) - [3]$ quiver theory up to the constant terms.

Note that the contribution of the bi-fundamental hypermultiplet 
in \eqref{FSptoSpx2} comes from the contribution in the rank $5$ antisymmetric representation of $SO(20)$ as in \eqref{Spx3map4}. The contributions was originally inferred from the analysis of the 5-brane web diagram. The duality to the $SU(2) \times SU(2)$ quiver theory shows that a part of the contribution is in fact the perturbative contributions in the quiver theory. Hence the terms in the rank $5$ antisymmetric representation are necessary to make the duality hold. 


\subsection{Consistency with Higgsing}

In section \ref{sec:rk2dualities} we have seen the complete prepotential \eqref{rk2FSp} reproduces the IMS prepotential in the weak coupling region of the dual theories. It is also possible to check the consistency with the Higgsing to the $Sp(1)$ gauge theory with $7$ flavors from the $Sp(2)$ gauge theory with $9$ flavors. The Higgsing can be carried out by tuning the gauge theory parameters as 
\begin{equation}\label{higgstuning}
a_2 = m_i = -m_j,
\end{equation} 
where $i$ and $j$ are one of $1, 2, \cdots, 9$ with $i \neq j$. Then inserting \eqref{higgstuning} into the complete prepotential \eqref{rk2FSp} should reproduce the complete prepotential \eqref{eq:invFforE8} of the $Sp(1)$ gauge theory with $7$ flavors up to terms which do not depend on the Coulomb branch modulus $\ia$.

In order to see which terms remain after inserting the tuning conditions \eqref{higgstuning}, we first need to determine the physical Coulomb branch moduli space of the $Sp(2)$ gauge theory with $9$ flavors. The explicit derivation will be done in section \ref{sec:Rank2GV-prep} and the we here use the final result which is given by 
\begin{align}
\ia_2 &\ge 0,\label{rk2SpCB1}
\\
\ia_1 + 3 \ia_2 &\ge w_{[0000000001]} \cdot \vec{m},\label{rk2SpCB2}
\\
\ia_1 + 2 \ia_2 &\ge w_{[0000000010]} \cdot \vec{m},\label{rk2SpCB3}
\\
\ia_1 - \ia_2 &\ge w_{[1000000000]} \cdot \vec{m},\label{rk2SpCB4}
\\
\ia_1 - 2 \ia_2 &\ge 0,\label{rk2SpCB5}
\\
2 \ia_1 + 3 \ia_2 &\ge w_{[0000001000]} \cdot \vec{m},\label{rk2SpCB6}
\\
2 \ia_1 + 2 \ia_2 &\ge w_{[0000010000]} \cdot \vec{m},\label{rk2SpCB7}
\\
2 \ia_1 &\ge w_{[0001000000]} \cdot \vec{m},\label{rk2SpCB8}
\\
2 \ia_1 - \ia_2 &\ge w_{[0010000000]} \cdot \vec{m},\label{rk2SpCB9}
\\
3 \ia_1 &\ge w_{[1000100000]} \cdot \vec{m},\label{rk2SpCB10}
\\
3 \ia_1 - 3 \ia_2 &\ge w_{[0000100010]} \cdot \vec{m}.\label{rk2SpCB11}
\end{align}
Here, the representation is expressed by the Dynkin label $[n_1, n_2,\cdots, n_{10}]$ of $SO(20)$. Especially, $[0000000010]$ is the spinor representation ${\bf 512}$ and 
$[0000000001]$ is the conjugate spinor representation $\overline{\bf  512}$. 

Let us then look at each term in \eqref{rk2FSp} together with the conditions \eqref{higgstuning}. We also relabel 
the mass parameters except for $m_i$ and $m_j$ appeared in \eqref{higgstuning} by $m_1, m_2, \cdots, m_7$ after inserting the tuning conditions \eqref{higgstuning}. The first line of \eqref{rk2FSp} with \eqref{higgstuning} imposed becomes 
\begin{align}
\frac{1}{6}(\ia_1^3 -2\ia_2^3) + \ia_1\ia_2^2 - \frac{1}{2}\sum_{k=0}^9m_k^2(\ia_1 + \ia_2) \to \frac{1}{6}\ia_1^3 - \frac{1}{2}\sum_{k=0}^7m_k^2\ia_1, \label{higgs1}
\end{align}
where the arrow indicates that we omit terms which do not depend on the Coulomb branch modulus $\ia_1$. 

Next we consider the first term in the second line of \eqref{rk2FSp} which is characterized by the linear combinations,
\begin{equation}\label{FSprk2}
\ia_1 \pm m_{i_1} \pm m_{i_2}.
\end{equation} 
When both $m_i$ and $m_j$ appear in \eqref{FSprk2}, \eqref{FSprk2} becomes
\begin{equation}
\ia_1 \pm \ia_2 \pm \ia_2,
\end{equation}
Depending on the signs in front of $\ia_2$ we have three types for the linear combinations which satisfy the following inequalities,
\begin{equation}
\ia_1 + 2\ia_2 \geq \ia_1 \geq \ia_1 - 2\ia_2 \geq 0,
\end{equation}
where the first two inequalities come from \eqref{rk2SpCB1} and the last inequality comes from \eqref{rk2SpCB5}. Therefore we conclude that $\relu{\ia_1 \pm \ia_2 \pm \ia_2} = 0$ in the physical Coulomb branch moduli space. Another possibility is that one of $m_i$ and $m_j$ appears in \eqref{FSprk2}. In this case \eqref{FSprk2} becomes
\begin{equation}
\ia_1 \pm \ia_2 \pm m_{i_2},
\end{equation}
where $m_{i_2} \neq m_i, m_j$. There are two types of the linear combinations depending on the sign in from of $\ia_2$ and the inequalities they satisfy are
\begin{equation}
\ia_1 + \ia_2 \pm m_{i_2} \geq \ia_1 - \ia_2 \pm m_{i_2} \geq 0.
\end{equation}
Here the first inequality again comes from \eqref{rk2SpCB1} and the last inequality comes from \eqref{rk2SpCB4}. Therefore we again have  $\relu{\ia_1 \pm \ia_2 \pm m_{i_2}} = 0$. The last possibility is that neither of $m_i$ nor $m_j$ appears in \eqref{FSprk2}. In this case, the physical Coulomb branch moduli space conditions do not fix the sign of the linear combination $\ia_1 \pm m_{i_1} \pm m_{i_2}$ and hence these terms remain after the Higgsing. As for the second terms in the second line in \eqref{rk2FSp}, they become terms which do not depend on $\ia_1$ and we ignore the contributions.

We then move onto the term in the third line in \eqref{rk2FSp}, which is characterized by the linear combinations,
\begin{equation}\label{FSpcs}
\ia_1 + \ia_2 + \frac{1}{2}\sum_{k=0}^9s_km_k.
\end{equation}
In this case $m_i$ and $m_j$ always appear in \eqref{FSpcs} and we have three types of the linear combinations \eqref{FSpcs} depending on the signs in front of $m_i$ and $m_j$. When the signs are both positive or both negative we have 
\begin{equation}
\ia_1 + \ia_2 + \frac{1}{2}\sum_{i=0}^7s_km_k,
\end{equation}
where the number of $s_k = +1$ is odd. Choosing a particular spinor weight for the condition in \eqref{rk2SpCB3} implies that 
\begin{equation}\label{FSpcs1}
\ia_1 + \ia_2 + \frac{1}{2}\sum_{k=0}^7s_km_k \geq 0,
\end{equation}
with odd numbers of $s_k = +1$'s. When the sign in front of $m_i$ is positive and the sign in front of $m_j$ is negative we have
\begin{equation}
\ia_1 + 2\ia_2 + \frac{1}{2}\sum_{k=0}^7s_km_k,
\end{equation}
where the number of $s_k = + 1$ is even. The condition \eqref{rk2SpCB2} with a particular conjugate spinor weight yields 
\begin{equation}\label{FSpcs3}
\ia_1 + 2\ia_2 + \frac{1}{2}\sum_{k=0}^7s_km_k \geq 0,
\end{equation}
with even numbers of $s_k = +1$'s. Finally when the sign in front of $m_i$ is negative and the sign in front of $m_j$ is positive the linear combination \eqref{FSpcs} is given by 
\begin{equation}
\ia_1 + \frac{1}{2}\sum_{k=0}^7s_km_k,
\end{equation}
where the number of $s_k = +1$ is even. The sign of this linear combination is not fixed by the physical Coulomb branch moduli condition and these terms remain after the Higgsing.

Finally we consider the last line in \eqref{rk2FSp}. Namely we consider the linear combinations,
\begin{equation}\label{FSprk5}
2\ia_1 + \ia_2 \pm m_{i_1} \pm m_{i_2} \pm m_{i_3} \pm m_{i_4} \pm m_{i_5}.
\end{equation}
When both $m_i$ and $m_j$ appear in \eqref{FSprk5}, \eqref{FSprk5} becomes
\begin{equation}
2\ia_1 + \ia_2 \pm \ia_{2} \pm \ia_{2} \pm m_{i_3} \pm m_{i_4} \pm m_{i_5}.
\end{equation}
Depending on the signs in front of $\ia_2$, we have three patterns. The three linear combinations satisfy the following inequalities,
\begin{equation}
2\ia_1 + 3\ia_2 \pm m_{i_3} \pm m_{i_4} \pm m_{i_5} \geq 2\ia_1 + \ia_2 \pm m_{i_3} \pm m_{i_4} \pm m_{i_5} \geq 2\ia_1 - \ia_2 \pm m_{i_3} \pm m_{i_4} \pm m_{i_5} \geq 0.
\end{equation}
The first two inequalities come from \eqref{rk2SpCB1} and the last inequality comes from \eqref{rk2SpCB9}. In the case when one of $m_i$ and $m_j$ appears in \eqref{FSprk5}, we have 
\begin{equation}
2\ia_1 + \ia_2 \pm \ia_{2} \pm m_{i_2} \pm m_{i_3} \pm m_{i_4} \pm m_{i_5},
\end{equation}
where $m_{i_2}, m_{i_3}, m_{i_4}, m_{i_5}$ are not equal to neither $m_i$ nor $m_j$. The different signs in front of $\ia_2$ yields two linear combinations and they satisfy the inequalities,
\begin{equation}
2\ia_1 + 2\ia_2 \pm m_{i_2} \pm m_{i_3} \pm m_{i_4} \pm m_{i_5} \geq 2\ia_1 \pm m_{i_2} \pm m_{i_3} \pm m_{i_4} \pm m_{i_5} \geq 0,
\end{equation}
where the first inequality comes from \eqref{rk2SpCB1} and the last inequality comes from \eqref{rk2SpCB8}. Lastly there are cases where neither $m_i$ nor $m_j$ appear in \eqref{FSprk5}. In this case, we can rewrite the linear combination \eqref{FSprk5} as
\begin{equation}
2\ia_1 + 2\ia_2 \pm m_{i_1} \pm m_{i_2} \pm m_{i_3} \pm m_{i_4} \pm m_{i_5} - m_i.
\end{equation}
Then the physical Coulomb branch moduli condition \eqref{rk2SpCB7} implies 
\begin{equation}
2\ia_1 + 2\ia_2 \pm m_{i_1} \pm m_{i_2} \pm m_{i_3} \pm m_{i_4} \pm m_{i_5} - m_i \geq 0.
\end{equation}
Therefore for all the cases the linear combinations \eqref{FSprk5} with the tuning condition \eqref{higgstuning} are positive and hence we have $\relu{2\ia_1 + \ia_2 \pm m_{i_1} \pm m_{i_2} \pm m_{i_3} \pm m_{i_4} \pm m_{i_5}} = 0$ for \eqref{higgstuning}.

Therefore, collecting all the terms which do not vanish after inserting the tuning conditions \eqref{higgstuning} into the complete prepotential \eqref{rk2FSp} yields 
\begin{equation}\label{higgsedE8}
\begin{aligned}
\mathcal{F}_{\text{Higgsed}} &= \frac{1}{6}\ia_1^3 - \frac{1}{2}\sum_{k=0}^7m_k^2\ia_1 \cr
&+ \frac{1}{6}\sum_{0\leq k < l \leq 7}\relu{\ia_1 \pm m_k \pm m_l}^3
 + \frac{1}{6}\sum_{\substack{\{ s_k = \pm 1 \} \\\text{even $+$}}}\relu{\ia_1 + \frac{1}{2}\sum_{k=0}^7s_km_k}^3,
\end{aligned}
\end{equation}
where we omitted terms which do not depend on $\ia_1$. We can see that the prepotential \eqref{higgsedE8} precisely reproduces the complete prepotential \eqref{eq:invFforE8} after identifying $\ia_1 = \ia$.

\subsection{Prepotential from partition function}
\label{sec:Rank2GV-prep}

Here, we derive the complete prepotential assuming that the 
observations made in section \ref{sec:Rank1GV-prep} also holds for this case.
As discussed in Appendix \ref{sec:app-Sp2-GV} the partition function for 5d $\N=1$ $Sp(2)$ gauge theory with $N_f=9$ flavors is given as
\begin{align}\label{eq:partfun_Sp2}
Z = \text{PE} \left( \sum_{k=0}^{\infty} \F_k \tilde{A}_1{}^k \right),
\end{align}
where each coefficient $\F_k$ is given as follows: \\
\noindent 
For $k=0$, 
\begin{align}
\F_0 =
\frac{1}{ ( t^{\frac{1}{2}} - t^{-\frac{1}{2}} ) ( q^{\frac{1}{2}} - q^{-\frac{1}{2}} ) }
\left([0,0] \tilde{A}_2 \chi_{\bf{20}} +   [0,\frac12] \tilde{A}_2{}^2 \right).
\end{align}
For $k=1$, 
\begin{align}
\F_1 
=& \frac{1}{( t^{\frac{1}{2}} - t^{-\frac{1}{2}} ) ( q^{\frac{1}{2}} - q^{-\frac{1}{2}} ) } 
\Biggl( 
\sum_{n=0}^{\infty} [0,n] \tilde{A}_2{}^{2n+1} \chi_{\overline{\bf 512}} 
+ \sum_{n=0}^{\infty} [0,n+\frac{1}{2}] \tilde{A}_2{}^{2n+2} \chi_{\bf 512} 
\cr
&\qquad
+ [0,0]  \left( \chi_{\bf 190} +1 \right)
+ [0,\frac{1}{2}] (\tilde{A}_2{}^{-1} + \tilde{A}_2) \chi_{\bf 20} 
+ [0,1] (\tilde{A}_2{}^{-2} + 1 + \tilde{A}_2{}^{2})
+ [\frac{1}{2}, \frac{1}{2}]
\Biggr).
\cr
\end{align}
For $k=2$, 
\begin{align}
\F_2 = 
\frac{1}{( t^{\frac{1}{2}} - t^{-\frac{1}{2}} ) ( q^{\frac{1}{2}} - q^{-\frac{1}{2}} ) } 
\sum_{R} c_{R} (\tilde{A}_2, t,q) \chi_{R} (m),
\end{align}
where the summation is over the representation of $SO(20)$.
Since the exact form for some of the $c_{R}$ are too complicated, we expand in terms of $\tilde{A}_2$ and
show the leading order results:
\begin{align}
& c_{[2000000000]} = [0,\frac12], 
\cr
& c_{[1000000010]} = [0,\frac12] \tilde{A}_2 + \mathcal{O} ( \tilde{A}_2{}^2),
\cr
& c_{[1000000001]} = \left( [0,0] + [0,1] \right) \tilde{A}_2{}^2 + \mathcal{O} ( \tilde{A}_2{}^3),
\cr
& c_{[0000000002]} = c_{[0000000020]} = [0,\frac52] \tilde{A}_2{}^6 + \mathcal{O} ( \tilde{A}_2{}^7),
\cr
& c_{[0000000011]} = [0,2] \tilde{A}_2{}^5 + \mathcal{O} ( \tilde{A}_2{}^6),
\cr
& c_{[0000000100]} = [0,\frac32] \tilde{A}_2{}^4 + \mathcal{O} ( \tilde{A}_2{}^5),
\cr
& c_{[0000001000]} = [0,1] \tilde{A}_2{}^3 + \mathcal{O} ( \tilde{A}_2{}^4),
\cr
& c_{[0000010000]} = [0,\frac12] \tilde{A}_2{}^2 + \mathcal{O} ( \tilde{A}_2{}^3),
\cr
& c_{[0000100000]} = [0,0] \tilde{A}_2 + \left( [0,0]+[0,1]+[\frac12,\frac32] \right) \tilde{A}_2{}^3
+ \mathcal{O} ( \tilde{A}_2{}^4),
\cr
& c_{[000100000]} = [0,\frac12] + \mathcal{O} ( \tilde{A}_2),
\cr
& c_{[001000000]} = [0,1] \tilde{A}_2{}^{-1} + \mathcal{O} (1),
\cr
& c_{[010000000]} = [0,\frac32] \tilde{A}_2{}^{-2} + \mathcal{O} ( \tilde{A}_2{}^{-1}),
\cr
& c_{[100000000]} = [0,2] \tilde{A}_2{}^{-3} + \mathcal{O} ( \tilde{A}_2{}^{-2}),
\cr
& c_{[0000000010]} = \left( [0,0] + [0,1] \right)  + \mathcal{O} ( \tilde{A}_2 ),
\cr
& c_{[0000000001]} = \left( [0,\frac12] + [0,\frac32] +  [\frac12,1] \right) \tilde{A}_2 + \mathcal{O} ( \tilde{A}_2{}^{2}),
\cr
& c_{[0000000000]} = [0,\frac52] \tilde{A}_2{}^{-4} + \mathcal{O} ( \tilde{A}_2{}^{-3} ).
\end{align}
%
\\
\noindent
For $k=3$, 
\begin{align}
\F_3 = 
\frac{1}{( t^{\frac{1}{2}} - t^{-\frac{1}{2}} ) ( q^{\frac{1}{2}} - q^{-\frac{1}{2}} ) } 
\sum_{R} C_{R} (\tilde{A}_2, t,q) \chi_{R} (m)
\end{align}
with $C_R$ begin given as follows.
Since the expression is even more complicated, 
we write only the order of $\tilde{A}_2$ 
except for the terms which we will focus later:
\begin{align}
&  C_{[2000000010]} = \mathcal{O} (\tilde{A}_2{}^2),
&&C_{[2000000001]} = \mathcal{O} (\tilde{A}_2{}^1),
&&C_{[1000000020]} = \mathcal{O} (\tilde{A}_2{}^5), 
\cr
&  C_{[1000000011]} = \mathcal{O} (\tilde{A}_2{}^4), 
&&C_{[1000000002]} = \mathcal{O} (\tilde{A}_2{}^5), 
&&C_{[1000000100]} = \mathcal{O} (\tilde{A}_2{}^3),
\cr
&  C_{[1000001000]} = \mathcal{O} (\tilde{A}_2{}^2), 
&&C_{[1000010000]} = \mathcal{O} (\tilde{A}_2{}^1), 
&
\cr
&C_{[1000100000]} = [0,1] + \mathcal{O} (\tilde{A}_2{}), 
\!\!\!\!\!\!\!\!\!\!\!\!\!\!\!\!\!\!\!\!\!\!\!\!\!\!
&&
&
\cr
&  C_{[100100000]} =\mathcal{O} (\tilde{A}_2{}^{-1}), 
&&C_{[101000000]} =\mathcal{O} (\tilde{A}_2{}^{-2}), 
&&C_{[110000000]} =\mathcal{O} (\tilde{A}_2{}^{-3}), 
\cr
&  C_{[2000000000]} =\mathcal{O} (\tilde{A}_2{}^{-4}), 
&&C_{[0000000030]} =\mathcal{O} (\tilde{A}_2{}^{8}), 
&&C_{[0000000021]} =\mathcal{O} (\tilde{A}_2{}^{7}), 
\cr
&  C_{[0000000012]} =\mathcal{O} (\tilde{A}_2{}^{8}), 
&&C_{[0000000003]} =\mathcal{O} (\tilde{A}_2{}^{9}), 
&&C_{[0000000110]} =\mathcal{O} (\tilde{A}_2{}^{6}), 
\cr
&  C_{[0000001010]} =\mathcal{O} (\tilde{A}_2{}^{5}), 
&&C_{[0000010010]} =\mathcal{O} (\tilde{A}_2{}^{4}), 
&
\cr
&C_{[0000100010]} = [0,1] \tilde{A}_2{}^3 + \mathcal{O} (\tilde{A}_2{}^4), 
\!\!\!\!\!\!\!\!\!\!\!\!\!\!\!\!\!\!\!\!\!\!\!\!\!\!
&&
&
\cr
&  C_{[0001000010]} =\mathcal{O} (\tilde{A}_2{}^{2}), 
&&C_{[0010000010]} =\mathcal{O} (\tilde{A}_2), 
&&C_{[0100000010]} =\mathcal{O} (1), 
\cr
&  C_{[1000000010]} =\mathcal{O} (\tilde{A}_2{}^{-1}), 
&&C_{[0000000101]} =\mathcal{O} (\tilde{A}_2{}^7), 
&&C_{[0000001001]} =\mathcal{O} (\tilde{A}_2{}^6), 
\cr
&  C_{[0000010001]} =\mathcal{O} (\tilde{A}_2{}^5), 
&&C_{[0000100001]} =\mathcal{O} (\tilde{A}_2{}^4), 
&&C_{[0001000001]} =\mathcal{O} (\tilde{A}_2{}^3), 
\cr
&  C_{[0010000001]} =\mathcal{O} (\tilde{A}_2{}^2), 
&&C_{[0100000001]} =\mathcal{O} (\tilde{A}_2), 
&&C_{[1000000001]} =\mathcal{O} (1), 
\cr
&  C_{[0000000002]} =\mathcal{O} (\tilde{A}_2{}^2), 
&&C_{[0000000011]} =\mathcal{O} (\tilde{A}_2{}^3), 
&&C_{[0000000002]} =\mathcal{O} (\tilde{A}_2{}^4), 
\cr
&  C_{[0000000100]} =\mathcal{O} (\tilde{A}_2{}^2),  
&&C_{[0000001000]} =\mathcal{O} (\tilde{A}_2),  
&&C_{[0000010000]} =\mathcal{O} (1), 
\cr
&  C_{[0000100000]} =\mathcal{O} (\tilde{A}_2{}^{-1}), 
&&C_{[0001000000]} =\mathcal{O} (\tilde{A}_2{}^{-2}), 
&&C_{[0010000000]} =\mathcal{O} (\tilde{A}_2{}^{-3}), 
\cr
&  C_{[0100000000]} =\mathcal{O} (\tilde{A}_2{}^{-4}), 
&&C_{[1000000000]} =\mathcal{O} (\tilde{A}_2{}^{-5}), 
&&C_{[0000000010]} =\mathcal{O} (\tilde{A}_2{}^{-2}), 
\cr
&  C_{[0000000001]} =\mathcal{O} (\tilde{A}_2{}^{-1}), 
&&C_{[0000000000]} =\mathcal{O} (\tilde{A}_2{}^{-6}).
&
\end{align}

We first derive physical Coulomb moduli from this partition function.
Following {\bf Observation 1} in section \ref{sec:Rank1GV-prep},
we focus on the particle with spin $(j_L, j_R) =(0, \frac12)$, and $(j_L, j_R) =(0, 1)$.
Corresponding to the term of the form 
\begin{align}
\frac{1}{( t^{\frac{1}{2}} - t^{-\frac{1}{2}} ) ( q^{\frac{1}{2}} - q^{-\frac{1}{2}} ) } 
[0,j_R] \chi_{[n_1,n_2, \cdots, n_{10}]} \tilde{A}_1{}^{N_1} \tilde{A}_2{}^{N_2}
\qquad (j_R = \frac12 \text{ or } j_R = 1),
\end{align}
inside PE in \eqref{eq:partfun_Sp2},
we obtain the following constraint for physical Coulomb moduli
\begin{align}
N_1 \ia_1 + N_2 \ia_2 + w_{[n_1, n_2, \cdots, n_{10}]} \cdot \vec{m} \ge 0,
\end{align}
where $w_{[n_1, n_2, \cdots, n_{10}]}$ is the arbitrary weights in the representation
labelled by the Dynkin label $[n_1, n_2, \cdots, n_{10}]$.
From all the constraints above, we find effective constraints that were listed in \eqref{rk2SpCB1}-\eqref{rk2SpCB11}. 
We propose that these give the physical Coulomb moduli for 5d $\N=1$ $Sp(2)$ gauge theory with $N_f=9$ flavors.
Using all these constraints, 
we have checked that 
the corresponding K\"ahler parameters are always positive 
for most of the terms.
Such terms does not contribute to the complete prepotential 
as discussed in section \ref{sec:Rank1GV-prep}.
Only exceptional terms are the following four
\begin{align}
&  \frac{[0,0] \tilde{A}_2 \chi_{\mathbf{20}}}{( t^{\frac{1}{2}} - t^{-\frac{1}{2}} ) ( q^{\frac{1}{2}} - q^{-\frac{1}{2}} ) } ,
&&
 \frac{[0,0] \tilde{A}_1 \chi_{\mathbf{190}}}{( t^{\frac{1}{2}} - t^{-\frac{1}{2}} ) ( q^{\frac{1}{2}} - q^{-\frac{1}{2}} ) },
\cr
& \frac{[0,0] \tilde{A}_1 \tilde{A}_2 \chi_{\overline{\mathbf{512}}}}{( t^{\frac{1}{2}} - t^{-\frac{1}{2}} ) ( q^{\frac{1}{2}} - q^{-\frac{1}{2}} ) },
&&
 \frac{ [0,0] \tilde{A}_1{}^2 \tilde{A}_2 \chi_{[0000100000]}}{( t^{\frac{1}{2}} - t^{-\frac{1}{2}} ) ( q^{\frac{1}{2}} - q^{-\frac{1}{2}} ) } .
\end{align}
All these terms correspond to the BPS particles with spin $(j_L,j_R)= (0,0)$,
which support {\bf Observation 2}.
From these terms, we obtain the complete prepotential
\begin{align}\label{eq:prep-part-Sp2}
\F_{\text{Complete}}
= &~ \F_{\text{CFT}}
+ \frac{1}{6} \sum_{w \in \mathbf{20}} \relu{ \ia_2 + w \cdot \vec{m}}^3
+ \frac{1}{6} \sum_{w \in \mathbf{190}} \relu{ \ia_1 + w \cdot \vec{m}}^3
\cr
& 
+ \frac{1}{6} \sum_{w \in \overline{ \mathbf{512} }} \relu{ \ia_1 + \ia_2 + w \cdot \vec{m}}^3
+ \frac{1}{6} \sum_{w \in \mathbf{15504}} \relu{ 2 \ia_1 + \ia_2 + w \cdot \vec{m}}^3,
\end{align}
where $\mathbf{15504}$ is the rank 5 antisymmetric tensor representation.
The last term can be rewritten in terms of the mass parameters more explicitly as
\begin{align}\label{eq:rank5-expression}
&\sum_{w \in \mathbf{15504}} \relu{ 2 \ia_1 + \ia_2 + w \cdot \vec{m}}^3
\cr
= &
\sum_{0\leq i_1 < i_2 < i_3 < i_4 < i_5 \leq 9}\relu{2\ia_1 + \ia_2 \pm m_{i_1} \pm  m_{i_2} \pm m_{i_3} \pm m_{i_4} \pm m_{i_5}}^3
\cr
& 
+ 7 \sum_{0\leq i_1 < i_2 < i_3  \leq 9}\relu{2\ia_1 + \ia_2 \pm m_{i_1} \pm  m_{i_2} \pm m_{i_3}}^3
+ 36 \sum_{i=0}^9 \relu{2\ia_1 + \ia_2 \pm m_{i} }^3.
\end{align}
We find that the last line in \eqref{eq:rank5-expression} vanishes in the physical Coulomb moduli \eqref{rk2SpCB1} - \eqref{rk2SpCB11}. 
Analogous discussion applies also to the term in the rank 2 antisymmetric tensor representation ${\bf 190}$.
Taking these into account, we find that \eqref{eq:prep-part-Sp2} reproduces the prepotential \eqref{rk2FSp}.

%

\subsection{Prepotential from Mori cone generators}\label{sec:rk2geometry}
Similar to the rank one case, we can identify holomorphic curves in the dual geometry which are related to one-loop corrections of the complete prepotential \eqref{rk2FSp} of the $Sp(2)$ gauge theory with $9$ flavors. For identifying such holomorphic curves we consider rational curves with the self-intersection number $-1$ in the generators of the Mori cone of the dual geometries as discussed in section \ref{sec:rk1geometry}. It turns out that it is not enough to focus on a K\"ahler moduli space of a local Calabi-Yau threefold  but we need to look at other phases related by flop transitions to find all such curves which account for the contributions given by $\relu{}$ to the complete prepotential \eqref{rk2FSp}. This will generically occur since the complete prepotential has been constructed so that it is valid in the extended K\"ahler cone. 

The geometry which yields the 5d $Sp(2)$ gauge theory with $9$ flavors is a local Calabi-Yau threefold with a base given by $dP_1 \cup \text{Bl}_9\mathbb{F}_5$ or the ones obtained by flop transitions 
from it. Namely, the base is given by two compact complex surfaces $dP_1$ and $\text{Bl}_9\mathbb{F}_5$ which are glued along a curve. The curves in $dP_1$ are generated by the hyperplane class $L$ and the exceptional curve $X'_0$ and the intersection numbers are given in \eqref{dPintersection}. One the other hand, the curves in $\text{Bl}_9\mathbb{F}_5$ are generated by the fiber curve $F$, the base curve $E$ and the exceptional curves $X_i\; (i=1, \cdots, 9)$. The intersection numbers are
\begin{equation} 
\begin{split}
&F\cdot F = 0, \qquad E\cdot E = -5, \qquad F\cdot E = 1,\\
&F \cdot X_i = 0, \qquad E \cdot X_i = 0, \qquad X_i \cdot X_j = -\delta_{ij},
\end{split}
\end{equation}
for $i, j = 1, \cdots, 9$. 
The gluing 
curve for the two surfaces is given by $2L - X'_0$ in $dP_1$ or $E$ in $\text{Bl}_9\mathbb{F}_5$ \cite{Jefferson:2018irk}.

We first determine the gauge theory parameterization for the volume of the curves in the two surfaces. The fiber classes which correspond to simple roots of the $Sp(2)$ gauge theory are $L- X'_0$ and $E + 7F - \sum_{i=1}^9X_i$ \cite{Jefferson:2018irk}. Hence we parameterize their volume as
\begin{align}\label{Sp2CB}
\text{vol}(L - X'_0) = a_1 - a_2,\qquad \text{vol}(E + 7F - \sum_{i=1}^9X_i) = 2a_2.
\end{align}

\begin{figure}[t]
\centering
\includegraphics[width=10cm]{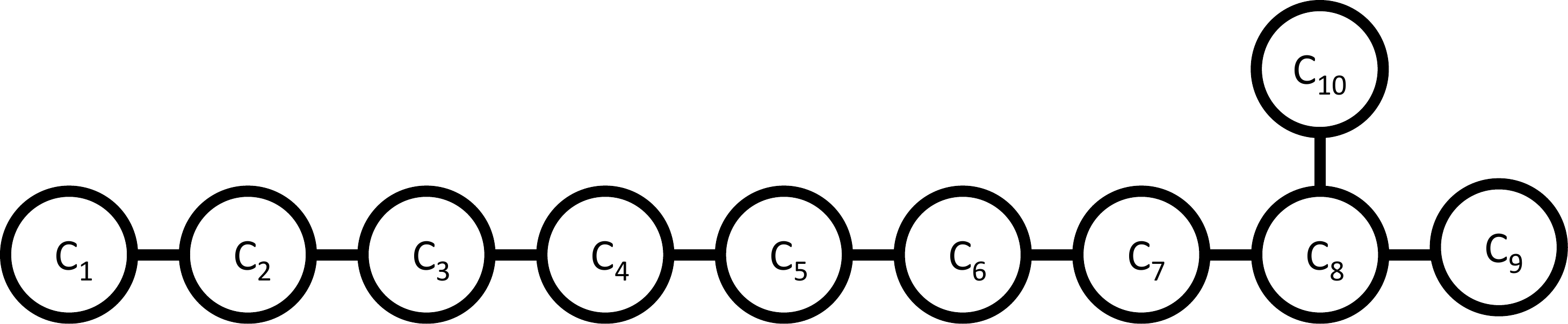}
\caption{The intersection structure of the curves in \eqref{SO20root}.}
\label{fig:D10dynkin}
\end{figure}

In order to determine the dependence on the mass parameters it is useful to flop the exceptional curve in $dP_1$ and to move on to a description by the geometry $\mathbb{P}^2 \cup \text{Bl}_{10}\mathbb{F}_6$. Due to this flop the second surface has one more exceptional curve $X_0$ which can be identified as $X_0 = F + X'_0$. Then all the exceptional curves are in the $\text{Bl}_{10}\mathbb{F}_6$ and the $SO(20)$ flavor symmetry can be seen from it. In other words, we may find the $SO(20)$ root lattice inside $H_2(\text{Bl}_{10}\mathbb{F}_6, \mathbb{Z})$. The curves $C$ correspond to the roots has the self-intersection number $-2$ and should not be charged under the $Sp(2)$ gauge group. Therefore, we may identify the roots of $SO(20)$ by the conditions, 
\begin{align}
C \cdot C = -2, \qquad C\cdot K_{\text{Bl}_{10}\mathbb{F}_6} = 0, \qquad C \cdot E = 0, \label{SO20root.condition}
\end{align}
where $K_{\text{Bl}_{10}\mathbb{F}_6}$ is the canonical divisor class and is given by
\begin{align}
K_{\text{Bl}_{10}\mathbb{F}_6} = -2E - 8F + \sum_{i=0}^9X_i.
\end{align}
Then the following curves satisfy \eqref{SO20root.condition} 
\begin{align}\label{SO20rootcurves}
\pm (X_i - X_j), \quad  \pm (E + 6F - X_1 - X_2 - \cdots - \check{X}_i - \cdots - \check{X}_j -\cdots - X_9),
\end{align}
for $0 \leq i < j \leq 9$, where $\check{X}_i$ means $X_i$ is absent. 
 The curves in \eqref{SO20rootcurves} are generated by
\begin{align}
C_i = X_{i-1} - X_i \; (i=1, \cdots, 9), \qquad C_{10} = E + 6F - \sum_{i=0}^7X_i, \label{SO20root}
\end{align}
and they form the $D_{10}$ Dynkin diagram as in Figure \ref{fig:D10dynkin}.
We can then identify the curves \eqref{SO20root} as the simple roots of $SO(20)$. Hence we associate the mass parameters $m_i \; (i=0, \cdots, 9)$ as 
\begin{equation}\label{Sp2mass}
\begin{split}
\qquad \text{vol}(C_{i}) = m_{i-1} - m_{i} \; (i=1, \cdots, 9), \qquad \text{vol}(C_{10}) = m_8 + m_9,
\end{split}
\end{equation}
where the sign of the mass parameters was chosen so that it agrees with the convention used in section \ref{sec:rank2prep}.

Solving \eqref{Sp2CB} and \eqref{Sp2mass} determines the relations between the volume of the curves in $\mathbb{P}^2 \cup \text{Bl}_{10}\mathbb{F}_6$ and we obtain
\begin{equation}\label{Sp2parameterization}
\begin{split}
\text{vol}(L)&= a_1 - 2a_2 + m_0, \quad \text{vol}(F) = a_1 + 2a_2 + \frac{1}{2}m_0 - \frac{1}{2}\sum_{i=1}^9m_i, \\
\text{vol}(X_i) &= a_1 + a_2 +\frac{1}{2}m_0 - \frac{1}{2}\sum_{j=1}^9m_j + m_i,
\quad (i=0, \cdots. 9).
\end{split}
\end{equation}

With this parameterization \eqref{Sp2parameterization}, we can compute the volume of 
rational curves with the self-intersection number $-1$ in the generators of the Mori cones in terms of the gauge theory parameters. In the case of $\mathbb{P}^2 \cup \text{Bl}_{10}\mathbb{F}_6$ the self-intersection number $-1$ rational curves exist only in $\text{Bl}_{10}\mathbb{F}_6$ and 
such curves in the generators of the Mori cone of $\text{Bl}_{10}\mathbb{F}_6$ are given by (for example see \cite{Jefferson:2018irk})
\begin{equation}\label{Mori.P2Bl10F6}
\begin{split}
&X_i, \quad F - X_i, \qquad E + 6F - \sum_{j=1}^7X_{i_j}, \quad E + 7F - \sum_{j=1}^9X_{i_j}, \quad 2E + 12F - \sum_{j=1}^52X_{i_j} - \sum_{j=1}^5X_{i_{j+5}}, \\
&3E + 18F - \sum_{j=1}^23X_{i_j} - \sum_{j=1}^7X_{i_{j+3}}, \quad 4E + 24F - 4X_{i_1} - \sum_{j=1}^9X_{i_{j+1}}.
\end{split}
\end{equation}
Expressing the volume of the curves in \eqref{Mori.P2Bl10F6} by the parameterization \eqref{Sp2parameterization}, we find
\begin{equation}\label{Mori1.P2Bl10F6}
\begin{split}
V_1 = \ia_1 + \ia_2 + \frac{1}{2}\sum_{i=0}^9s_im_i \qquad V_2 = \ia_2 \pm m_i\; (i=0, \cdots, 9),
\end{split}
\end{equation}
where $\ia_1 = a_1 + m_0$ and $\ia_2 = a_2$. $s_i = \pm 1$ and the number of the plus signs is odd. M2-branes wrapped on the curves in \eqref{Mori1.P2Bl10F6} exactly reproduce the BPS particle contributions which are in the vector and the conjugate spinor representations of $SO(20)$ which appeared in the complete prepotential \eqref{rk2FSp}.

Rational curves with the self-intersection number $-1$ which yield BPS particles in the other representations contributing to the complete prepotential \eqref{rk2FSp} can be also found 
in the generators of the Mori cones of flopped geometries. We consider performing flops with respect to curves in the followings,
\begin{equation}\label{flopcurve}
C_{i} = F - X_i, \qquad C_{i+10} = E + 7F - (X_0  + \cdots + \check{X}_i + \cdots + X_9), \qquad i = 0, \cdots, 9.
\end{equation}
The volume of the curves is 
\begin{equation}
\begin{split}
&\text{vol}(C_i) = a_2 - m_i,  \quad \text{vol}(C_{i + 10}) = a_2 + m_i,
\end{split}
\end{equation}
for $i=0, \cdots, 9$. Note that each curve in \eqref{flopcurve} intersects with the gluing curve $E$ with the intersection number one and also $C_i \cdot C_{j} = 0$ when $i \neq j$ modulo $10$. 
For example, performing a flop transition with respect to a curve in \eqref{flopcurve} yields $dP_1 \cup \text{Bl}_9\mathbb{F}_5$. 

Let us first flop two curves $C_i, C_j$ where $j \neq i$ and $j \neq i + 10$ 
from $\mathbb{P}^2\cup \text{Bl}_{10}\mathbb{F}_6$. The flopped geometry is then $dP_2 \cup \text{Bl}_8\mathbb{F}_4$ and 
one of the rational curves with the self-intersection number $-1$ in the generator of the Mori cone of $dP_2$ is $L + C_i + C_j$. In total, we have $180$ choices to select $C_i$ and $C_j$ and the volume of the curves can be summarized as 
\begin{align}\label{rk2antisym}
V_3 = \ia_1 \pm m_i \pm m_j,
\end{align}
where $0 \leq i < j \leq 9$. 
M2-branes wrapped on the curves in \eqref{rk2antisym} give rise to the contributions which are in the rank 2 antisymmetric representation of $SO(20)$ in \eqref{rk2FSp}. 

We then consider flop five curves $C_{i_1}, C_{i_2}, C_{i_3}, C_{i_4}, C_{i_5}$ where each subscript is different from each other modulo $10$. The flopped geometry is $dP_5 \cup \text{Bl}_5\mathbb{F}_1$ which can be also written as $\text{Bl}_4\mathbb{F}_1 \cup dP_6$. In this case the gluing curve $2L + C_{i_1} + C_{i_2} + C_{i_3} + C_{i_4} + C_{i_5}$ becomes a 
rational curve with the self-intersection number $-1$ in the generator of the Mori cone of $dP_6$. There are $8064$ choices to choose $C_{i_1}, C_{i_2}, C_{i_3}, C_{i_4}, C_{i_5}$ and the volume of the curves can be written as 
\begin{align}\label{rk5antisym}
V_4 = 2\ia_1 + \ia_2 \pm m_{i_1} \pm m_{i_2} \pm m_{i_3} \pm m_{i_4} \pm m_{i_5},
\end{align}
where $0\leq i_1 < i_2 < i_3 < i_4 < i_5 \leq 9$. The curves in \eqref{rk5antisym} 
are related to the contributions in the rank 5 antisymmetric representation of $SO(20)$ in \eqref{rk2FSp}. 
Combining \eqref{Mori.P2Bl10F6}, \eqref{rk2antisym} and \eqref{rk5antisym}, we are able to identify all the holomorphic curves which give rise to the contributions of type $\relu{}$ to the complete prepotential \eqref{rk2FSp}. 


\section{Prepotential for Rank-2 theories: $Sp(2)$ gauge theory with 1 antisymmetric and 7 flavors}\label{sec:rank2AS}

In previous section, we studied the complete prepotential for the $Sp(2)$ gauge theory with 9 flavors. In this section, we propose the complete prepotential for the $Sp(2)$ gauge theory with 7 flavors and 1 hypermultiplet in antisymmetric tensor representation. This theory is known to correspond to rank 2 SCFT with $E_8 \times SU(2)$ global symmetry.

\subsection{Complete prepotential}
\label{sec:rank2ASprep}
Similar to previous sections, we first start from IMS prepotential of the $Sp(2)$ gauge theory with $7$ flavors and 1 antisymmetric tensor, which is given by \cite{Intriligator:1997pq}
\begin{align}\label{rk2ASIMS}
\F_{\text{IMS}} 
= & ~
\frac{1}{2} m_0(a_1^2 + a_2^2) 
+ \frac{1}{12} \sum_{I=1}^2 \sum_{J=1}^2 |\pm a_I \pm  a_J|^3  
\cr
& \qquad  - \frac{1}{12} \sum_{I=1}^2 \sum_{i=1}^7 | \pm a_I + m_i|^3 
- \frac{1}{12} |\pm a_1 \pm a_2 + m_{\text{AS}}|^3 
\cr
= &~ \F_{\text{CFT}} 
+ \frac{1}{6} \sum_{I=1}^2 \sum_{i=1}^7 \relu{a_I \pm m_i}^3 
+ \frac{1}{6} \relu{a_1 \pm a_2 \pm m_{\text{AS}}}^3
\end{align}
in the Weyl chamber of $Sp(2)$ gauge group $a_1 \ge a_2 \ge 0$.
Here, we denote
\begin{align}\label{eq:rk2ASCFT}
\F_{\text{CFT}} 
= \frac{1}{6}(a_1^3 + a_2^3) 
+ \frac{1}{2}m_0 (a_1^2 + a_2^2) 
- \frac{1}{2}\sum_{i=1}^7 m_i^2 (a_1+a_2) 
- m_{\text{AS}}{}^2 a_1,
\end{align}
where we omitted the terms which do not depend on the Coulomb branch parameters.
In the parameter region $a_1, a_2 \gg |m_f|$ with $f=1,\cdots 7$, the IMS prepotential \eqref{rk2ASIMS} reduces to this $\F_{\text{CFT}} $. 
We can compute the effective coupling of this phase by taking the second derivative of this expression as
\begin{align}
\frac{\partial^2 \F_{\text{CFT}} }{\partial a_1^2} = a_1 + m_0,
\qquad 
\frac{\partial^2 \F_{\text{CFT}} }{\partial a_1 \partial a_2} = 0,
\qquad 
\frac{\partial^2 \F_{\text{CFT}} }{\partial a_2^2} = a_2 + m_0.
\end{align}
This implies that the invariant Coulomb moduli parameters should be defined as
\begin{align}\label{eq:rk2ASinvCB}
\ia_1 = a_1 + m_0, \qquad \ia_2 = a_2 + m_0
\end{align}
and that \eqref{eq:rk2ASCFT} is indeed valid at the CFT phase $a_1, a_2 \gg |m_i|$ ($i=0,1,\cdots, 7$). 
The prepotential at the CFT phase \eqref{eq:rk2ASCFT} can be rewritten in terms of the invariant Coulomb moduli parameters
\eqref{eq:rk2ASinvCB} as
\begin{align}\label{eq:CFTrk2AS}
\F_{\text{CFT}} = \frac{1}{6} (\ia_1{}^3 + \ia_2{}^3) - m_{\text{AS}}{}^2 \ia_1 - \frac{1}{2} \sum_{i=0}^7m_i{}^2 (\ia_1 + \ia_2).
\end{align}

Now, we add more terms to the IMS prepotential in such a way that it becomes invariant under the Weyl reflections 
of the enhanced global symmetry $E_8 \times SU(2)$. 
The 7 flavor masses $m_f$ ($f=1,\cdots 7$) and the instanton mass $m_0$ can be interpreted as $E_8$ fugacity similar to the rank one case while $m_{\text{AS}}$ is the fugacity for $SU(2)$ part.
Similar to the discussion in section \ref{sec:rank1}, the terms 
\begin{align}
\frac{1}{6} \sum_{I=1}^2 \sum_{i=1}^7 \relu{a_I \pm m_i}^3 
= \frac{1}{6} \sum_{I=1}^2 \sum_{i=1}^7 \relu{\ia_I - m_0 \pm m_i}^3 
\end{align}
in \eqref{rk2ASIMS} indicates that we need the terms in the 248 dimensional representation of $E_8$
\begin{align}\label{eq:248rk2}
\F_{\bf(248,1)} = \frac{1}{6} 
\sum_{I=1}^2 \sum_{w \in \mathbf{248} \times \mathbf{1}} \relu{ \ia_I + w \cdot \vec{m} }.
\end{align}
Also, the term 
\begin{align}
 \frac{1}{6} \relu{a_1 + a_2 \pm m_{\text{AS}}}^3
 =  \frac{1}{6} \relu{\ia_1 + \ia_2 + 2 m_0 \pm m_{\text{AS}}}^3
\end{align}
in \eqref{rk2ASIMS} indicates that we need the terms in the 3875 dimensional representation of $E_8$
\begin{align}\label{eq:3875rk2}
\F_{\bf(3875, 2)} = \frac{1}{6} 
\sum_{w \in \mathbf{3875} \times \mathbf{2}} \relu{ \ia_1 + \ia_2 + w \cdot \vec{m} }.
\end{align}
because $2m_0$ corresponds to the highest weight state of this representation as mentioned in \eqref{eq:high-mass-E8}
and the term $\pm m_{\text{AS}}$ corresponds the 2 dimensional representation of the $SU(2)$.
On the contrary, the term 
\begin{align}\label{eq:1-2rk2}
\F_{(\mathbf{1}, \mathbf{2})}
 = \frac{1}{6} \relu{a_1 - a_2 \pm m_{\text{AS}}}^3
 =  \frac{1}{6} \relu{\ia_1 - \ia_2 \pm m_{\text{AS}}}^3
\end{align}
in \eqref{rk2ASIMS} is already invariant under the global symmetry and thus do not need further terms.
On top of that, it turns out that we need two more types of terms 
\begin{align}\label{eq:1-3rk2}
\F_{(\mathbf{1}, \mathbf{3})} 
 =  \frac{1}{6} \relu{\ia_1 \pm 2 m_{\text{AS}}}^3
\end{align}
and 
\begin{align}\label{eq:30380rk2}
\F_{(\mathbf{30380}, \mathbf{3})} 
 =  \frac{1}{6} \sum_{w \in \mathbf{30380} \times \mathbf{3}}
 \relu{2 \ia_1 + \ia_2 +  w \cdot \vec{m} }.
\end{align}
The necessity of these two terms will be discussed in the following subsections.

In summary, we propose that the complete prepotential is given by
\begin{align}\label{eq:prep:rk2AS}
\F = \F_{\text{CFT}}  + \F_{\bf(248,1)} + \F_{\bf(3875, 2)} + \F_{(\mathbf{1}, \mathbf{2})} 
+ \F_{(\mathbf{1}, \mathbf{3})}  + \F_{(\mathbf{30380}, \mathbf{3})} 
\end{align}
where each term is defined in \eqref{eq:CFTrk2AS}, \eqref{eq:248rk2}, \eqref{eq:3875rk2}, 
\eqref{eq:1-2rk2}, \eqref{eq:1-3rk2}, and in \eqref{eq:30380rk2}.

Here, we consider the case $m_{\text{AS}} = 0$ to give a consistency check.
In this case, it is known that the Seiberg-Witten curve of the rank 2 $E_n$ theory is factorized into the two curves for rank one theory \cite{Kim:2014nqa}.
The partition function is also known to be factorized \cite{Gadde:2015tra}. 
These indicate that our complete prepotential should also become the sum of two prepotentials for rank one case.
That is,
\begin{align}\label{eq:mAS0}
\F(m_{\text{AS}}=0) = 
\frac{1}{6} (\ia_1{}^3 + \ia_2{}^3) - \frac{1}{2} m_i{}^2 (\ia_1 + \ia_2)
+ \frac{1}{6} \sum_{I=1}^2 \sum_{i=1}^7 \relu{\ia_I - m_0 \pm m_i}^3.
\end{align}
Furthermore, it is expected that the physical Coulomb branch is given as the copies of the rank one case \eqref{eq:physCB-E8}, 
\begin{align}\label{eq:physCB-rk2E8}
&2 \ia_I - w \cdot \vec{m} \ge0 , \quad \forall w \in \mathbf{3875}
\qquad (I=1,2)
\cr
&3 \ia_I - w \cdot \vec{m} \ge0 , \quad \forall w \in \mathbf{147250}
\qquad (I=1,2)
\end{align}
together with
\begin{align}\label{eq:rk2ASa1a2}
\ia_1 \ge \ia_2
\end{align}
coming from the choice of the Weyl chamber of the $Sp(2)$ gauge group.
Analogous to \eqref{eq:ineq-E8}, we see that 
\begin{align}\label{eq:ineq-rk2E8}
&\ia_I \ge 0 \qquad (I=1,2)
\cr
&3 \ia_I - w \cdot \vec{m} \ge0 , \quad \forall w \in \mathbf{30380}
\qquad (I=1,2)
\end{align}
are satisfied in the expected physical Coulomb moduli \eqref{eq:physCB-rk2E8}.
All these are enough to show that
$\F_{\bf(3875, 2)}$, $\F_{(\mathbf{1}, \mathbf{2})}$, $\F_{(\mathbf{1}, \mathbf{3})}$,
and $\F_{(\mathbf{30380}, \mathbf{3})}$ 
indeed vanish in the expected physical Coulomb moduli \eqref{eq:physCB-rk2E8} and \eqref{eq:rk2ASa1a2}
if $m_{\text{AS}} = 0$. Therefore, we have shown that our prepotential 
\eqref{eq:prep:rk2AS} indeed reduces to \eqref{eq:mAS0} 
in the expected physical Coulomb moduli if $m_{AS} = 0$.
Moreover, the analogous computation in Appendix \ref{sec:PhysCB} indicates that 
\begin{align}
\frac{\partial \F (m_{\text{AS}}=0)}{\partial \ia_I} \ge 0 \qquad (I=1,2)
\end{align}
is satisfied in the expected physical Coulomb moduli while the equality is saturated at the boundary of \eqref{eq:physCB-rk2E8}.
This implies that the expected physical Coulomb moduli \eqref{eq:physCB-rk2E8} and \eqref{eq:rk2ASa1a2} 
are indeed the correct physical Coulomb moduli space for $m_{\text{AS}} = 0$.

\subsection{Consistency with dualities}
\label{sec:rank2ASdual}

As in section \ref{sec:rk2dualities}, we can give support for the complete prepotential of the $Sp(2)$ gauge theory with $7$ flavors and a hypermultiplet in the antisymmetric representation given in \eqref{eq:prep:rk2AS} by reproducing the IMS prepotentials of the dual theories. The rank 2 $E_8$ theory has two dual gauge theory descriptions. One is the $SU(3)$ gauge theory with $8$ flavors and the CS level $\kappa = \pm 2$ and the other is the quiver gauge theory $[1] - SU(2) - SU(2) - [5]$,
where $[n]$ is the $n$ hypermultiplets in fundamental representation.
 We look at the prepotentials in the dual descriptions one by one. 

\subsubsection*{Duality between $Sp(2)+1AS+7F$ and $SU(3)_2+8F$}
First we consider the duality to the $SU(3)$ gauge theory with $8$ flavors and the CS level $\kappa = 2$. The duality map for the marginal theories by adding one flavor on the both sides has been obtained in \cite{Hayashi:2018lyv}. It is straightforward to obtain the duality map for the current case and it is given by
\begin{align}
m_0^{Sp} &= \frac{5}{4}m_0^{SU} - \frac{1}{4}\sum_{i=1}^7m_i^{SU} - \frac{3}{4}m^{SU}_8,\label{rk2E8SUv1}\\
m_{\text{AS}}^{Sp} &= -\frac{1}{2}m_8^{SU} + \frac{1}{2}m_0^{SU} - \frac{1}{2}\sum_{i=1}^7m_i^{SU},\\
m_i^{Sp}&= m_i^{SU} + \frac{1}{4}m_0^{SU} - \frac{1}{4}\sum_{i=1}^7m_i^{SU} + \frac{1}{4}m_8^{SU}, \qquad (i=1,2,\cdots 7) \\
a_I^{Sp} &=a_I^{SU} + \frac{1}{4}m_0^{SU} - \frac{1}{4}\sum_{i=1}^7m_i^{SU} + \frac{1}{4}m_8^{SU}. \qquad (I=1,2)
\label{rk2E8SUv5}
\end{align}
Therefore, inserting the duality map \eqref{rk2E8SUv1}-\eqref{rk2E8SUv5} into the prepotential \eqref{eq:prep:rk2AS} yields the complete prepotential for the $SU(3)$ gauge theory with $8$ flavors and the CS level $\kappa = 2$. 

Since we are interested in the weak coupling region of the $SU(3)$ gauge theory 
we 
consider the region $m_0^{SU} \gg |m_f^{SU}|, |a_I^{SU}|$.
Then various terms in the prepotential are simplified in the following way. 
\begin{align}
\F_{\bf(248,1)}
& \to  \frac{1}{6}\sum_{I=1}^2\sum_{i=1}^{8}\relu{ a_I^{SU} - m_i^{SU}}^3 \\
\F_{\bf(3875, 2)}
& \to 
\frac{1}{6} \sum_{i=1}^8 \relu{ a_1^{SU} + a_2^{SU} + m_i^{SU} }^3
\cr
&=\frac{1}{6} \sum_{i=1}^8 \relu{- a_3^{SU}  + m_i^{SU} }^3\\
\F_{\bf(1, 2)}
& \to \frac{1}{6} \left(a^{SU}_1 + a^{SU}_2  - \frac{1}{2}m_0^{SU} + \frac{1}{2}\sum_{i=1}^8m_i^{SU} \right)^3 \label{rk2E8SUFv3}
\\
\F_{\bf(1, 3)}
& \to 0
\\
\F_{(\mathbf{30380}, \mathbf{3})} 
& \to 0.
\end{align}

Then, the prepotential in the weak coupling region 
reduces to
\begin{equation}\label{rk2ESpSU}
\begin{split}
\mathcal{F}_{\text{weak}}^{SU(3)} 
&= \frac{1}{6}\left(2\left(a_1^{SU}\right)^3 - 3\left(a_1^{SU}\right)^2a_2^{SU} - 3a_1^{SU}\left(a_2^{SU}\right)^2 \right) + \frac{1}{2}m_0^{SU}\left(\left(a_1^{SU}\right)^2 + a_1^{SU}a_2^{SU} + \left(a_2^{SU}\right)^2\right)\cr
&\qquad -\frac{1}{2}a_1^{SU}a_2^{SU}\sum_{i=1}^8m_i^{SU} -\frac{1}{2}\left(a_1^{SU} + a_2^{SU}\right)\sum_{i=1}^8\left(m_i^{SU}\right)^2,\cr
&\qquad +\frac{1}{6}\sum_{I=1}^2\sum_{i=1}^{8}\relu{ a_I^{SU} - m_i^{SU}}^3  + \frac{1}{6} \sum_{i=1}^8 \relu{-a_3^{SU} + m_i^{SU} }^3,
\end{split}
\end{equation}
where we ignored the terms which do not depend on the Coulomb branch moduli. 
This is exactly the IMS prepotential 
\begin{equation}\label{rk2ESUIMS}
\begin{split}
\mathcal{F}_{\text{IMS}}^{SU(3)}
&= \frac{1}{4}m_0^{SU}\left((a_1^{SU})^2 + (a_2^{SU})^2 + (a_3^{SU})^2\right) + \frac{1}{3}\left((a_1^{SU})^3 + (a_2^{SU})^3 + (a_3^{SU})^3\right) \\
&\quad+\frac{1}{6}\left((a_1^{SU} - a_2^{SU})^3 + (a_1^{SU} - a_3^{SU})^3 + (a_2^{SU} - a_3^{SU})^3\right) ,\\
&\quad-\frac{1}{12}\sum_{i=1}^3\sum_{j=1}^8\left| a_i^{SU} - m_j^{SU} \right|^3
\end{split}
\end{equation}
for the $SU(3)$ gauge theory with $8$ flavors and the CS level $\kappa =2$.
Also, note that the prepotential of the form $\F_{\bf(1, 2)}$ was important to perform one flop transition in order to move 
from the CFT region
to the weak coupling region of the $SU(3)$ gauge theory.

\subsubsection*{Duality between $Sp(2)+1AS+7F$ and $[1] - SU(2)-SU(2) - [5]$}

\begin{figure}[t]
\centering
\subfigure[]{\label{fig:SU3-8}
\includegraphics[width=7cm]{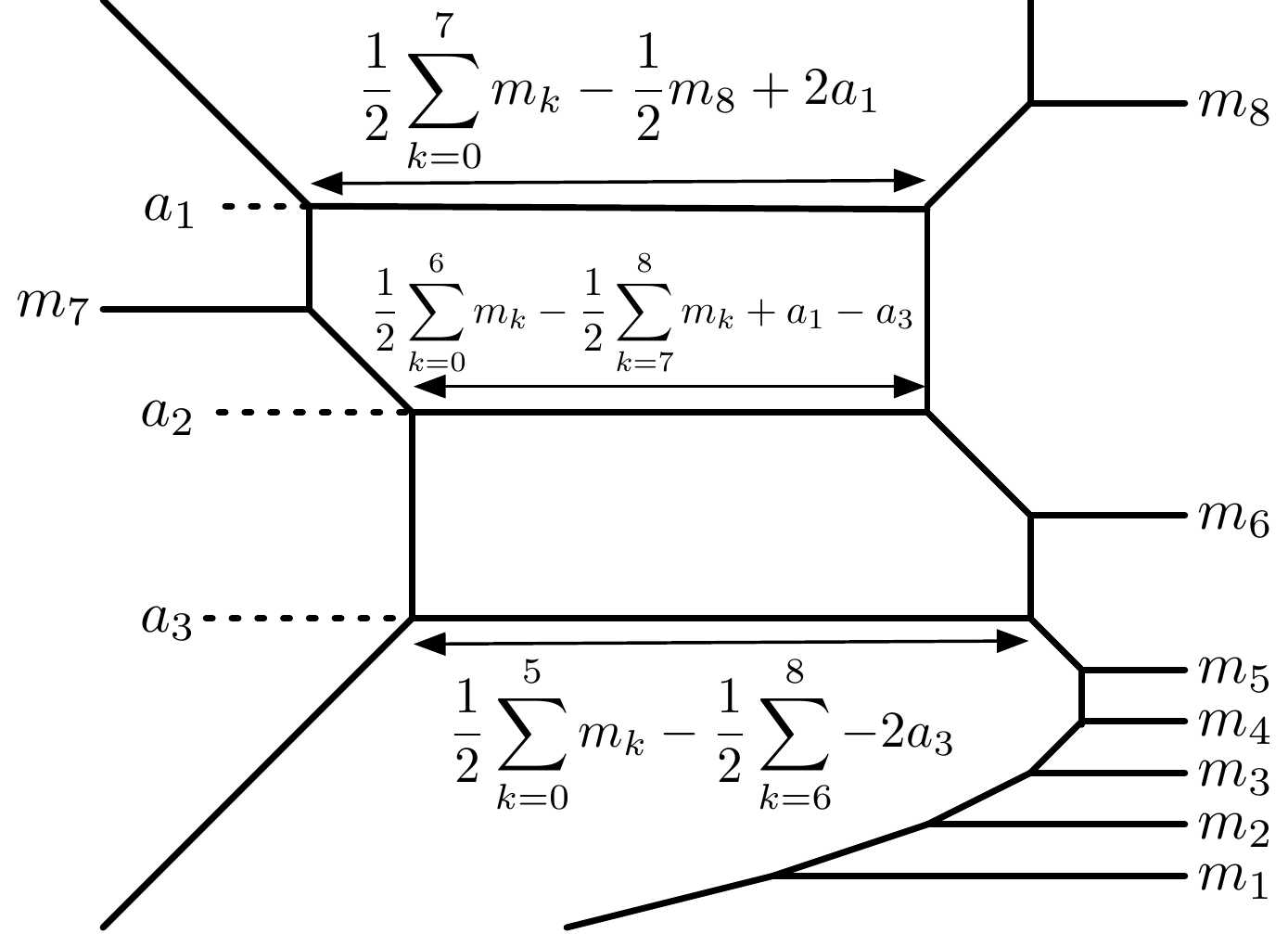}}
\subfigure[]{\label{fig:SU3-8HW}
\includegraphics[width=7.5cm]{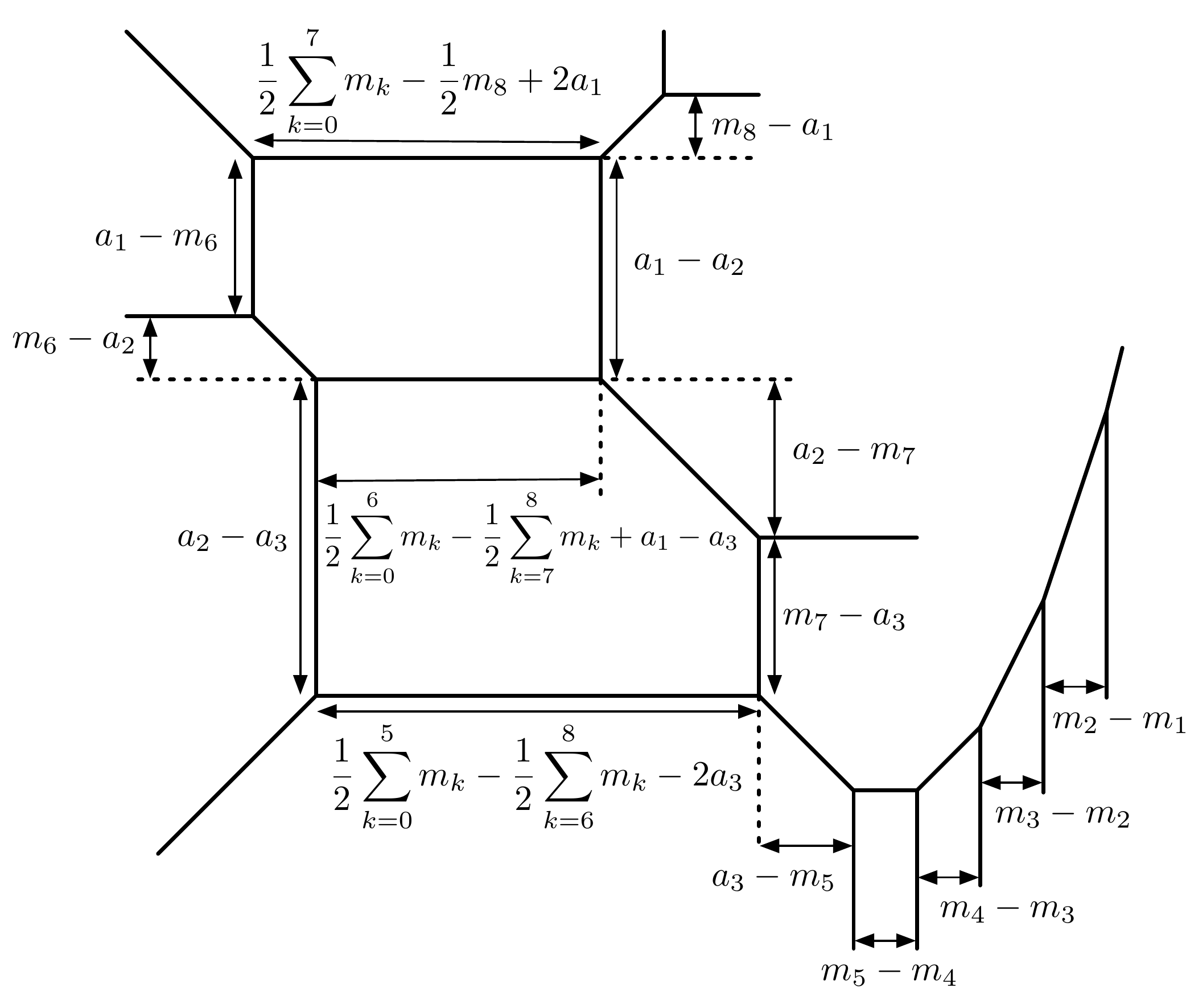}}
\subfigure[]{\label{fig:1-SU2-SU2-5}
\includegraphics[width=8cm]{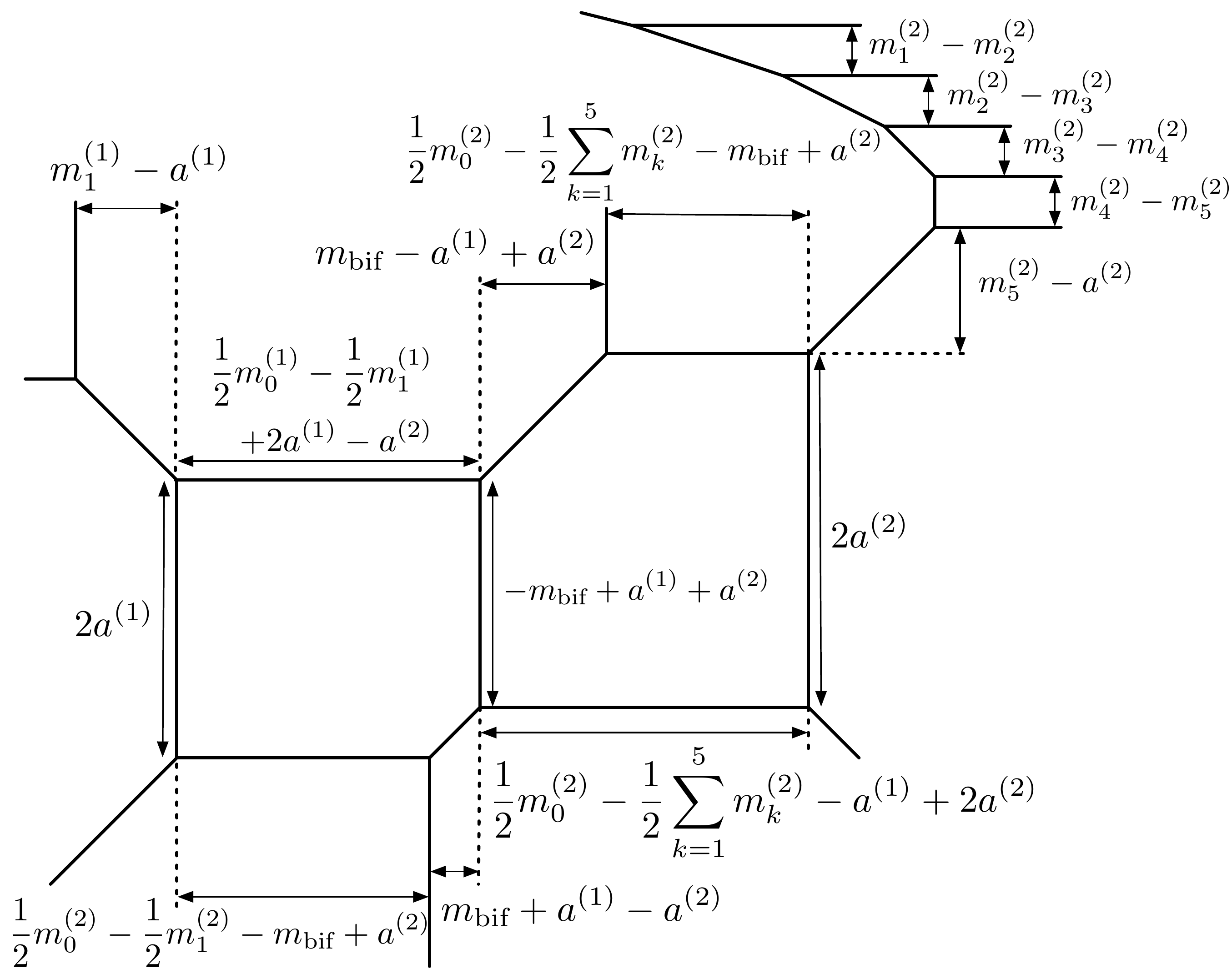}}
\caption{(a). 5-brane web for 5d $\N =1$ $SU(3)$ gauge theory with Chern-Simons level 2 and with  $N_f=8$ flavors. 
(b). 5-brane web after sequence of Hanany-Witten transition.
(c). 5-brane web for 5d $\N =1$ $SU(2)$ quiver gauge theory.}
\label{fig:SU3-8-duality}
\end{figure}

The $Sp(2)$ gauge theory with 7 flavors and with 1 antisymmetric tensor is dual to the quiver gauge theory 
$
[1] - SU(2)^{(1)}-SU(2)^{(2)} - [5].
$
In order to obtain the duality map between them,
we first consider the duality between the $SU(3)$ gauge theory considered above and this $SU(2)$ quiver gauge theory
through the 5-brane web as in Figure \ref{fig:SU3-8-duality}. 
Figure \ref{fig:SU3-8} is a 5-brane web for the 5d $\N =1$ $SU(3)$ gauge theory with Chern-Simons level 2 and with $N_f=8$ flavors .
Figure \ref{fig:SU3-8HW} is a 5-brane web obtained from \ref{fig:SU3-8} by a sequence of Hanany-Witten transitions.
Figure \ref{fig:1-SU2-SU2-5} is a 5-brane web for 5d $\N =1$ $SU(2)$ quiver gauge theory.
The diagram in \ref{fig:SU3-8HW} and in \ref{fig:1-SU2-SU2-5} are related by the S-duality transformation and can be used to compare the parameters in the two gauge theories.
This 5-brane analysis shows that the duality map 
between the $SU(3)$ gauge theory and the $SU(2)$ quiver gauge theory is given by
\begin{align}\label{eq:Map-SU3-SU2SU2}
m_0^{SU} 
& =  
\Lambda_1 + \Lambda_2 + m_1^{(1)} + m_0^{(2)} 
\cr
m_i^{SU} 
& = \Lambda_1 - m_{i}^{(2)},
\quad (i=1,2,\cdots, 5)
\cr
m_i^{SU} 
& = 
- \Lambda_1 - \Lambda_2 + (-1)^{i} m_{\text{bif}},
\quad
(i=6,7)
\cr
m_8^{SU} 
& = \Lambda_2 + m_{1}^{(1)},
\cr
a_1^{SU}
&= 
\Lambda_2
+a^{(1)},
\cr
a_2^{SU}
&= 
- \Lambda_1 -\Lambda_2  - a^{(1)} + a^{(2)},
\cr
a_3^{SU}
&= 
\Lambda_1 - a^{(2)},
\end{align}
where we introduced 
\begin{align}
\Lambda_1 &\equiv - \frac{1}{6} m_0^{(1)} + \frac{1}{6} m_1^{(1)}
-\frac{1}{3} m_0^{(2)} + \frac{1}{3} \sum_{k=1}^5 m_k^{(2)},
\cr
\Lambda_2 & \equiv \frac{1}{3} m_0^{(1)} - \frac{1}{3} m_1^{(1)}
+ \frac{1}{6} m_0^{(2)} - \frac{1}{6} \sum_{k=1}^5 m_k^{(2)} 
\end{align}
to simplify the expression.
The upper indices with bracket $(n)$ is the label to indicate 
the $SU(2)^{(n)}$ gauge group of the quiver gauge theory.

Then, combining \eqref{eq:Map-SU3-SU2SU2} with \eqref{rk2E8SUv1}-\eqref{rk2E8SUv5}, we obtain the duality map between $Sp(2)$ gauge theory and and the quiver gauge theory. 
After transformation of the Weyl group of $E_8$, the duality map is given as
\begin{align}
m_i^{Sp}  & = m_i^{(2)}, \qquad ( i=0,1, \cdots, 5 )
\cr
m_j^{Sp} & = m_{\text{bif}}
+ (-1)^j \left( 
\frac{1}{2} m_0^{(1)} + \frac{1}{2} m_1^{(1)} + m_0^{(2)} 
\right),
\qquad ( j = 6, 7 )
\cr
m_{\text{AS}}^{Sp} & = 
\frac{1}{2} m_0^{(1)}  - \frac{1}{2} m_1^{(1)} + m_0^{(2)} ,
\cr
\ia_1
&= a^{(1)} + 2 m_0^{(2)} + m_0^{(1)},
\cr
\ia_2 
&= a^{(2)} - a^{(1)} 
+ \frac{1}{2} m_0^{(1)} + \frac{1}{2} m_1^{(1)} + 2 m_0^{(2)},
\end{align}
where we note $\ia_I=a_I^{Sp}+m_0^{Sp}$ $(I=1,2)$.
After substituting them into the prepotential \eqref{eq:prep:rk2AS}, 
we consider the weak coupling region $m_0^{(1)}, m_0^{(2)} \gg |m_f^{(1)}|,|m_f^{(2)}|,  |a^{(1)}|, |a^{(2)}|$. 
Then, only small number of terms remains as follows:
\begin{align}\label{eq:weak-SU2-SU2}
\F_{\bf(248,1)}
& \to 
\frac{1}{6} (a^{(2)} - a^{(1)} - \frac{1}{2} m_0^{(1)} - \frac{1}{2} m_1^{(1)}  )^3
+ \frac{1}{6} \relu{ a^{(1)} - m_1^{(1)} }^3
\cr
& \qquad
+ \frac{1}{6} \relu{ a^{(2)} - a^{(1)} \pm m_{\text{bif}} }^3
\cr
\F_{\bf(3875, 2)}
& \to 
\frac{1}{6} \sum_{i=1}^5 \relu{ a^{(2)} \pm m_i^{(2)} }^3
\cr
\F_{\bf(1, 2)}
& \to \frac{1}{6} ( 2 a^{(1)} - a^{(2)} - m_0^{(2)} )^3
\cr
\F_{\bf(1, 3)}
& \to \frac{1}{6} \relu{a^{(1)} + m_1^{(1)} }^3
\cr
\F_{(\mathbf{30380}, \mathbf{3})} 
& \to \frac{1}{6} \relu{a^{(1)} + a^{(2)} \pm m_{\text{bif}}}^3.
\end{align}
Combined with $\F_{\text{CFT}}$, 
we find that the complete prepotential reduces to 
\begin{align}
\F_{\text{weak}}
&=  \frac{7}{6} ( a^{(1)} )^3  -  ( a^{(1)} )^2 a^{(2)}  + \frac{1}{6} ( a^{(2)} )^3 
+ \frac{1}{2} \sum_{I=1}^2 m_0^{(I)} ( a^{(I)} )^2 
\cr
& \quad 
+ \frac{1}{6} \relu{ a^{(1)} \pm  m_1^{(1)}}^3
+ \frac{1}{6} \sum_{i=1}^5 \relu{ a^{(2)} \pm m_i^{(2)} }^3
+ \frac{1}{6} \relu{ a^{(1)}  \pm a^{(2)} \pm  m_{\text{bif}} }^3
\end{align}
in the weak coupling region,
where we ignored the constant terms independent of Coulomb branch parameters $a^{(1)}$ and $a^{(2)}$.
We find that it agrees with the IMS prepotential 
\begin{align}
\F_{\text{IMS}}
&= \sum_{I=1}^2 \left( \frac{4}{3} ( a^{(I)} )^3 + \frac{1}{2} m_0^{(I)} ( a^{(I)} )^2 \right)
\cr
& \quad 
- \frac{1}{12} | a^{(1)} \pm  m_1^{(1)} |^3
- \frac{1}{12} \sum_{i=1}^5 |  a^{(2)} \pm m_i^{(2)} |^3
- \frac{1}{12} | a^{(1)}  \pm a^{(2)} \pm  m_{\text{bif}}|^3
\end{align}
for the quiver gauge theory $[1] - SU(2)^{(1)} - SU(2)^{(2)} - [5]$ up to the constant terms.

Here, we make two comments. 
First, note that the terms $\F_{(\mathbf{1}, \mathbf{3})}$  and $\F_{(\mathbf{30380}, \mathbf{3})} $ in \eqref{eq:prep:rk2AS}, which we did not give derivation there,
turn out to be necessary to reproduce this IMS prepotential correctly. 
Second, we find from \eqref{eq:weak-SU2-SU2} that
\begin{align}
\F_{\text{CFT}} 
= \F_{\text{IMS}} 
- \frac{1}{6} (a^{(2)} - a^{(1)} - \frac{1}{2} m_0^{(1)} - \frac{1}{2} m_1^{(1)}  )^3
-  \frac{1}{6} ( 2 a^{(1)} - a^{(2)}  - m_0^{(1)} )^3
\end{align}
in the CFT region. 
This indicates that,
at the region $a^{(1)}, a^{(2)} > 0$, 
$m_j^{(1)}=0$ $(j=0,1)$, $m_i^{(2)}=0$ $(i=0,1,\cdots, 5)$,
the IMS prepotential $\F_{\text{IMS}}$
 is different from the correct prepotential $\F_{\text{CFT}}$.
In this massless region, 
 $\F_{\text{IMS}}$ agree with $\F_{\text{CFT}}$ only in the subspace of the Coulomb moduli, $a^{(1)} = 0$.

\subsection{Consistency with RG flows}
\label{sec:rank2ASRG}
It is known that $Sp(2)$ gauge theory with an antisymmetric and 7 flavors ($Sp(2)+1{\bf AS}+7{\bf F}$) has three different RG flows \cite{Jefferson:2018irk}: (i) $Sp(2)+1{\bf AS}+6{\bf F}$ with $E_7\times SU(2)$ global symmetry, (ii) $Sp(2)+7{\bf F}$ with $SO(14)\times U(1)$ global symmetry, and (iii) the quiver theory $SU(2)_0{\bf-}SU(2){\bf-}[5]$ with $E_8$ global symmetry. We take these three decoupling limits and show that our complete prepotential for $Sp(2)+1{\bf AS}+7{\bf F}$ naturally produces all the complete prepotentials for  three theories with the expected enhanced global symmetries. 

\paragraph{\underline{Flow to $Sp(2)+1{\bf AS}+6{\bf F}$}} 
Let us first take the limit to $Sp(2)$ gauge theory with an antisymmetric and 6 flavors. It is achieved by decoupling a flavor. Namely, we take one of mass of flavors of $\F^{Sp(2)+1{\bf AS}+7{\bf F}}$ to infinity, say $m_7$, together with $m_0^{Sp(2)+1{\bf AS}+7{\bf F}}$,  
\begin{align}\label{eq:decouple7To6}
	m_7, ~m^{Sp(2)+1{\bf AS}+7{\bf F}}_0 \to \infty, ~ {\rm while}~~m^{Sp(2)+1{\bf AS}+7{\bf F}}_0\!-m_7 \equiv m_0^{Sp(2)+1{\bf AS}+6{\bf F}}~~{\rm fixed.}
\end{align} 
From here on, we drop the superscripts for ${Sp(2)+1{\bf AS}+6{\bf F}}$ so that $m_0$ refers to $m_0^{Sp(2)+1{\bf AS}+6{\bf F}}$. 
As shown in the previous subsection, the structure of the complete prepotential for $Sp(2)+1{\bf AS}+7{\bf F}$ takes the form
\begin{align}
\F_{E_8\times SU(2)} &= \F_{\rm CFT}+\F_{\bf (248,1)}+\F_{\bf (3875,2)}+\F_{\bf (1,2)}+\F_{\bf (1,3)}+\F_{\bf (30380,3)}.
\end{align}
As we are decoupling a flavor, the $SU(2)$ global symmetry part coming from an antisymmetric is not altered. It is therefore similar to the decoupling of the rank-1 theory, $SU(2)+N_f{\bf F}$. Recall that the prepotential become a sum of two $Sp(1)$ theories in the massless limit of antisymmetric matter. It then follows that the invariant Coulomb branch parameters of $Sp(2)+1{\bf AS} + 6{\bf F}$ theory  become
\begin{align}
\ia_{I} = a_{I}+\frac12 m_0. 
\end{align}
%
We note that, as in the rank-1 case, in the decoupling limit \eqref{eq:decouple7To6},  $\F^{Sp(2)+1{\bf AS}+7{\bf F}}_{\bf(248,1)}$ gives rise to the terms associated with the fundamental representation of $E_7$ and also terms which contribute to $\F^{Sp(2)+1{\bf AS}+6{\bf F}}_{\rm CFT}$, 
\begin{align}\label{eq:248-1}
\F^{E_8\times SU(2)}_{\bf(248,1)}&\to\F^{E_7\times SU(2)}_{\bf(56,1)} + \frac16\sum_{I=1}^2 (a_I -m_7 )^3,
\end{align}
where  
\begin{align}\label{eq:E756SU2sing}
 \F^{E_7\times SU(2)}_{\bf (56,1)}&=\frac16 \sum_{I=1}^2\bigg( \sum_{i=1}^6 \relu{\ia_I \pm \frac12 m_0 \pm m_i}^3 +\!\!\! \sum_{\substack{\{s_i=\pm1\}\\{\rm even +}}}\!\! \!\relu{\ia_I +\frac12\sum_{k=1}^6 s_k m_k}^3\bigg).
\end{align}
Here we have neglected constant terms. As in the $N_f=7$ case where $E_8$  is realized as the embedding of $E_8\supset SO(16)\supset SO(14)\times U(1)$, the $E_7$ symmetry is also realized through the following embedding
\begin{align}
	E_7 &\supset SO(12)\times U(1)\crcr
\bf56 & = {\bf12}_{1}+{\bf12}_{-1}+{\bf32}_{0},
\end{align}
where the first term of \eqref{eq:E756SU2sing} represents the fundamental representation of $SO(12)$ with the opposite $U(1)$ charges  appearing as the coefficients of $m_0$, and  the second term of \eqref{eq:E756SU2sing} represents the spinor representation of $SO(12)$ without  $U(1)$ charges.  

 
With the second term in the RHS of \eqref{eq:248-1}, one finds that the decoupling a flavor yields
\begin{align}
\F^{E_8\times SU(2)}_{\bf(248,1)}+ \F^{E_8\times SU(2)}_{\rm CFT}
~\to~ \F^{E_7\times SU(2)}_{\bf(56,1)}  +\F^{E_7\times SU(2)}_{\rm CFT} 
,
\end{align}
where 
\begin{align}
\F^{E_7\times SU(2)}_{\rm CFT}=& -m_{AS}^2 \,\ia_1 +\sum_{I=1}^2\Big[\frac13\,\ia_I^3- \frac14 m_0^2\ia_I- \frac12\sum_{k=1}^{6} m_k^2 \ia_I\Big] .
\end{align}

In a similar fashion, one can show that the decoupling limit \eqref{eq:decouple7To6} on the complete prepotential for  $Sp(2)+1{\bf AS}+7{\bf F}$ theory leads to the complete prepotential for $Sp(2)+1{\bf AS}+6{\bf F}$ theory with the expected properties, $E_7\times SU(2)$ global symmetry and the factorized prepotential form as the sum of two rank-1 theories in the massless $m_{AS}$ limit, 
\begin{align}\label{eq:Sp21AS76to1AS6F}
\F^{Sp(2)+1{\bf AS}+7{\bf F}}_{E_8\times SU(2)} &\to 
\F_{\rm CFT}^{E_7\times SU(2)}+\F_{\bf (56,1)}^{E_7\times SU(2)}+\F_{\bf (133,2)}^{E_7\times SU(2)}+\F_{\bf (1,2)}^{E_7\times SU(2)}+\F_{\bf (56,3)}^{E_7\times SU(2)}	\crcr	
&=\F^{Sp(2)+1{\bf AS}+6{\bf F}}_{E_7\times SU(2)}, 
\end{align}
whose explicit forms are given 
in Appendix \ref{sec:app-rk2-1ASPrep}.\\



\paragraph{\underline{Flow to $Sp(2)+7{\bf F}$}} 
Now consider the RG flow to $Sp(2)+7{\bf F}$. It is achieved by decoupling the antisymmetric matter contribution from the complete prepotential for $Sp(2)+1{\bf AS}+7{\bf F}$ given in \eqref{eq:prep:rk2AS}. We take 
\begin{align}\label{eq:decoupleASTo7F}
	m_{AS},~ m_0^{Sp(2)+1{\bf AS}+7{\bf F}}\to\infty, ~~{\rm while}~~
	m^{Sp(2)+1{\bf AS}+7{\bf F}}_0-2m_{AS} \equiv m^{Sp(2)+7{\bf F}}_0~{\rm fixed}.
\end{align}
In this decoupling limit, one finds that the prepotentials for $Sp(2)$ gauge theory with 7 flavors is given by
\begin{align}\label{eq:Sp27FpreFrom1AS}
	\mathcal{F}^{Sp(2)+7{\bf F}} = &
	~ \frac12\,a_1^3+ a_1 a_2^2+\frac16 a_2^3 +\frac12 m_0 (a_1^2+a_2^2)  -\frac12 \sum_{i=1}^{7} m_i^2 (a_1+a_2)\cr
	&+\frac16\relu{a_1 \pm m_0}^3+\frac16
\sum_{I=1}^{2} \sum_{i=1}^{7}\relu{a_I \pm m_i}^3\crcr
&+\frac16
\sum_{\substack{\{s_i=\pm1\},\\\rm odd+}} \relu{a_1+a_2 + \frac12m_0 + \frac12\sum_{i=1}^{7} s_i\, m_i }^3\crcr
	&+\frac16
 \sum_{1\le i< j\le 7} \relu{2 a_1+ a_2 +m_0\pm m_{i}\pm m_{j}}^3.
\end{align}
Notice that terms in the prepotential \eqref{eq:Sp27FpreFrom1AS} are organized as the representations of $SO(14) \times U(1)_I$, where the $SO(14)$ symmetry is associated with 7 mass parameters $m_i$ and the $U(1)$ is associated with the instanton mass $m_0$ of $Sp(2)+7{\bf F}$ theory. Namely, the complete prepotential for $Sp(2)+7{\bf F}$ theory \eqref{eq:Sp27FpreFrom1AS} is of the form
\begin{align}
	\mathcal{F}^{Sp(2)+7{\bf F}}_{SO(14)\times U(1)} = &~\F_{\rm CFT}+ \F_{\rm singlet}+\F_F+\F_{S}+\F_{\rm rk\text{-}2},
\end{align}
where $\F_{\rm CFT}$ refers to the first line of \eqref{eq:Sp27FpreFrom1AS}, and $\F_{\rm singlet}+\F_F+\F_{S}+\F_{\rm rk\text{-}2}$ refer the singlet, fundamental, spinor, and rank-2 antisymmetric representations of $SO(14)$. 
We note that this complete prepotential \eqref{eq:Sp27FpreFrom1AS} agrees with the one obtained by a successive decouplings of two flavors from the complete prepotential for $Sp(2)$ gauge theory with 9 flavors, given in \eqref{rk2FSp}. (See also Appendix \ref{sec:app-rk2Prep} for the details.)
Hence, we confirm that the decoupling limit \eqref{eq:decoupleASTo7F} yields $\F^{Sp(2)+1{\rm AS}+7{\bf F}}_{E_8\times SU(2)}\to \F^{Sp(2)+7{\bf F}}_{SO(14)\times U(1)}$.
\\

\paragraph{\underline{Flow to $SU(2)_0{\bf-}SU(2){\bf-}[5]$}} 
The RG flow to $SU(2)_0{\bf-}SU(2){\bf-}[5]$ is achieved as follows: As $Sp(2)+1{\bf AS}+7{\bf F}$ is dual to the quiver $[1]{\bf-}SU(2)^{(1)}{\bf-}SU(2)^{(2)}{\bf-}[5]$, this quiver theory also enjoy an $E_8\times SU(2)$ global symmetry. We can then decouple the flavor associated with the $SU(2)^{(1)}$ gauge theory. As before, the masses of the first $SU(2)^{(1)}$ are labeled as $m_0^{(1)}$ $m_1^{(1)}$ and the masses of the second $SU(2)^{(2)}$ are labeled as $m_0^{(2)}$ and $m_i^{(2)}$  ($i=1,\cdots, 5$). We then take
\begin{align}
	m_0^{(1)}\to \infty\quad {\rm and} \quad m_1^{(1)}\to-\infty,\quad  {\rm while}~ m_0^{(1)}+m_1^{(1)}~~{\rm  fixed}.
\end{align} Or equivalently, from $Sp(2)+1{\bf AS}+7{\bf F}$, we take the limit 
\begin{align}
m_{AS}\to\infty \quad {\rm and} \quad a_1\to\infty,\quad {\rm while}\quad  a_1-m_{AS}~~{\rm fixed}.	
\end{align}
Now take the limit from \eqref{eq:prep:rk2AS}. With new Coulomb moduli parameter 
$\ia^{\rm new}_1 := \ia_1 - m_{AS}$, 
we find the complete prepotential for $SU(2)_0{\bf-}SU(2){\bf-}[5]$ shows an $E_8$ global symmetry \cite{Apruzzi:2019opn}\footnote{We checked that the 7-brane analysis based on the 5-brane web \cite{Hayashi:2018lyv} for the quiver theory $SU(2)_0{\bf-}SU(2){\bf-}[5]$ also supports the $E_8$ global symmetry of the quiver theory, whose 7-brane configuration in one of 5-brane loop as ${\bf \sf A^8 X_{[2,-1]} C}$, where $\bf \sf A, X_{[2,-1]}$, and ${\bf \sf C}$ represent $[1,0], [2,-1]$, and $[1,1]$ 7-branes, respectively \cite{DeWolfe:1998pr}.} as follows:
\begin{align}
	6\,\F^{SU(2)_0{\bf-}SU(2){\bf-}[5]}_{E_8}
	=&~ \ia_1^3 + \ia_2^3 -3\sum_{k=0}^7 m_k^2 ( \ia_1^2+\ia_2^2) +\relu{\ia_1-\ia_2}^3 \cr
	&+ \sum_{\vec{w} \in {\bf 248}}\relu{\ia_2+\vec{w}\cdot \vec{m}}^3 +\sum_{\vec{w} \in {\bf 3875}}\relu{\ia_1+\ia_2+\vec{w}\cdot \vec{m}}^3\crcr
	&
	+ \sum_{\vec{w} \in {\bf 30380}}\relu{2\ia_1+\ia_2+\vec{w}\cdot \vec{m}}^3,
\end{align}
where we have omitted irrelevant constant part and dropped the superscript, $\ia_1^{\rm new}\to \ia_1$.

We note here that one can also take $m_0^{(1)},m_1^{(1)}\to \infty$, while $m_0^{(1)}-m_1^{(1)}$ fixed. Or equivalently, from $Sp(2)+1{\bf AS}+7{\bf F}$, we take $m_{0}, m_3\to\infty$, while $m_0-m_{3}$ fixed. This decoupling limit hence leads to 
\begin{align}
	SU(2)_\pi{\rm -}SU(2){\rm -}[5] \qquad{\rm or}\qquad Sp(2)+1{\bf AS}+6{\bf F}.
\end{align}

\section{Conclusion}\label{sec:conclusion}
In the paper, we proposed ``complete" prepotential for 5d $\mathcal{N}=1$ gauge theories which covers all the parameter regions of the theories. 
This new prepotential is defined over the extended K\"ahler cone of the 5d theories. Hence, it includes non-perturbative contributions  
and, of course, reproduces the Intriligator-Morrison-Seiberg's (perturbative) prepotential in the weak coupling limit. 
Global symmetry enhancement and UV-duality of 5d $\mathcal{N}=1$ gauge theories are also captured in our complete prepotential. For that, we introduced the invariant Coulomb branch moduli which are shifted Coulomb branch parameters with the instanton mass parameter, so that the Coulomb branch moduli are invariant under the Weyl  reflections of enhanced global symmetry. 
The complete prepotential manifests enhanced global symmetry, when written in terms of the invariant Coulomb branch parameters

To obtain the complete prepotential, we mainly used 5-brane web configurations of a given 5d theory to keep track of possible flop transitions in the 5-brane web. It can be systematized by taking into account the Weyl reflections 
of enhanced global symmetry. We also used Gopakumar-Vafa invariants or Nekrasov partition functions defined on $\mathbb{R}^{3,1}\times S^1$, to obtain the same complete prepotential by taking the decompactification limit where the radius $R$ of an $S^1$ are very large. Along the way, we observed that the boundary of the physical Coulomb branch where monopole string tensions vanish, corresponds to the gauge bosons and string junctions being massless. 
 We observed that the contributions to the complete prepotential come from some of BPS hypermultiplets in the Gopakumar-Vafa invariant. From geometry, the same complete prepotential was obtained by considering volume of Mori cone generators.

As concrete examples, we discussed 5d gauge theories of rank-1 and -2, and worked out the form of the complete prepotentials explicitly in the main text and also in Appendix. In particular, 
the complete prepotentials for two pure $SU(2)=Sp(1)$ gauge theories with distinct discrete theta angles, show not only distinct global symmetries $E_1=SU(2)$ and $\widetilde{E}_1=U(1)$, respectively, but also different flop structure. Pure $SU(2)_\pi$ gauge theory has a flop which allows an RG flows to the non-Lagrangian $E_0$ theory. For rank-2 gauge theories, we obtained the complete prepotentials for $Sp(2)$ gauge theories of $N_f\le 9$. For the $N_f=9$ case, the enhanced global symmetry is $SO(20)$, and when expressed in terms of the invariant Coulomb branch parameters, the corresponding complete prepotential is $SO(20)$ manifest as expected. 
As it is the prepotential invariant under the enhanced global symmetry, it is also the complete prepotential for its dual theories. In other words, it is the prepotential for $SU(3)_\frac12$ gauge theory of $N_f=9$ flavors and also for $[3]{\rm-}SU(2){\rm -}SU(2){\rm-}[4]$ quiver theory, as they are all UV-dual to the $Sp(2)$ gauge theory with $N_f=9$ flavors. As another rank-2 example, we presented the complete prepotential for the $Sp(2)$ gauge theories of one antisymmetric and $N_f\le 7$ flavors, which is also known as 5d rank-2 $E_{N_f+1}$ theories. The complete prepotentials for these rank-2 theories are expressed in terms of polynomials that are invariant under Weyl group symmetry of the enhanced global symmetry and also tend to involve higher dimensional representations of $E_{N_f+1}$ symmetry.  
As a consistency check, we discussed RG flows that lead to less number of hypermultiplets and we cross-checked the prepotentials by considering different decoupling limits to reach the same theories of less number of hypermultiplets. 

Though we did not discuss all the rank-2 theories, one can straightforwardly apply our method to obtain the complete prepotential for those rank-2 theories which are not discussed here. In principle, it is applicable to other 5d higher rank superconformal theories. 
If both 5-brane webs and enhanced global symmetry of the 5d gauge theories are known, our method is exhaustive. 
For those theories whose 5-brane configurations are not yet known, one may still apply our method to find the corresponding prepotential, but there could be some missing terms that disappear in the weak coupling. On the other hand, our method might be useful for constructing or finding a 5-brane configuration for such theories which have yet to be found. As it suggests existence of possible flops which a proper 5-brane web should have, our method thus provides some consistency checks in attempting to construct new/unknown 5-brane configurations.  
\acknowledgments
We thank Joonho Kim, Hee-Cheol Kim, Xiaobin Li, Jaemo Park and Yuji Sugimoto for useful discussions. HH is grateful to APCTP for hospitality during the final stage of this work. 
SSK thank the hospitality of Seoul National University, POSTECH, and KIAS where part of this work is done, and also APCTP for hosting the Focus program “Strings, Branes and Gauge Theories 2019.”
FY thank the international conference ``Operators, Functions, and Systems of Mathematical Physics Conference” at Khazar University and ``Mini-workshop on Symmetry and Interactions'' at Shing-Tung Yau Center of Southeast University. 
The work of HH is supported in part by JSPS KAKENHI Grant Number JP18K13543. SSK is supported by the UESTC Research Grant A03017023801317. 
KL is supported in part by the National Research Foundation of Korea Grant NRF-2017R1D1A1B06034369. 
FY is supported by the NSFC grant No. 11950410490, by Fundamental Research Funds for the Central Universities A0920502051904-48, by Start-up research grant A1920502051907-2-046, 
in part by NSFC grant No. 11501470 and No. 11671328, and by Recruiting Foreign Experts Program No. T2018050 granted by SAFEA.
\appendix 
\section{Convention for representation of $E_8$} 
\label{sec:conventionE8}
Throughout the paper,
we use the convention that the simple roots $\alpha_k$ ($k=1,2, \cdots, 8$) of $E_8$ are expressed in terms of orthonormal basis 
$e_i$ ($i=0,1,\cdots, 7$) as 
\begin{align}\label{eq:simple-root}
& \alpha_1 = \frac{1}{2} (e_0 + e_1 - e_2- e_3 - e_4 - e_5 - e_6 - e_7  ),
\cr
& \alpha_2 = e_1 + e_2,
\cr
& \alpha_{i} = e_{i-1} - e_{i-2}  \qquad (i=3,4,5,6,7,8).
\end{align}
The Weyl reflection on a weight $w$ is generated by 
\begin{align}
w \to w - (w \cdot  \alpha_i  ) \alpha_i \qquad (i=1,2,\cdots ,8).
\end{align}
Introducing the mass vector as 
\begin{align}\label{eq:massvec}
\vec{m} = \sum_{i=0}^7 m_i e_i,
\end{align}
we can identify a weight $w$ with the linear combination of masses $\vec{m} \cdot w$.
Then, the Weyl reflection 
can be reinterpreted as a transformation on masses:
\begin{align}
& \alpha_1: \left\{
\begin{array}{l l}
m_i \to m_i - \displaystyle\frac{1}{4} \bigg(m_0 + m_1 - \sum_{j=2}^7 m_j\bigg) , & \qquad i=0,1 \\
m_i \to m_i + \displaystyle\frac{1}{4} \bigg(m_0 + m_1 - \sum_{j=2}^7 m_j\bigg) , & \qquad i=2,3,4,5,6,7 
\end{array}
\right.
\\
& \alpha_2: m_1 \leftrightarrow -m_2 \qquad 
\\
& \alpha_i: m_{i-1} \leftrightarrow m_{i-2}   \qquad (i=3,4,5,6,7,8).
\end{align}
Or, combining these transformations, we write the $E_8$ Weyl reflection as 
\begin{align}
&m_i \leftrightarrow m_j 
\\
& m_i \leftrightarrow - m_j  \qquad (\text{for } \,\, i \neq j)
\\
& m_i \to m_i - \frac{1}{4} \sum_{k=0}^7 m_k \qquad \text{(Simultaneously for all }i ) .
\end{align}

We take the Weyl chambers by imposing 
\begin{align}\label{eq:Weyl-ch}
\vec{m} \cdot \alpha_i \ge 0 \qquad (i=1,2, \cdots, 8),
\end{align}
which can be written explicitly as
\begin{align}
&\vec{m} \cdot \alpha_1 = m_0 + m_1 - \sum_{k=2}^7 m_k \ge 0, 
\cr
&\vec{m} \cdot \alpha_2 = m_1 + m_2 \ge 0, 
\cr
&\vec{m} \cdot \alpha_i = m_{i-1} - m_{i-2} \ge 0 \quad (i=3,4,\cdots 8),
\end{align}
or equivalently,
\begin{align}\label{eq:E8chamber}
& m_7 \ge m_6 \ge m_5 \ge m_4 \ge m_3 \ge m_2 \ge |m_1|,
\cr
&m_0 \ge \sum_{k=2}^7 m_k - m_1 .
\end{align}

The fundamental weights 
$\mu_i$ $(i=1,2, \cdots, 8)$ are specified by imposing that their corresponding highest weight $w_{\mu_i}$ satisfy
\begin{align}
w_{\mu_i} \cdot \alpha_j = \delta_{ij}.
\end{align}
More explicitly, we have 
\begin{align}\label{eq:high-mass-E8}
&\vec{m} \cdot w_{\mu_1}\! = 2m_0,
\quad
\vec{m} \cdot w_{\mu_2} \! = \frac{5}{2} m_0 +  \frac{1}{2} \sum_{i=1}^7 m_i, 
\quad
\vec{m} \cdot w_{\mu_3} \! = \frac{7}{2} m_0 - \frac{1}{2} m_1 +  \frac{1}{2} \sum_{i=2}^7 m_i, 
\cr
&\vec{m} \cdot w_{\mu_4} \! = 5m_0 + \sum_{k=3}^7 m_k,
\quad
\vec{m} \cdot w_{\mu_5}\!  = 4m_0 +  \sum_{k=4}^7 m_k,
\quad
\vec{m} \cdot w_{\mu_6} \! = 3m_0 +  \sum_{k=5}^7 m_k,
\cr
&\vec{m} \cdot w_{\mu_7}  = 2m_0 + m_7 +  m_6,
\qquad
\vec{m} \cdot w_{\mu_8}  = m_0 + m_7.
\end{align}
Since any weight can be obtained by subtracting the simple roots from the highest weight,
\eqref{eq:high-mass-E8} are the largest among all the weights 
in each representation in the Weyl chamber defined as in \eqref{eq:Weyl-ch}:
\begin{align}
\vec{m} \cdot w_{\mu_i} \ge \vec{m} \cdot w \qquad \forall w \in \mu_i .
\end{align}
We introduce the characters 
for each fundamental representation $\mu_i$ as
\begin{align}
{ \chi_{\mu_i}} (\vec{m}) = \sum_{w \in \mu_i} \exp \left[ - R w \cdot \vec{m} \right]. 
\end{align}
In the massless limit, they give
the dimension of each representation:
\begin{align}
	&\chi_{\mu_1}(0)= 3875, 
	&&\chi_{\mu_2}(0)= 147250, 
	&& \chi_{\mu_3}(0)= 6696000, 
		\crcr
	& \chi_{\mu_4}(0)=6899079264, 	
	&&\chi_{\mu_5}(0)= 146325270, 
	&& \chi_{\mu_6}(0)= 2450240, 
	\crcr
	& \chi_{\mu_7}(0)= 30380,
	 &&\chi_{\mu_8}(0)=248.
	 &&
\end{align}
Based on these dimensions of the representations, we also use the notation $\mathbf{248}$ for $\mu_8$, $\mathbf{3875}$ for $\mu_1$ and so on.
%
%
%
Decomposing the representation {\bf 248} of $E_8$ into the representations of the 
subgroup $E_8\supset SO(16) $,
\begin{align}
\mathbf{248} \to \mathbf{120} + \mathbf{128}.
\end{align}
The weights $w \in \mathbf{120}$ of $SO(16)$ are explicitly given as
\begin{align}
\vec{m} \cdot w = \pm\, m_i \pm \,m_j \quad (0 \le i\neq j \le 7) ,
\end{align}
which are in fact 112 in total, but we also took into account 8 Cartans which are of zero weights $\vec{m} \cdot w = 0$.
The weights $w \in \mathbf{128}$ are explicitly given as
\begin{align}
\vec{m} \cdot w = \frac{1}{2} \sum_{i=0}^7 s_i m_i
\qquad (s_i = \pm 1),
\end{align}
where even number of $s_i$ take the value $s_i = 1$ 
while the remaining $s_i$ take the value $s_i = -1$. 

\section{Prepotentials for rank-1 theories}
\label{sec:app-rk1Prep}
Here, we list the complete prepotentials for the rank-1 $SU(2)=Sp(1)$ gauge theories of $N_f$ flavors ($SU(2)+N_f\mathbf{F}$) associated with $E_{N_f+1}$ enhanced global symmetry. (We also include the prepotential for $E_0$ theory.)  We denote by $\ia$ the invariant Coulomb branch parameter and by $a$ the usual Coulomb branch parameter which appears in the IMS prepotential,
$\ia=a+\frac{m_0}{8-N_f}$. $m_0$ are instanton mass and $m_i$ masses of $N_f$ flavors.   \\

\noindent\underline{$SU(2)+7{\bf F}$}: $E_8$ global symmetry, $\ia= a+ m_0$.
\begin{align}
	6\,\mathcal{F}_{E_8} = &~ \ia^3 -3\sum_{k=0}^7 m_k^2\, \ia +  \sum_{0\le i< j\le 7}
	 \relu{\ia \pm m_i \pm m_j }^3\crcr
	&
	+\sum_{\substack{\{s_i=\pm1\},\\\text{even}+ }}\relu{\ia + \frac12 \sum_{k=0}^7 s_k\, m_k }^3\ .
\end{align}\\

\noindent\underline{$SU(2)+6{\bf F}$}: $E_7$ global symmetry, $\ia= a+\frac12 m_0$.
\begin{align}
6\,\mathcal{F}_{E_7} = &~2\,\ia^3 -3\Big( \frac12 m_0^2+\sum_{k=1}^6 m_k^2\Big)\, \ia 
+  \sum_{i=1}^{6} \relu{\ia \pm \frac{1}{2} m_0 \pm \,m_i }^3\crcr
	&
	+\sum_{\substack{\{s_i=\pm1\},\\
	\text{even}+}}\relu{\ia + \frac12 \sum_{k=1}^6 s_k\, m_k }^3\ .
\end{align}\\

\noindent\underline{$SU(2)+5{\bf F}$}: $E_6$ global symmetry, $\ia= a+\frac13 m_0$.
\begin{align}
	6\,\mathcal{F}_{E_6} = &~3 \,\ia^3 -\Big( m_0^2+3\sum_{k=1}^5 m_k^2\Big)\, \ia +  \sum_{i=1}^5 \relu{\ia - \frac13m_0 \pm \,m_i }^3\crcr
	&
	+\sum_{\substack{\{s_i=\pm1\},\\\text{even}+ }}\relu{\ia + \frac16 m_0 +\frac12 \sum_{k=1}^5 s_k\, m_k }^3+\relu{\ia +\frac23 m_0 }^3 \ .
\end{align}\\
%

\noindent\underline{$SU(2)+4{\bf F}$}: $E_5=SO(10)$ global symmetry, $\ia= a+\frac14 m_0$.
\begin{align}
	6\,\mathcal{F}_{E_5} = &~4 \,\ia^3 -3 \sum_{k=1}^5 x_k^2\, \ia +  \sum_{\substack{\{ s_i=\pm1 \},\\
	\text{odd}+}}\relu{\ia + \frac12 \sum_{i=1}^5\,s_i\,  x_i}^3 \ ,
\end{align}
where 
\begin{align}\label{eq:E5chemical}
	x_1=&\frac1{2}m_0, \crcr
	x_2=&\, \frac12 (-m_1+m_2+m_3+m_4),\quad
	x_3=\frac12 (m_1-m_2+m_3+m_4),\crcr	
	x_4=\, &\frac12 (m_1+m_2-m_3+m_4),\quad 
	x_5= \frac12 (m_1+m_2+m_3-m_4), 
\end{align}%
with $\displaystyle\sum_{k=1}^{5} x_k^2= \frac1{4}m_0^2+ \sum_{k=1}^4 m_k^2$.\\

\noindent\underline{$SU(2)+3{\bf F}$}: $E_4=SU(5)$ global symmetry, $\ia= a+\frac15 m_0$.
\begin{align}
	6\,\mathcal{F}_{E_4} = &~5\, \ia^3 -3 \sum_{k=1}^5 x_k^2\, \ia +  \sum_{1\le i<j\le 5}\relu{\ia + x_i+ x_j}^3 \ ,
\end{align}
where 
\begin{align}\label{eq:E4chemical}
	x_1=&\frac2{5}m_0, \crcr
	x_2=&\, -\frac1{10} m_0-\frac12 (m_1+m_2+m_3),\quad
	x_3= -\frac1{10}m_0-\frac12 (m_1-m_2-m_3),\crcr	
	x_4=\, &-\frac1{10}m_0-\frac12 (-m_1+m_2-m_3),\quad 
	x_5= -\frac1{10} m_0-\frac12 (-m_1-m_2+m_3),
\end{align}
subject to $\displaystyle\sum_{k=1}^5 x_k =0$. \\
%

\noindent\underline{$SU(2)+2{\bf F}$}: $E_3=SU(3)\times SU(2)$ global symmetry, $\ia= a+\frac16 m_0$.

\begin{align}
	6\,\mathcal{F}_{E_3} = &~6\,\ia^3 -3 \Big(\sum_{k=1}^3 x_k^2 + 2y^2\Big) \ia +\sum_{i=1}^3 \relu{\ia + x_i \pm y}^3\ ,
\end{align}
where 
\begin{align}\label{eq:E3chemical}
	x_1=&~ \frac13 m_0\ , \quad x_2=-\frac16 m_0 +\frac12 (m_1-m_2)\ , \quad x_3= -\frac16 m_0 -\frac12 (m_1-m_2) \ , \crcr
	y =&~ \frac12 (m_1+m_2)\ .
\end{align}\\
%

\noindent\underline{$SU(2)+1{\bf F}$}: $E_2=SU(2)\times U(1)$ global symmetry, $\ia= a+\frac17 m_0$.

\begin{align}
	6\,\mathcal{F}_{E_2} = &\,7\, \ia^3 -
	 6\big(x^2+\frac17\, y^2\big) \ia 
	 +\relu{\ia+ \frac47 y}^3+ \relu{\ia \pm x -\frac37 y}^3\ ,
\end{align}
where 
\begin{align}\label{eq:E2chemical}
	x= \frac14 m_0 + \frac14 m_1\ ,\qquad 	
	y=- \frac14 m_0 + \frac74 m_1\ ,
\end{align}
and the last term is an $SU(2)$ doublet, invariant under $x\leftrightarrow -x$.\\
%

\noindent\underline{pure $SU(2)_0$}: $E_1=SU(2)_I$ global symmetry, $\ia= a+\frac18 m_0$.
\begin{align}
	6\,\mathcal{F}_{E_1} =8\, \ia^3 -\frac3{8}\, m_0^2\, \ia\ .
\end{align}\\
\noindent\underline{pure $SU(2)_\pi$}: $\widetilde{E}_1=U(1)_I$ global symmetry, $\ia= a$.
\begin{align}\label{eq:appE1t}
	6\, \mathcal{F}_{\widetilde{E}_1} = 8\, a^3+ 3 \,m_0^2 \,a +\relu{ a+\frac12 m_0 }^3  .
\end{align}
\\

These complete prepotentials for $SU(2)$ gauge theory with $N_f$ flavors can be summarized as
\begin{align}\label{eq:app-rk1gene}
	6\,\mathcal{F}_{E_{N_f+1}} =&~(8-N_f)\, \ia^3 -3\Big( \frac{1}{8-N_f} m_0^2 +\sum_{k=1}^{N_f} m_k^2\Big) \ia + \sum_{\vec{w}\in \mu_{N_f+1}} \relu{ \ia + \vec{w}\cdot \vec{m}}^3\ , 	
\end{align}
where 
\begin{align}
	\ia = a + \frac{1}{8-N_f}m_0\ ,
\end{align}
and $\mu_{N_f+1}$ is the fundamental weight of the Lie algebra of the $E_{N_f+1}$ enhanced global symmetry ($\mu_8= {\bf 248}$, $\mu_7= {\bf 56}$, $\mu_6= \overline{\bf 27}$, $\mu_5= {\bf 16}$, $\mu_4= {\bf 10}$, $\mu_3= ({\bf 3}, {\bf 2})$, $\mu_2= {\bf 2}_{-\frac{3}{7}} + {\bf 1}_{\frac{4}{7}}$).
In terms of the Coulomb branch parameter, $a$, the complete prepotentials \eqref{eq:app-rk1gene} are expressed as 
\begin{align}
	6\mathcal{F}_{E_{N_f+1}}\!=&~3 m_0a^2 +(8-N_f) a^3 -3 \sum_{k=1}^{N_f} m_k^2 a + \sum_{\vec{w}\in \mu_{N_f+1}} \relu{ a + \frac{1}{8-N_f}m_0+\vec{w}\cdot \vec{m}}^3.
\end{align}
%
%

Finally, we can take a decoupling limit from the $\widetilde{E}_1$ theory to get the $E_0$ theory which is non-Lagrangian,\\ 
\noindent\underline{$E_0$ theory}
\begin{align}
	\mathcal{F}_{E_0} = \frac32 a^3_{E_0}\ ,  
\end{align}
where $a_{E_0}$ is defined in terms of the parameters of the $\widetilde{E}_1$ theory in \eqref{eq:appE1t}, 
\begin{align}
a_{E_0}\equiv a_{\widetilde{E}_1} +\frac16 m_0{}_{\widetilde{E}_1} \ .	
\end{align} 

\section{Prepotentials for $Sp(2)+(N_f\le 9)\, {\bf F}$}\label{sec:app-rk2Prep}
As in rank-1 case, we list the complete prepotentials for 5d $Sp(2)$ gauge theory with $N_f\le 9$ flavors, which include the prepotential for pure $Sp(2)_0$, $Sp(2)_\pi$, $SU(3)_4$ gauge theories, and also for non-Lagrangian theory of the geometry $\mathbb{F}_6\cup \mathbb{P}^2$.   \\

\noindent\underline{$Sp(2)+9{\bf F}$}: $SO(20)$ global symmetry, $\ia_1= a_1+ m_0$ and $\ia_2= a_2$.
\begin{align}
	\mathcal{F}_{SO(20)} = &~\F_{\rm CFT}+ \F_{\bf 20}+\F_{\bf 190}+\F_{\bf \overline{512}}+\F_{\bf 15504},
\end{align}
where
\begin{align}	
	6\,\F_{\rm CFT}=&
	~ \ia_1^3- 2\ia_2^3 + 6 \,\ia_1 \,\ia_2^2 -3 \sum_{i=0}^9 m_i^2 (\ia_1+\ia_2)\ , \\
	6\,\F_{\bf 20}=&
 \sum_{i=0}^9\relu{\ia_2 \pm m_i}^3,\\
	6\,\F_{\bf 190}=&
 \sum_{0\le i< j\le 9}\relu{\ia_1 \pm m_i\pm m_j }^3,\\
	6\,\F_{
	\bf \overline{512}}=&
\sum_{\substack{\{s_i=\pm1\},\\\rm odd+}} \relu{\ia_1+\ia_2 + \frac12\sum_{i=0}^9 s_i\, m_i }^3,\\
	6\,\F_{\bf 15504}=&
 \sum_{0\le i_i< i_2<\cdots<i_5\le 9}\relu{2\ia_1+\ia_2 \pm m_{i_1}\pm m_{i_2}\pm m_{i_3}\pm m_{i_4}\pm m_{i_5} }^3.
\end{align}

\noindent\underline{$Sp(2)+8{\bf F}$}: $SO(16)\times SU(2)_I$ global symmetry, $\ia_1= a_1+ \frac12 m_0$ and $\ia_2= a_2$.
\begin{align}
\mathcal{F}_{SO(16)\times SU(2)} = &~\F_{\rm CFT}+ \F_{(\bf 16,2)}+\F_{(\bf 120,1)}+\F_{(\bf \overline{128},1)}
	+\F_{(\bf 560,1)},
\end{align}
where
\begin{align}	
	6\,\F_{\rm CFT}=&
	~ 2\ia_1^3+ 6\, \ia_1\, \ia_2^2 - \frac32 m_0^2\ia_1  -3 \sum_{i=1}^8 m_i^2 (\ia_1+\ia_2)\ , \\
	6\,\F_{(\bf 16,2)}
	=&
 \sum_{i=1}^8\relu{\ia_1 \pm \frac12 m_0\pm m_i}^3,\\
	6\,\F_{(\bf 120,1)}
	=&
 \sum_{i=1}^{8}\relu{\ia_2 \pm m_i}^3,\\
	6\,\F_{(\bf \overline{128},1)}
	=&
\sum_{\substack{\{s_i=\pm1\},\\\rm odd+}} \relu{\ia_1+\ia_2 + \frac12\sum_{i=1}^8 s_i\, m_i }^3,\\
	6\,\F_{(\bf 560,1)}=&
 \sum_{1\le i_i< i_2<i_3\le 8}\relu{2\ia_1+\ia_2 \pm m_{i_1}\pm m_{i_2}\pm m_{i_3}}^3.
\end{align}

For $N_f\le7$, there is no global symmetry enhancement and global symmetry remains as  $SO(2N_f)\times U(1)_I$, and thus the invariant Coulomb branch parameters are not required. It is then convenient to define terms which frequently appear in the
complete prepotential,  
\begin{align}
	6\,\F^{}_{\rm CFT}=&
	~ (10-N_f)\,a_1^3+ 6 \,a_1 \,a_2^2+(8-N_f)\,a_2^3 \crcr
	& +3\, m_0 (a_1^2+a_2^2)  -3 \sum_{i=1}^{N_f} m_i^2 (a_1+a_2)\ ,\\
	6\,\F_{\rm singlet}=&~
\relu{a_1 \pm m_0}^3,\\
	6\,\F_F=&~
\sum_{I=1}^{2} \sum_{i=1}^{N_f}\relu{a_I \pm m_i}^3,\\
	6\,\F_{C}=&
\sum_{\substack{\{s_i=\pm1\},\\\rm odd+}} \relu{a_1+a_2 + \frac12m_0 + \frac12\sum_{i=1}^{N_f} s_i\, m_i }^3,\\
	6\,\F_{\rm rk\text{-}\ell}=&
 \sum_{1\le i_i< i_2<\cdots <i_\ell\le N_f} \relu{2 a_1+ a_2 +m_0\pm m_{i_1}\pm m_{i_2}\pm \cdots \pm m_{i_\ell}}^3.
\end{align}
Here the subscript $F$ or $C$ denotes the fundamental or the conjugate spinor representation of $G_F$, respectively, and the index ${\rm rk\text{-}\ell}$ of $\F_{{\rm rk\text{-}}\ell}$ to denote the rank-$\ell$ representation of the global symmetry group $G_{F}$. 
Then the complete prepotentials for $Sp(2)$ gauge theory with $N_f\le 7$ flavors are given as follows:\\
\noindent\underline{$Sp(2)+7{\bf F}$}: 
\begin{align}
	\mathcal{F}_{SO(14)\times U(1)} = &~\F_{\rm CFT}+ \F_{\rm singlet}+\F_F+\F_{C}+\F_{\rm rk\text{-}2},
\end{align}

\noindent\underline{$Sp(2)+6{\bf F}$}: 
\begin{align}
	\mathcal{F}_{SO(12)\times U(1)} = &~\F_{\rm CFT}+\F_F+\F_{C}+\F_{\rm rk\text{-}1},
\end{align}

\noindent\underline{$Sp(2)+5{\bf F}$}: 
\begin{align}
	\mathcal{F}_{SO(10)\times U(1)} = &~\F_{\rm CFT}+\F_F+\F_{C}+\F_{\rm rk\text{-}0},
\end{align}

\noindent\underline{$Sp(2)+(1 \le N_f\le 4)\,{\bf F}$}:
\begin{align}
	\mathcal{F}_{SO(2N_f)\times U(1)} = &~\F_{\rm CFT}+\F_F+\F_{C}.
\end{align}

\noindent\underline{$Sp(2)_\pi$}: $Sp(2)$ gauge theory with the discrete theta angle $\theta=\pi$.
\begin{align}\label{eq:sp2pi}
	\mathcal{F}_{\theta=\pi} = &~\F_{\rm CFT} =
	~ \frac53\,a_1^3+ a_1 a_2^2+\frac43\,a_2^3 
	+\frac12\, m_0 \,(a_1^2+a_2^2)  .
\end{align}

\noindent\underline{$Sp(2)_0$}: $Sp(2)$ gauge theory with the discrete theta angle $\theta=0$.
\begin{align}\label{eq:sp20}
	\mathcal{F}_{\theta=0} =& ~\F_{\rm CFT}+\F_{C}\crcr
	=&~ \frac53\,a_1^3+ a_1 a_2^2+\frac43\,a_2^3 
	+\frac12\, m_0 \,(a_1^2+a_2^2) +\frac16 \relu{a_1+a_2 +\frac12 m_0}^3 .
\end{align}

\noindent\underline{$SU(3)_4$}: Recall that $Sp(2)+1{\bf F}$ gauge theory also has a decoupling limit to $SU(3)_4$ gauge theory, which is achieved by taking $a_I \to \infty$ and $m_1,m_0\to -\infty$ with $m_0=m_1$. The resulting  prepotential is the complete prepotential for $SU(3)_4$ gauge theory: 
\begin{align}
	\F = \frac53 a_1^3 +\frac12 a_1^2 a_2 + \frac32 a_1 a_2^2 +\frac43 a_2^3+m_0(a_1^2+ a_1a_2+a_2^2) + \frac16\sum_{I=1}^2 \relu{a_I+m_0}^3\ .
\end{align}
\noindent Here one can take further decoupling \cite{Jefferson:2018irk} from either $Sp(2)_0$ or $SU(3)_4$ gauge theory to obtain a non-Lagrangian theory of the geometry $\mathbb{F}_6\cup\mathbb{P}^2$. From  $Sp(2)_0$ gauge theory, it is achieved by taking $m_0\to -\infty$ and $a_I\to \infty$, while $b_I\equiv a_I +\frac18 m_0$ are kept finite. From  $SU(3)_4$ gauge theory, it is achieved by taking $m_0\to -\infty$ and $a_I\to \infty$, while $b_I\equiv a_I +\frac14 m_0$ are kept finite. The prepotential for the theory of geometry  $\mathbb{F}_6\cup\mathbb{P}^2$ is then given by\\
\noindent\underline{$\mathbb{F}_6\cup\mathbb{P}^2$}:
\begin{align}\label{eq:F6P6}	
\F^{\mathbb{F}_6\cup\mathbb{P}^2} = \frac{11}{6} b_1^3 +\frac12 b_1^2 b_2 +\frac32 b_1 b_2^2 + \frac32 b_2^3\ .
\end{align}

\section{Prepotentials for $Sp(2)+1{\bf AS}+(N_f\le7){\bf F}$}\label{sec:app-rk2-1ASPrep}
We list the complete prepotential for the $Sp(2)$ gauge theories with one antisymmetric and $N_f$ flavors ($Sp(2)+1{\bf AS}+N_f{\bf F}$) associated with $E_{N_f+1}\times SU(2)$ enhanced global symmetry. (The prepotential for $SU(3)_6$ or rank-2 $E_0$ theory is also included.) 
Like rank-1 cases, the invariant Coulomb branch parameters for each flavor is given by 
\begin{align}
\ia_I=a_I+\frac{1}{8-N_f}m_0.
\qquad (I=1,2)
\end{align}
In the limit where $m_{AS}\to 0$, the complete prepotential becomes a sum of two copies of rank-1 prepotentials $\F(\ia_1, \ia_2)\to\F_{E_{N_f+1}}(\ia_1)+\F_{E_{N_f+1}}(\ia_2)$ \cite{Kim:2014nqa,Gadde:2015tra}. \\

\noindent\underline{$Sp(2)+1{\bf AS}+7{\bf F}$}: 
\begin{align}
\F_{E_8\times SU(2)} &= \F_{\rm CFT}+\F_{\bf (248,1)}+\F_{\bf (3875,2)}+\F_{\bf (1,2)}+\F_{\bf (1,3)}+\F_{\bf (30380,3)},
\end{align}
where
{\allowdisplaybreaks
\begin{align}
6\, \F_{\rm CFT}=& 
 -6\,m_{AS}^2 \,\ia_1 +\sum_{I=1}^2\Big[\ia_I^3-3\sum_{k=0}^{7} m_k^2 \ia_I\Big] , \\
6\,\mathcal{F}_{\bf(248,1)} = &~ \sum_{I=1}^2  
\sum_{w_{E_8}\in \bf{248}} \relu{\ia_I + \vec{w}_{E_8}\cdot \vec{m}}^3\\
=&\sum_{I=1}^2  \Bigg[
\sum_{0\le i< j\le 7}\relu{\ia_I \pm m_i\pm \,m_j }^3
	+\sum_{\substack{\{s_i=\pm1\},\\
	{\rm even+} }}\relu{\ia_I+ \frac12 \sum_{k=0}^7 s_k\, m_k }^3\Bigg]  ,\nonumber \\
6\,\F_{\bf(3875,2)} = &~ \sum_{w_{E_8\times SU(2)}\in \bf{(3875,2)}} \relu{\ia_1+\ia_2+ \vec{w}\cdot \vec{m}}^3\crcr
=& 
\sum_{0\le i\le j\le 7}\relu{\ia_1+\ia_2 \pm m_i\pm m_j \pm m_{AS}}^3\crcr
&+\sum_{0\le i_1<i_2<i_3<i_4\le 7}\relu{\ia_1+\ia_2 +\pm m_{i_1} \pm m_{i_2} \pm m_{i_3} \pm m_{i_4}\pm m_{AS}}^3\crcr
&	+\sum_{\substack{\{s_i=\pm1\},\\
	 {\rm odd~+} }}\sum_{i=0}^7\relu{\ia_1+\ia_2+ s_im_i\pm m_{AS}+\frac12 \sum_{k=0}^7 s_k\, m_k }^3,\\
6\,\F_{\bf(1,2)} = &~ \sum_{w_{E_8\times SU(2)}\in {\bf (1,2)}} \relu{\ia_1-\ia_2+ \vec{w}\cdot \vec{m}}^3 = \relu{\ia_1-\ia_2\pm m_{AS}}^3,\\
6\,\F_{\bf(1,3)} = &~ \sum_{w_{E_8\times SU(2)}\in {\bf(1,3)}} \relu{\ia_1+ \vec{w}\cdot \vec{m}}^3 = \relu{\ia_1\pm m_{AS}}^3,\\
6\,\F_{\bf(30380,3)} = &~ \sum_{w_{E_8\times SU(2)}\in {\bf(30380,3)}} \relu{2\ia_1+\ia_2+ \vec{w}\cdot \vec{m}}^3\crcr
=& 
\sum_{0\le i< j\le 7}\relu{2\ia_1+\ia_2 \pm m_i\pm m_j \pm2 m_{AS}}^3
\crcr
&	+\sum_{\substack{\{s_i=\pm1\},\\
	{\rm odd~+} }}\sum_{k=0}^7\relu{2\ia_1+\ia_2+ s_km_k\pm 2m_{AS}+\frac12 \sum_{i=0}^7 s_i\, m_i }^3
\crcr	
&	+\sum_{0\le i_1<i_2<i_3<i_4\le 7}\relu{2\ia_1+\ia_2 \pm m_{i_1}\pm m_{i_2} \pm m_{i_3}\pm m_{i_4}\pm 2 m_{AS}}^3
\crcr
&	+\sum_{k=0}^7\sum_{\substack{0\le i_1<i_2\le 7,\\
j_1\neq k\neq j_2}}\relu{2\ia_1+\ia_2 \pm 2m_k\pm m_{i_i}\pm m_{i_2}\pm 2 m_{AS}}^3
\\
& 
+\sum_{0\le i_1<\cdots< i_6\le 7}\relu{2\ia_1+\ia_2 \pm m_{i_1}\pm\cdots\pm m_{i_6} \pm2 m_{AS}}^3
\crcr
&	+\sum_{\substack{\{s_i=\pm1\}\\
	{\rm even~+} }}\sum_{0\le k_1,k_2\le7}\relu{2\ia_1+\ia_2+ s_{k_1}m_{k_1}+ s_{k_2}m_{k_2}\pm 2m_{AS}+\frac12 \sum_{i=0}^7 s_i\, m_i }^3.\qquad\nonumber
\end{align}
}
We have used the decomposition
\begin{align}
	E_8 &\supset SO(16)\crcr
\bf248 & = \bf120+128\crcr
\bf3875 & = \bf135+1820+1920\crcr
\bf30380 & = \bf120+1920+7020+8008+13312.
\end{align}\\

\noindent\underline{$Sp(2)+1{\bf AS}+6{\bf F}$}: 
\begin{align}\label{eq:Sp21AS6FApp}
\F_{E_7\times SU(2)} &= \F_{\rm CFT}+\F_{\bf (56,1)}+\F_{\bf (133,2)}+\F_{\bf (1,2)}+\F_{\bf (56,3)},
\end{align}
where
\begin{align}
6\, \F_{\rm CFT}=& -6\,m_{AS}^2 \,\ia_1 +\sum_{I=1}^2\Big[2\,\ia_I^3-3\Big( \frac12 m_0^2+\sum_{k=1}^{6} m_k^2\Big) \ia_I\Big] ,\\
6\, \F_{\bf (56,1)}=& \sum_{I=1}^2\bigg( \sum_{i=1}^6 \relu{\ia_I \pm \frac12 m_0 \pm m_i}^3 + \sum_{\substack{\{s_i=\pm1\}\\ {\rm even +} }} \relu{\ia_I +\frac12\sum_{k=1}^6 s_k m_k}^3\bigg),\\
6\, \F_{\bf (133,2)}=& \sum_{\substack{\{s_i=\pm1\}\\
	{\rm odd +}}} \relu{2\ia_1+\ia_2 +\frac12\sum_{k=1}^6 s_k m_k \pm  m_{AS}}^3
	 \crcr
	&+\sum_{i\le i <j \le 6} \relu{\ia_1+\ia_2 \pm m_i\pm m_j \pm m_{AS}}^3\crcr
	&+\relu{\ia_1 +\ia_2\pm m_0 \pm m_{AS}}^3,\\
6\, \F_{\bf (1,2)}=&  \relu{\ia_1-\ia_2\pm m_{AS}}^3,
	\\
6\, \F_{\bf (56,3)}=&  \sum_{i=1}^6 \relu{2\ia_1+\ia_2 \pm \frac12 m_0 \pm m_i\pm2m_{AS}}^3\crcr
& + \sum_{\substack{\{s_i=\pm1\}\\{\rm even +}}} \relu{2\ia_1+\ia_2 +\frac12\sum_{k=1}^6 s_k m_k \pm 2 m_{AS}}^3.
\end{align}\\
We have used the decomposition
\begin{align}
	E_7 &\supset SO(12)\times U(1)\crcr
\bf56 & = {\bf12}_{1}+{\bf12}_{-1}+{\bf32}_{0}\crcr
\bf133 & = {\bf32}_{1}+{\bf32}_{-1}+{\bf66}_{0}+{\bf1}_{-2}+{\bf1}_{0}+{\bf1}_{2}.
\end{align}\\

\noindent\underline{$Sp(2)+1{\bf AS}+5{\bf F}$}: 
\begin{align}
\F_{E_6\times SU(2)} &= \F_{\rm CFT}+\F_{\bf (\overline{27},1)}+\F_{\bf (27,2)}+\F_{\bf (1,2)}+\F_{\bf (1,3)},
\end{align}
where 
\begin{align}
6\, \F_{\rm CFT}&= -6\,m_{AS}^2 \,\ia_1 +\sum_{I=1}^2\Big[3\,\ia_I^3-3\Big( \frac13 m_0^2+\sum_{k=1}^{5} m_k^2\Big) \ia_I\Big] , \\
6\, \F_{\bf (\overline{27},1)}=&  \sum_{I=1}^2\bigg( \relu{\ia_I +\frac23 m_0 }^3 + \sum_{i=1}^5 \relu{\ia_I -\frac13 m_0\pm m_i}^3\crcr
& + \sum_{\substack{\{s_i=\pm1\}\\
	{\rm even +}}} \relu{\ia_I+\frac16 m_0+\frac12\sum_{k=1}^5 s_k m_k }^3\bigg),\\
6\, \F_{\bf (27,2)}=&  \relu{\ia_1+\ia_2 - \frac23 m_0 \pm m_{AS}}^3 +\sum_{i=1}^5 \relu{\ia_1+\ia_2 + \frac13 m_0\pm m_i \pm m_{AS}}^3\crcr
& + \sum_{\substack{\{s_i=\pm1\}\\
	{\rm odd +}}} \relu{\ia_1+\ia_2 -\frac16 m_0+\frac12\sum_{k=1}^5 s_k m_k \pm  m_{AS}}^3,\\
6\, \F_{\bf (1,2)}=&\relu{\ia_1-\ia_2 \pm m_{AS}}^3,\\
6\, \F_{\bf (1,3)}=&\relu{2\ia_1+\ia_2 \pm 2m_{AS}}^3.
\end{align}
We have used the decomposition
\begin{align}
	E_6 &\supset SO(10)\times U(1)\crcr
\bf\overline{27} & = {\bf10}_{-\frac13}+{\bf\overline{16}}_{\frac16}+{\bf1}_{\frac23}\crcr
\bf27 & = {\bf10}_{\frac13}+{\bf{16}}_{-\frac16}+{\bf1}_{-\frac23}.
\end{align}\\

\noindent\underline{$Sp(2)+1{\bf AS}+4{\bf F}$}: 
\begin{align}
\F_{E_5\times SU(2)} &= \F_{\rm CFT}+\F_{\bf (16,1)}+\F_{\bf (10,2)}+\F_{\bf (1,2)},
\end{align}
where, with the $E_5$ chemical potential $x_i$ given in \eqref{eq:E5chemical},
\begin{align}
6\, \F_{\rm CFT}&= -6\,m_{AS}^2\, \ia_1 +\sum_{I=1}^2\Big[4\,\ia_I^3-3 \sum_{k=1}^{5} x_k^2\, \ia_I\Big] , \\
6\,\F_{\bf (16,1)} &= \sum_{I=1}^2\sum_{\substack{\{s_i=\pm1\},\\
	{\rm odd +} }}\relu{\ia_I  +\frac12 \sum_{k=1}^5 s_i\, x_i }^3, \\
6\,\F_{\bf (10,2)} &= \sum_{i=1}^{5} \relu{\ia_1+\ia_2 \pm x_i \pm m_{AS}}^3, \\
6\,\F_{\bf (1,2)} &= \relu{\ia_1-\ia_2\pm m_{AS}}^3.
\end{align}\\

\noindent\underline{$Sp(2)+1{\bf AS}+3{\bf F}$}: 
\begin{align}
\F_{E_4\times SU(2)} &= \F_{\rm CFT}+\F_{\bf (10,1)}+\F_{\bf (\overline{5},2)}+\F_{\bf (1,2)},
\end{align}
where, with the chemical potential $x_i$ given in \eqref{eq:E4chemical}, 
\begin{align}
6\, \F_{\rm CFT}&= -6\,m_{AS}^2\, \ia_1 +\sum_{I=1}^2\Big[5\ia_I^3-3 \sum_{k=1}^{5} x_k^2\, \ia_I\Big] , \\
6\, \F_{\bf (10,1)} &= \sum_{I=1}^2\sum_{i\le i<j\le 5} \relu{\ia_I +x_i +x_j}^3,\\
6\,\F_{\bf (\overline{5},2)}&=\sum_{i=1}^{5} \relu{\ia_1+\ia_2 -x_i \pm m_{AS}}^3,\\
6\, \F_{\bf (1,2)} &=\relu{\ia_1-\ia_2 \pm m_{AS}}^3.
\end{align}\\

\noindent\underline{$Sp(2)+1{\bf AS}+2{\bf F}$}: 
\begin{align}
\F_{E_3\times SU(2)} &= \F_{\rm CFT}+\F_{\bf (3,2,1)}+\F_{\bf (\overline{3},1,2)}+\F_{\bf (1,1,2)},
\end{align}
where
\begin{align}
6\, \F_{\rm CFT}&= -6\,m_{AS}^2 \,\ia_1 +\sum_{I=1}^2\Big[6\,\ia_I^3-3 \Big(\sum_{k=1}^{3} x_k^2+2y^2\Big) \ia_I\Big] , \\
6\, \F_{\bf (3,2,1)} & = \sum_{I=1}^2 \sum_{i=1}^3\relu{\ia_I +x_i \pm y}^3,\\
6\,\F_{\bf (\overline{3},1,2)}&=  \sum_{I=1}^2\sum_{i=1}^3\relu{\ia_I -x_i \pm m_{AS}}^3,\\
6\,\F_{\bf (1,1,2)} & = \relu{\ia_1-\ia_2 \pm m_{AS}}^3.
\end{align}
The $SU(3)$ chemical potentials $x_1,x_2,x_3$, subject to $\sum_{k=1}^3x_k=0$, and the $SU(2)$ chemical potential $y$ are given in \eqref{eq:E3chemical}.\\

\noindent\underline{$Sp(2)+1{\bf AS}+1{\bf F}$}: 
\begin{align}
\F_{E_2\times SU(2)} &= \F_{\rm CFT}+\F_{ ({\bf1}_{4/7}+{\bf2}_{-3/7},{\bf1})}+\F_{ ({\bf2}_{1/7},{\bf2})}+\F_{ ({\bf1}_0,{\bf2})},
\end{align}
where
\begin{align}
6\, \F_{\rm CFT}&= -6\,m_{AS}^2 \,\ia_1 +\sum_{I=1}^2\Big[7\,\ia_I^3-3 \Big(x^2+\frac17y^2\Big) \ia_I\Big] ,\\
6\, \F_{ ({\bf1}_{4/7}+{\bf2}_{-3/7},{\bf1})}&=
\sum_{I=1}^2 \Big( \relu{\ia_I +\frac47 y}^3+ \relu{\ia_I\pm x -\frac37 y}^3\Big),\\
6\,\F_{ ({\bf2}_{1/7},{\bf2})} &=
\relu{\ia_1+\ia_2\pm x +\frac17 y\pm m_{AS}}^3,\\
6\, \F_{ ({\bf1}_0,{\bf2})} &= \relu{\ia_1-\ia_2\pm m_{AS}}^3,
\end{align}
The $SU(2)$ chemical potential $x$ and the $U(1)$ chemical potential $y$ are defined in \eqref{eq:E2chemical}\\

\noindent\underline{$Sp(2)_0+1{\bf AS}$}: $\ia_I = a_I +\frac18 m_0$. 
\begin{align}
6\,\F_{E_1\times SU(2)} 
=& -6\,m_{AS}^2 \,\ia_1 +\sum_{I=1}^2\Big[8\,\ia_I^3-\frac38\,m_0^2\, \ia_I\Big] \cr
&+\relu{\ia_1+\ia_2\pm \frac14 m_0\pm m_{AS}}^3
+\relu{\ia_1-\ia_2\pm m_{AS}}^3.
\end{align}
Decoupling one antisymmetric by taking taking 
$m_{AS}, m_0\to \infty$ while $m_0^{Sp(2)_0}\equiv m_0-2m_{AS} $ kept finite, we  obtain the complete prepotential for pure $Sp(2)_0$ gauge theory \eqref{eq:sp20}.\\

\noindent\underline{$Sp(2)_\pi+1{\bf AS}$}: $\ia_I=a_I$. 
\begin{align}
6\,\F_{\widetilde{E}_1\times SU(2)} 
=& -6\,m_{AS}^2\, a_1 +\sum_{I=1}^2\Big[8\,a_I^3+3\,m_0 \,a_I^2\Big]
\crcr
& 
+\sum_{I=1}^2\relu{a_I+ \frac12 m_0}^3
+\relu{a_1\pm a_2\pm m_{AS}}^3.
\end{align}
If we decouple an antisymmetric by taking 
$m_{AS}, m_0\to \infty$ while $m_0^{Sp(2)_\pi}\equiv m_0-2m_{AS} $ kept finite, then 
we get the complete prepotential for pure $Sp(2)_\pi$ gauge theory \eqref{eq:sp2pi}. On the other hand, if we decouple an antisymmetric by taking 
$m_0\to -\infty, a_I\to\infty$ while $a_I+\frac16 m_0\equiv b_I $ kept finite, then one can see that $m_{AS}$ is naturally identified as $\frac12 m_0$ of $SU(3)_6$, and it is straightforward to obtain the complete prepotential for pure $SU(3)_6$ gauge theory \cite{Jefferson:2018irk, Hayashi:2018lyv} as follows.\\
\noindent\underline{$SU(3)_6$}:
\begin{align}\label{eq:SU3_6}
\F^{SU(3)_6}= -\frac14 m_{0}^2\, b_1 + \frac32 b_1^3+ \frac32 b_2^3 +\frac16 \relu{b_1-b_2\pm \frac12 m_{0}}^3.
\end{align}
Notice that the prepotential \eqref{eq:SU3_6} has $\relu{\cdots}^3$, which implies a further flop. It is, in fact, possible to take a flop transition on the web diagram for the pure $SU(3)_6$ gauge theory, as depicted in Figure 87(b) in \cite{Hayashi:2018lyv}. 
We note that if we take the infinite coupling limit 
$m_0\to 0$, then the complete prepotential for pure $SU(3)_6$ gauge theory reduces to two copies of that for $E_0$ theory, as expected:
\begin{align}
\F^{SU(3)_6}\,\to\frac32\, b_1^3+ \frac32\, b_2^3\,=\,\F_{E_0}(b_1) +\F_{E_0}(b_2) \ .
\end{align}
\section{Derivation of physical Coulomb moduli for rank-1 $E_8$ SCFT}\label{sec:PhysCB}

In this appendix, we derive the physical Coulomb moduli for rank-1 $E_8$ CFT, 
which corresponds to $SU(2)=Sp(1)$ gauge theory with $N_f=7$ flavors.  
The physical Coulomb moduli are defined as the region of Coulomb moduli where the monopole tensions are positive:
\begin{align}\label{eq:E8 positive}
\frac{\partial \F_{E_8}}{\partial  a} = \frac{\partial  \F_{E_8}}{\partial  \ia} \ge 0.
\end{align}

In order to compute such region, we first concentrate on the specific Weyl chamber given in \eqref{eq:E8chamber}.
In this chamber, it is straightforward to show that the term in \eqref{eq:invFforE8} satisfies
\begin{align}
\ia + \frac{1}{2} \sum_{i=0}^7 s_i m_i \ge 0\ ,
\qquad \text{ unless } ~ s_0 = s_1 = \cdots = s_7 = -1 .
\end{align}
 This indicates that, due to the definition \eqref{eq:doubleAbs}, the following term in \eqref{eq:invFforE8} simplifies as
\begin{align}
\sum_{\substack{\{s_i=\pm1\},\\
	\text{even}+}} \relu{ \ia + \frac12 \sum_{k=0}^7 s_k\, m_k }^3\ 
= \relu{ \ia - \frac12 \sum_{k=0}^7 m_k }^3\ 
= \relu{ a +\frac12 m_0 - \frac12 \sum_{k=1}^7 m_k } ^3\ ,
\end{align}
in the considered Weyl chamber.
Analogously, one can show that the term in \eqref{eq:invFforE8} satisfies
\begin{align}
\ia + s_1 m_i + s_2 m_j \ge 0
\qquad \text{ unless } ~
\left\{
\begin{array}{llll}
~i=0, & j=1, & s_0 = -1, & \qquad  \\ \qquad \text{ or }\\
~i=0, & j \ge 2, & s_0 = -1, &  s_j =-1.
\end{array}
\right.
\end{align}
 This indicates that the following term in \eqref{eq:invFforE8} simplifies as
\begin{align}
\sum_{0\le i< j\le 7}&
\relu{ \ia \pm m_i \pm \,m_j } ^3\crcr
= &~ \relu{ \ia -  m_0 \pm m_1 } ^3
+ \sum_{j=2}^7 \relu{ \ia -  m_0 - \,m_j }^3
\cr
= &~ \relu{ a + \,m_1 }^3 + \sum_{i=1}^7 \relu{a - \,m_i }^3.
\end{align}
In summary, in the Weyl chamber given in \eqref{eq:E8chamber}, 
the prepotential \eqref{eq:invFforE8} simplifies as
\begin{align}\label{eq:reducedE8}
\F_{E_8}
= &~\frac16 (a+m_0)^3 -\frac12\sum_{k=0}^7 m_k^2\, (a+m_0) 
\cr
& 
+ \frac16 \relu{ a + \,m_1 }^3 + \frac16\sum_{i=1}^7 \relu{ a - \,m_i }^3
+ \frac16 \relu{ a +\frac12 m_0 - \frac12 \sum_{k=1}^7 m_k }^3.\quad
\end{align}

Now, starting from the region with large $a$, we gradually reduce the value of $a$.
In the region $a \ge |m_1|$, the structure is quite simple:
\begin{align}
\frac{\partial \F_{E_8} }{\partial a}
= & \left\{
\begin{array}{lllll}
\frac{\partial\F_{E_8}^{\,\,0} }{\partial a}& \quad a \ge m_7, \\
\frac{\partial\F_{E_{k+1}}^{\,\,0} }{\partial a}& \quad m_{k-1} \le a \le m_k  & \quad (k=3,\cdots, 7), \\
\frac{\partial\F_{E_{2}}^{\,\,0} }{\partial a}& \quad |m_{1}| \le a \le m_2,   &  \\
\end{array}
\right.
\end{align}
where we defined
\begin{align}
	\F_{E_{n+1}}^{\,\, 0} 
	\equiv &~\frac12  
	\left( m_0 - \sum_{k=n+1}^{7} m_k \right) \,a^2 +\frac{8-n}{6} a^3 -\frac12 \sum_{k=1}^{n} m_k^2 a.
\end{align}
As discussed in Appendix 
\ref{sec:app-rk1Prep}, we observe that this is the prepotential in the CFT phase of the $E_{n+1}$ theory when we identify the combination
$
m_0 - \sum_{k=n+1}^{7} m_k 
$
to be the instanton mass parameter for the $E_{n+1}$ theory.
This can be interpreted that reducing $a$ is equivalent to decoupling the flavor.
By considering all the region above one by one, it is also straightforward to show that 
\eqref{eq:E8 positive} is satisfied for $a \ge |m_1|$.

In order to investigate the remaining parameter region 
$a \le |m_1|$, 
we consider the following three cases:
\begin{align}
\begin{array}{lll}
\text{Case }1: & m_1 \le 0, & \\
\text{Case }2: & m_1 \ge 0, & m_0 - \displaystyle\sum_{i=1}^7 m_i \ge 0, \\
\text{Case }3: & m_1 \ge 0, & m_0 - \displaystyle\sum_{i=1}^7 m_i \le 0 .
\end{array}
\nonumber
\end{align}
For Case 1, the prepotential \eqref{eq:reducedE8} at the remaining parameter region is given by
\begin{align}\label{eq:case1}
\frac{ \partial \F_{E_8}^{\text{Case 1}} }{\partial a}= \frac{ \partial \F_{E_1}^{\,\,0} }{\partial a}
\ge 0
\quad \text{ for } \quad 0 \le a \le - m_1
\end{align}
with the inequality saturated at $a=0$.
For Case 2, the prepotential \eqref{eq:reducedE8} at the remaining parameter region is given by
\begin{align}\label{eq:case1}
\frac{ \partial \F_{E_8}^{\text{Case 1}} }{\partial a}= \frac{ \partial \F_{\tilde{E}_1}^{\,\,0} }{\partial a} \ge 0
\quad \text{ for }  \quad 0 \le a \le m_1
\end{align}
with the inequality saturated at $a=0$,
where actually $ \F_{\widetilde{E}_{1}}^{\,\,0} = \F_{E_{1}}^{\,\,0}$.
Since $m_0 - \displaystyle\sum_{i=1}^7 m_i \ge 0$ is automatically satisfied in Case 1 due to \eqref{eq:E8chamber},
the results in Case 1 and Case 2 can be summarized that the physical Coulomb moduli is given by
\begin{align}\label{eq:summary-case12}
a \ge 0  \quad \text{ if } \quad -\frac16 \left( m_0 - \sum_{i=1}^7 m_i \right) \le 0.
\end{align}
%
For Case 3, the parameter region should be further divided into two parts as 
\begin{align}\label{eq:case1}
\frac{\partial \F_{E_8}^{\text{Case 3}} }{\partial a}= 
\left\{
\begin{array}{lllll}
\frac{ \partial \F_{\tilde{E}_{1}}^{\,\,0} }{\partial a }
 & \quad  -\frac12 \left( m_0 - \displaystyle\sum_{i=1}^7 m_i \right) \le a \le m_1,   \\
\frac{ \partial \F_{E_{0}}^{\,\,0}}{\partial a} & \quad 
a \le  -\frac12 \left( m_0 - \displaystyle\sum_{i=1}^7 m_i \right). &\\
\end{array}
\right.
\end{align}
In this case, a non-trivial condition 
is obtained 
\begin{align}
\frac{ \partial \F_{E_8}^{ \text{Case 3} } }{\partial a} \ge 0 \quad \Leftrightarrow
\quad a \ge -\frac16 \left( m_0 - \sum_{i=1}^7 m_i \right).
\end{align}
Rephrasing this result for Case 3, 
the physical Coulomb moduli is
\begin{align} \label{eq:summary-case3}
a \ge -\frac16 \left( m_0 - \sum_{i=1}^7 m_i \right)  \quad \text{ if } \quad -\frac16 \left( m_0 - \sum_{i=1}^7 m_i \right) \ge 0.
\end{align}
Combining the results  \eqref{eq:summary-case12} and \eqref{eq:summary-case3},
we conclude that the physical Coulomb moduli is given by
\begin{align}
a \ge 0 \quad \text{ and } \quad
a \ge -\frac16 \left( m_0 - \sum_{i=1}^7 m_i \right),
\end{align}
or equivalently, 
\begin{align}\label{eq:phyCB-ch}
2 \ia \ge 2 m_0 \quad \text{ and } \quad
3 \ia \ge \frac52 m_0 + \frac12 \sum_{i=1}^7 m_i 
\end{align}
in the Weyl chamber \eqref{eq:E8chamber}.

Finally, we discuss the conditions for physical Coulomb moduli in the different Weyl chambers of the $E_8$.
Such conditions can be obtained by acting the Weyl reflection
 to \eqref{eq:phyCB-ch}
by taking into account that $\ia$ is invariant under the Weyl group symmetry.  
From \eqref{eq:high-mass-E8}, we observe that the right hand side of \eqref{eq:phyCB-ch} is the highest weight of 
the representation $\mu_1$ and $\mu_2$, respectively.
By acting the Weyl reflection, 
the weights in each representation appear.%
\footnote{Not all the weights can be obtained by the Weyl reflection 
of the highest weight.
However, the conditions corresponding to such weights can be obtained from the other conditions corresponding to the weights obtained from the Weyl reflection 
to the highest weight. Therefore, the results does not depend on whether or not we include such weights.} 
Thus, the conditions \eqref{eq:phyCB-ch} can be extended to the whole parameter regions as
\begin{align}\label{eq:phCB}
2 \ia \ge w \cdot \vec{m}\,\, \text{ for }\,\, \forall w \in \mu_1,
 \qquad 
 \qquad
3 \ia \ge w \cdot \vec{m}\,\, \text{ for } \,\, \forall w \in \mu_2.
\end{align}
This is the physical Coulomb moduli for rank-1 $E_8$ theory.

In this physical Coulomb moduli, we can derive several non-trivial inequalities.
For example, from \eqref{eq:phyCB-ch}, we can deduce that 
\begin{align}\label{eq:hw78}
\ia \ge 0,
\qquad
2 \ia \ge m_0 + m_7,
\qquad
3 \ia \ge 2m_0 + m_7 + m_6,
\end{align}
are satisfied in the Weyl chamber \eqref{eq:E8chamber}.
From \eqref{eq:high-mass-E8}, we observe that the right hand side in \eqref{eq:hw78} is the highest weight of representation $\mathbf{1}$, $\mu_8$ and $\mu_7$, respectively.
Again, by acting the Weyl reflection, 
the weights in each representation appear, and thus,
\begin{align}\label{eq:physCBineq}
\ia \ge 0,
\qquad \qquad
2 \ia \ge w \cdot \vec{m}\,\, \text{ for }\,\, \forall w \in \mu_8,
 \qquad 
 \qquad
3 \ia \ge w \cdot \vec{m}\,\, \text{ for } \,\, \forall w \in \mu_7.
\end{align}
Combining all the inequalities in \eqref{eq:phCB} and  \eqref{eq:physCBineq}, we reproduce \eqref{eq:physCB-E8}.

\section{Partition function for 5d $Sp(2)+9{\bf F}$ 
from elliptic genus}
\label{sec:app-Sp2-GV}

There are several ways to obtain partition function for 5d $\mathcal{N}=1$ $Sp(2)$ gauge theory with $N_f=9$ flavors. 
One of the methods \cite{Hwang:2014uwa} is based on ADHM quantum mechanics using the Jeffrey-Kirwan method. It would be also possible to compute it based on 5-brane web of the type discussed in \cite{Aharony:1997bh, Benini:2009gi} and apply the (refined) topological vertex method \cite{Aganagic:2003db, Iqbal:2007ii, Awata:2008ed}.
Here, we derive it from the elliptic genus and express it in terms of Gopakumar-Vafa invariants as in \eqref{eq:GV-general}.
As discussed in \cite{Hayashi:2016abm, Yun:2016yzw}, the elliptic genus for 6d $\mathcal{N}=1$ $Sp(1)$ gauge theory with $N_f=10$ flavors and with a tensor reproduces the partition function for 5d $\mathcal{N}=1$ $Sp(2)$ gauge theory with $N_f=10$ flavors up to duality map. 
Then, the partition function for $N_f=9$ flavors can be obtained by decoupling limit from them.

The elliptic genus $Z_k$ for strings are given as follows \cite{Kim:2015fxa, Hayashi:2016abm, Yun:2016yzw}:

\noindent
For $k=0$, 
\begin{align}\label{eq:Z0}
&Z_0 = \text{PE} ( \F_0 ),
\cr
&\F_0 = \frac{[0,\frac12]_{t,q} ( e^{ - 2 \alpha} + e^{ 2 \alpha}  ) + [0,0]_{t,q}  (e^{ - \alpha} +e^{ \alpha}) \sum_{i=0}^9 (e^{-m_i} + e^{m_i}) }{ ( t^{\frac{1}{2}} - t^{-\frac{1}{2}} ) ( q^{\frac{1}{2}} - q^{-\frac{1}{2}} ) } 
\left( \frac{e^{2 \pi i \tau}}{1-e^{2 \pi i \tau}} + \frac{1}{2} \right) .\qquad
\end{align}
%
%
%
%
For $k=1$,
\begin{align}\label{ell-Sp1}
Z_1 = - \frac{\eta^2}{\theta(\epsilon_{1,2})}
\sum_{I=1}^4 \frac{\eta^2}{\theta_{I}(\epsilon_+ \pm \alpha )} \prod_{l=1}^{10} \frac{\theta_{I}(m_l)}{\eta},
\end{align}
with
$
\epsilon_{\pm} = \frac{1}{2} ( \epsilon_1 \pm \epsilon_2 ),
$
$
\theta_i(x) := \theta_i(x; \tau).
$
%
%
%
%
%
\\
\noindent
For $k=2$, 
\begin{align}
Z_{2}
&=
\sum_{s=1}^2 
\frac{ \eta^6 }{2 \theta_1(\epsilon_1)\theta_1(\epsilon_2)
\theta_1(\epsilon_1 \pm 2( (-1)^s \epsilon_+ +  \alpha)) \theta_1(\epsilon_2 \pm 2( (-1)^s \epsilon_+ +  \alpha)) }
\cr
&
\qquad\qquad\qquad\qquad
 \cdot
\frac{\eta^2}{
\theta_1(- 2 \alpha )
\theta_1(2 (-1)^s \epsilon_+ +  2 \alpha )}
\cdot \prod_{l=1}^{10}\frac{\theta_1(m_l \pm ( (-1)^s \epsilon_+ + \alpha))}{\eta^2}
\cr
& \quad + 
\sum_{I=1}^4 \sum_{s=1}^2
\frac{\eta^4}{
4 \theta_1(\epsilon_1)\theta_1(\epsilon_2) \theta_1(2 \epsilon_s)  
\theta_1(2 \epsilon_+ - 2\epsilon_s) }
\frac{\eta^4}{\theta_I(\epsilon_+\pm \alpha \pm \frac{\epsilon_s}{2})}
\cdot
\prod_{l=1}^{10}\frac{
\theta_I(m_l \pm \frac{\epsilon_s}{2})}{\eta^2} 
\cr
& \quad + \sum_{(I,J,K) \in S}
\frac{
\eta^8 \theta_I(0)\theta_I(2\epsilon_+)
}{
4 \theta_1(\epsilon_1) ^2 \theta_1(\epsilon_2) ^2
\theta_I(\epsilon_1)  \theta_I(\epsilon_2)
\theta_J(\epsilon_+\pm \alpha)
\theta_K(\epsilon_+\pm \alpha)
}
\cdot\prod_{l=1}^{10}\frac{
\theta_J(m_l) \theta_K (m_l)
}{\eta^2}
\cr
\end{align}
where $S = \{ (2,2,1), (3,3,1), (4,4,1), (2,3,4),(3,4,2),(4,2,3) \}$. \\
\noindent For $k=3$, 
\begin{align}
Z_{3}
= &
- \sum_{i=1}^4 \sum_{s=1}^2
\frac{
\theta_{J_{Ii}} (\pm \frac{\epsilon_{s}}{2})
\theta_{J_{Ii}} (2 \epsilon_+ \pm \frac{\epsilon_{s}}{2})
\prod_{l=1}^{10}
\theta_I(m_l)
\theta_i (m_l \pm \frac{\epsilon_{s}}{2})
}{
2 \eta^{18} \theta_1(\epsilon_{1,2})^2 
\theta_1(2 \epsilon_s)  \theta_1(2 \epsilon_+ - 2 \epsilon_s) 
\theta_{J_{Ii}} (\epsilon_{1,2} \pm \frac{\epsilon_{s}}{2})
\theta_I(\epsilon_+ \pm \alpha )
\theta_i (\epsilon_+ \pm \alpha \pm \frac{\epsilon_{s}}{2})
}
\cr
& 
 - \sum_{s=1}^2 
 \frac{
\theta_{1} (2 \epsilon_+ \pm \epsilon_{s})
\theta_{1} ( \pm \epsilon_{s})
\prod_{l=1}^{10}  \theta_I(m_l) \theta_I (m_l \pm \epsilon_{s})
}{
\eta^{18} 
\theta_1(\epsilon_{1,2})^2 
\theta_{1} (2 \epsilon_s) 
\theta_{1} (2 \epsilon_+ - 2 \epsilon_s)
\theta_1(\epsilon_{1,2} \pm 2\epsilon_s) 
\theta_I(\epsilon_+ \pm \alpha )
\theta_I (\epsilon_+ \pm \alpha \pm \epsilon_s)
}
\cr
& - \sum_{s=1}^2 
 \frac{
\theta_I(2 \epsilon_+ \pm (\epsilon_{+} + (-1)^s \alpha ))
\theta_I(\pm (\epsilon_{+} + (-1)^s \alpha ))
}{
 \eta^{18} \theta_1(\epsilon_{1,2})^2 \theta_I(\epsilon_+ \pm \alpha )
\theta_I(\epsilon_{1,2} \pm (\epsilon_{+} + (-1)^s \alpha )) 
\theta_1(\epsilon_{1,2} \pm 2(\epsilon_{+} + (-1)^s \alpha ))
}
\cr
&  \qquad \qquad \times 
\frac{
\prod_{l=1}^{10} 
\theta_I(m_l) \theta_1(m_l \pm (\epsilon_{+} + (-1)^s \alpha ))
}{
\theta_{1} (2 \epsilon_+ + 2(-1)^s \alpha) 
\theta_{1}( 2 (-1)^{s+1} \alpha )
}
\cr
&  - \frac{1}{\eta^{18}} \prod_{k=2}^4
\frac{
\theta_{k}(0) \theta_{k}(2 \epsilon_+) 
}{
\theta_{k}(\epsilon_{1,2}) 
}
\prod_{I \neq I'}
\frac{
\prod_{l=1}^{10}  \theta_{I} (m_l)
}{
\theta_1(\epsilon_{1,2}) 
\theta_{I} (\epsilon_+ \pm \alpha)
}
\end{align}
with
\begin{align}
J 
= \left(
\begin{array}{cccc}
1&2&3&4 \\
2&1&4&3  \\
3&4&1&2 \\
4&3&2&1
\end{array}
\right).
\end{align}
Then, the partition function is give as
\begin{align}
Z = Z_0 \sum_{k=1}^{\infty} Z_k \phi^k.
\end{align}
Here, $\phi$ is the tensor branch moduli, which is related to the invariant Coulomb moduli of 5d gauge theory $\tilde{A}_1$ as
\begin{align}
\phi 
= \tilde{A}_1 e^{-\pi i \tau}
\end{align}
Also, we identify 
\begin{align}
e^{-\alpha} = \tilde{A}_2.
\end{align}

Suppose we rewrite the partition function in the form
\begin{align}
Z = \text{PE} \left( \sum_{k=0}^{\infty} \F_k \tilde{A}_1{}^k \right), 
\end{align}
with $\F_0$ given in \eqref{eq:Z0}.
$\F_k$ ($k=1,2,3$) can be written in terms of $Z_k$ ($k=1,2,3$) as
\begin{align}\label{eq:FkZk}
\F_1 &= Z_1 e^{- \pi i \tau},
\cr
\F_2 & = \left( Z_2 -\frac12 Z_1{}^2 - \frac12 Z_1 (* \to 2*) \right) e^{- 2 \pi i \tau},
\cr
\F_3 & = \left( 
Z_3 + \frac13 Z_1{}^3 - Z_1 Z_2 - \frac13 Z_1(* \to 3*)
\right) e^{- 3 \pi i \tau}.
\end{align}
Here, $(* \to n*)$ means the following simultaneous rescaling of all the variables
\begin{align}
(\alpha, m_i, \epsilon_s,\tau) \to (n \alpha, nm_i, n\epsilon_s,n\tau).
\end{align}
We find that elliptic genus $Z_k$ $(k\ge 1 )$ can be expanded in terms of $e^{\pi i \tau}$ as 
\begin{align}\label{eq:Zkn}
Z_k = \sum_{n=0}^{\infty} Z_k^{(2n-k)} e^{(2n-k)\pi i \tau} 
\end{align}
Therefore, $\F_k$ can be also expanded in terms of $e^{\pi i \tau}$ as 
\begin{align}\label{eq:Fkn}
\F_k = \sum_{n=0}^{\infty} \F_k^{(2n-k)} e^{(2n-2k)\pi i \tau} 
\end{align}

The decoupling limit is 
\begin{align}
e^{\pi i \tau} \to 0
\end{align} 
while fixing $\tilde{A}_1$ (instead of $\phi$).
Before taking this limit,
the terms with negative power of $e^{\pi i \tau}$ in $\F_k$ should be converted into 
the terms with positive power by using the transformation
\begin{align}\label{eq:floptr}
\text{PE} \left( \frac{[j_L,j_R]_{t,q} }{ ( t^{\frac{1}{2}} - t^{-\frac{1}{2}} ) ( q^{\frac{1}{2}} - q^{-\frac{1}{2}} ) } Q \right) 
\to 
\text{PE} \left( \frac{[j_L,j_R]_{t,q} }{ ( t^{\frac{1}{2}} - t^{-\frac{1}{2}} ) ( q^{\frac{1}{2}} - q^{-\frac{1}{2}} ) } Q^{-1} \right),
\end{align}
which is interpreted as performing the flop transition.
Then, the decoupling limit ends up with picking up the constant term:
\begin{align}
\F_k \to \F_k^{(0)}.
\end{align}
%
Therefore, the partition function for 5d $\mathcal{N}=1$ $Sp(2)$ gauge theory with $N_f=9$ flavors is given as
\begin{align}
Z = \text{PE} \left( \sum_{k=0}^{\infty} \F_k^{(0)} \tilde{A}_1{}^k \right).
\end{align}
The each coefficient $\F_k^{(0)}$ is given in terms of $Z_k^{(n)}$ introduced in \eqref{eq:Zkn} as 
\begin{align}
\F_1^{(0)} &= Z_1^{(1)},
\cr
\F_2^{(0)} &= Z_2^{(2)} - \frac12 ( Z_1^{(1)} )^2 - Z_1^{(-1)} Z_1^{(3)} - \frac12 Z_1^{(1)} (* \to 2*) , 
\cr
\F_3^{(0)} &= Z_3^{(3)} +
\left( \frac13 (Z_1^{(1)})^3 + 2 Z_1^{(-1)} Z_1^{(1)} Z_1^{(3)} + (Z_1^{(-1)})^2 Z_1^{(5)} \right)
\cr
& \qquad - \left( Z_1^{(-1)} Z_2^{(4)} + Z_1^{(1)} Z_2^{(2)} + Z_1^{(3)} Z_2^{(0)} + Z_1^{(5)} Z_2^{(-2)} \right)
- \frac13 Z_1^{(1)} (* \to 3*),\quad
\end{align}
where we used \eqref{eq:FkZk}. 
Also, $\F_0^{(0)}$ can be read off from \eqref{eq:Z0} as
\begin{align}
\F_0^{(0)} 
&= 
\frac{[0,\frac12]_{t,q} ( e^{ - 2 \alpha} + e^{ 2 \alpha}  ) +  [0,0]_{t,q}  (e^{ - \alpha} +e^{ \alpha}) \sum_{i=0}^9 (e^{-m_i} + e^{m_i}) }{ 2 ( t^{\frac{1}{2}} - t^{-\frac{1}{2}} ) ( q^{\frac{1}{2}} - q^{-\frac{1}{2}} ) } 
\cr
& \to
\frac{[0,\frac12]_{t,q} \tilde{A}_2{}^2 + [0,0]_{t,q} \tilde{A}_2 \chi_{\bf{20}} }{ ( t^{\frac{1}{2}} - t^{-\frac{1}{2}} ) ( q^{\frac{1}{2}} - q^{-\frac{1}{2}} ) } ,
\end{align}
where we used the flop transition \eqref{eq:floptr}.
From these expressions, we obtain the partition function for 5d $\N=1$ $Sp(2)$ gauge theory with $N_f=9$ flavors given in section \ref{sec:Rank2GV-prep}.

\bibliographystyle{JHEP}
\bibliography{ref}
\end{document}